\documentclass[12pt,a4paper]{JHEP3}
\usepackage{amsmath,epsfig}
\usepackage{amssymb,amsfonts}
\usepackage{latexsym}
\usepackage[latin1]{inputenc}
\usepackage{slashed}
\usepackage{empheq}
\numberwithin{equation}{section}
\usepackage{cancel}
\usepackage{overpic}
\usepackage{subcaption}
\usepackage{subeqnarray}
\usepackage{xcolor}
\let\normalcolor\relax
\usepackage{longtable}
\usepackage{color}
\usepackage{multirow}
\usepackage{epstopdf}
\usepackage{caption,graphicx,color}
\usepackage{amsmath,epsfig}
\usepackage{amssymb,amsfonts}
\usepackage{latexsym}
\usepackage[latin1]{inputenc}
\usepackage{float}
\usepackage{overpic}
\usepackage{slashed}
\usepackage{caption}
\usepackage{subcaption}
\usepackage{xcolor}
\let\normalcolor\relax
\usepackage{graphicx}
\usepackage{longtable}
\def\qfor{~~{\rm for}~~}
\relax
\renewcommand{\theequation}{\arabic{section}.\arabic{equation}}

\def\be{\begin{equation}}
\def\ee{\end{equation}}

\newcommand{\bear}{\begin{eqnarray}}
\newcommand{\bea}{\begin{eqnarray}}
\newcommand{\eear}{\end{eqnarray}}
\newcommand{\eea}{\end{eqnarray}}

\def\bsq{\begin{subequations}}
\def\esq{\end{subequations}}
\def\hri#1#2{\href{http://arxiv.org/abs/#1}{[ArXiv:#1]#2}}
\def\hre#1#2{\href{http://arxiv.org/abs/#1/#2}{[ArXiv:#1/#2]}}

\def\hrj#1#2{\href{www.doi.org/#1}{#2}}
\def\hree#1#2{\href{https://doi.org/#1}{#2}}
\newbox\pippobox

\def\II{\relax{\rm I\kern-.18em I}}

\def\e{\epsilon}
\def\l{\lambda}
\def\m{\mu}
\def\n{\nu}
\def\r{\rho}

\def\s{\sigma}
\def\pa{\partial}

\def\sp{\;\;\;,\;\;\;}
\def\spp{\,,\quad }

\def\f{\varphi}

\def\a{\alpha}
\def\b{\beta}

\def\k{\kappa}

\def\ar{{~~~\Rightarrow~~~}}

\def\nn{\nonumber}

\title{Holographic CFTs on $AdS_d\times S^n$ and conformal defects}
\author{ Ahmad Ghodsi$^a$, Elias Kiritsis$^{b,c}$ and Francesco Nitti$^{b}$ \\

$^a$
 Department of Physics, Faculty of Science,	Ferdowsi University of Mashhad,  Mashhad, Iran.
~\\

$^b$ \href{http://www.apc.univ-paris7.fr}{Universit\'e Paris Cit\' e, CNRS, Astroparticule et Cosmologie}, F-75013 Paris, France.
~\\

$^c$ \href{http://hep.physics.uoc.gr}{Crete Center for Theoretical Physics}, Institute for Theoretical and Computational Physics,
Department of Physics\\
University of Crete, Heraklion, Greece
}

\preprint{CCTP-2023-6\\ITCP-2023/6}

\abstract{We consider ($d+n+1$)-dimensional solutions of Einstein gravity with constant negative curvature.
Regular solutions of this type are expected to be dual to the ground states of ($d+n$)-dimensional holographic CFTs on $AdS_d\times S^n$.
Their only dimensionless parameter is the ratio of radii of curvatures of $AdS_d$ and $S^n$. The same solutions  may also be dual to $(d-1)$-dimensional conformal defects in holographic QFT$_{d+n}$.
We solve the gravity equations with an associated conifold ansatz, and we classify all solutions both singular and regular by a combination of analytical and numerical techniques.
There are no solutions, regular or singular, with two boundaries along the holographic direction.
Out of the infinite class of regular solutions, only one is diffeomorphic to $AdS_{d+n+1}$ and another to $AdS_d\times AdS_{n+1}$.
For the regular solutions, we compute the on-shell action as a function of the relevant parameters.
}

\begin{document}


\section{Introduction, results and outlook}

Quantum field theories are usually considered in flat background space-time.
They can be studied, however, in background space-times that have non-zero curvature. Space-time curvature is irrelevant in the UV, as at short distances any regular manifold is flat. However, the curvature is relevant in the IR and can affect the low-energy structure of the QFT.

There are several reasons to consider QFT in curved backgrounds.
\begin{itemize}

\item Partition functions of QFTs on compact manifolds (spheres), are important elements in the study of the monotonicity of the RG Flow and the definition of generalized C-functions, especially in odd dimensions, \cite{myers,J,F}.

\item Many observables in CFTs and other massless QFTs (supersymmetric indices are examples) are well-defined when a  mass gap is introduced. This can be generated by putting the theory on a positive curvature manifold, like a sphere.
     Sphere compactifications have been used in calculating supersymmetric indices in CFTs, \cite{komar}. They have also been used as regulators of  IR divergences of perturbation theory in QFT, \cite{Adler,JR,CW} and string theory, \cite{KK}.

\item Curvature in QFT, although UV-irrelevant is IR-relevant and importantly affects the IR physics. It can drive (quantum) phase transitions in the QFT, \cite{Bu,C}.

\item The ground-states of holographic QFTs on curved manifolds lead to constant (negative) curvature metrics sliced by curved slices.
    The Fefferman-Graham theorem indicates that such regular metrics exist near the asymptotically $AdS$ boundary, \cite{PG}. However, it is not known whether such solutions can be extended to globally regular solutions in the Euclidean case. If yes, then there may exist associated Minkowski signature solutions with horizons\footnote{Such metrics have been discussed in section 5 of \cite{CdL}.}. The few (mathematical) facts that are known can be found in \cite{WY,Ander}.

Holography suggests that because we can put any holographic CFT on any manifold we choose, there should be dual regular saddle point solutions. This argument has, however, a loophole: it may be that for a regular solution to exist, more bulk fields need to be turned-on (spontaneously), via asymptotically vev solutions\footnote{A milder version of this phenomenon associated with spontaneous symmetry breaking of a parity-like $Z_2$ symmetry has been observed in \cite{S2xS2}.}.

\item Cosmology has always given a motivation to study QFT in curved space-time, \cite{Fulling,BD}.
    In particular, QFT in de Sitter or almost de Sitter space is expected to describe early universe inflation as well as the current acceleration of the universe.

\item The issue of quantum effects in approximate de Sitter backgrounds is a controversial issue even today, \cite{Mottola}--\cite{GKNW2}.

\item Partition functions of holographic QFTs on curved manifolds are important building blocks in the no-boundary proposal of the wave-function of the universe, \cite{Hertog,hh}. They serve to determine probabilities for various universe geometries.

\end{itemize}

{Many examples of holographic QFTs living on non-trivial geometries have been already discussed in the past.}

The simplest case of $(S^1)^n$ has already been systematically studied in the case where all circles have the same radius as well as when there are two different radii, \cite{exotic,mateos,Gur}.

The case of $S^1\times S^{d-1}$ has been studied extensively but not systematically. It contains $AdS_{d+1}$ in global coordinates, {as well as (Euclidean) Schwarzschild-$AdS$},   and some RG flows have been analyzed in this case.

A systematic analysis of curved space-time holographic RG flows in Einstein-dilaton theories has been initiated in \cite{C}, when the boundary field theory is defined on an Einstein space with positive or negative curvature. For positive curvature, the RG flow pattern is not very different from that of flat space field theories. The main difference is that curvature dominates in the  IR and provides a gap to the theory before the deep IR regime is reached. On the other hand, many quantum phase transitions appear, driven by the positive curvature.

The general problem where the boundary is a product of constant (positive) curvature manifolds and the QFT is a CFT has been addressed in \cite{aharony}.  Phase transitions were found, generalizing the Hawking-Page transition (which is relevant in the $S^1\times S^{d-1}$ case), \cite{HP}. Efimov resonances were also found that were explored in \cite{Horo} to generate a class of associated black hole solutions.
The general case of QFTs on $S^2\times S^2$ was addressed in \cite{S2xS2}. Among other things, it was found that a $Z_2$ parity-like symmetry that exists when the two spheres have the same size is always spontaneously broken by quantum effects. Therefore the vacuum is always doubly degenerate.

In the case where the boundary has negative curvature, however, the holographic QFT interpretation of the solutions is {\em very} different from that of a standard  RG flow.  The reason is that, when the bulk is foliated by constant negative curvature $d$-dimensional slices, the solution has {\em two} asymptotically $AdS_{d+1}$ boundaries.  This corresponds to two UV CFTs that are interacting through the bulk.

Solutions in string theory, with asymptotic boundary metrics being $AdS$,  have been studied for some time, \cite{Bak}--\cite{ads}.
They have two (apparently) distinct conformal boundaries at the two end-points of the holographic coordinate. However, as the slices involve a non-compact manifold, which has also a conformal boundary, the two boundaries are connected. This results in a single conformal boundary.

If the bulk is $d+1$ dimensional, and the slices are $AdS_d$, the total boundary
is conformal to two pieces of $S^d$ separated by an overlap on the equator\footnote{In the context of holography, this description is most appropriate when the bulk $AdS_{d+1}$ is written in global coordinates, see \cite{ads}. }  $S^{d-1}$. The two endpoints of the flow can have different sources, the two holographically-dual theories can have different couplings and they are separated by an interface, justifying the name ``Janus solutions".
A similar class of solutions contains a single boundary and is delimited in the bulk by a brane that ends on ``the boundary of the boundary". They are also $AdS$-sliced and a prototypical example was discussed in \cite{KR}. They have been proposed as holographic duals of boundary CFTs, \cite{BCFT1,BCFT2}. Related holographic RG flows have been considered in  \cite{Gutperle:2012hy, Arav:2020asu}.

There is another incarnation of such solutions. In Euclidean cases, where the slice manifold is a constant negative curvature manifold with finite volume and no boundary,  such a solution is an example of a Euclidean wormhole. This is an object that still holds mysteries for the holographic correspondence, \cite{MM,BKP,SR,BKP2}.
The holographic interpretation of such solutions is still debated and for this reason, their occurrence is also an interesting datum.

$AdS$-sliced solutions were studied systematically in \cite{ads} with three purposes
\begin{itemize}

\item The holographic construction of QFTs on $AdS$ manifolds.

\item The exploration of the space of holographic interfaces.\

\item The study of ``proximity of QFTs" defined by which ones can be connected by wormholes.

\end{itemize}

A specific potential landscape was fully analyzed by a combination of analytical and numerical methods.  It was found that the solution space contained many exotic RG flow solutions that realized unusual asymptotics, as boundaries of different regions in the space of solutions. Phenomena like ``walking" flows and the generation of extra boundaries via ``flow fragmentation" were found.

The purpose of the present paper is to
 pursue the research program started in \cite{exotic} and \cite{C}, and
to study a further example along similar lines: holographic CFTs on product manifolds of the type\footnote{All our results are valid if we replace $AdS_d$ with any $d$-dimensional negative constant curvature manifold, with or without finite volume. A similar statement holds for $S^n$.} $AdS_d\times S^n$.
Such manifolds are interesting as they combine a piece that has constant negative curvature and one that has constant positive curvature.

 {Moreover, these geometries are also interesting since upon (generalized) dimensional reduction on $S^n$ they give rise to the infrared region of confining field theories defined on $AdS_d$ \cite{GK,Gouteraux:2011qh}. This connection, and the space of solutions of the reduced theory,  will be thoroughly analyzed in a forthcoming work.}

\subsection{Results}

We consider an Einstein theory with a negative cosmological constant in $d+n+1$ dimensions. The ansatz used is a conifold ansatz that contains a holographic (radial) coordinate and a product of a $d$-dimensional constant negative curvature manifold and an $n$-dimensional constant positive curvature manifold.
\be\label{metric}
ds^2=du^2+e^{2A_1(u)} ds^2_{AdS_d} + e^{2A_2(u)}ds^2_{S^n}\,.
\ee
 The solutions should have a  $d+n$-dimensional conformal boundary, where a holographic CFT lives.
In this context, we obtain and solve the equations of motion and compute the scalar curvature invariants, which are necessary ingredients to check the regularity/singularity of the solutions.

\begin{itemize}

\item {\bf{Classification of the solutions:}}
 We classify the solutions according to their ``end-points,'' which we define as limiting values of the radial coordinate of the conifold. A detailed analysis shows that we have four classes of end-points:
\begin{enumerate}
\item An $AdS$-like boundary where the scale factors of $AdS$ and the sphere diverge. We shall denote this end-point as {\bf B}.

\item A regular end-point where the scale factor of the sphere shrinks to zero sizes while the $AdS$ factor asymptotes to a constant value.
We shall denote this end-point as {\bf R}.

\item A singular end-point in which the size of $AdS$ vanishes while the size of the sphere diverges. We shall denote this end-point as {\bf A}.

\item A singular end-point in which the size of the sphere vanishes while the size of $AdS$ diverges.  We shall denote this end-point as {\bf S}.
\end{enumerate}

Only the first two of the four end-points correspond to a regular geometry.
A solution is characterized by its two end-points along the radial (holographic) direction. We denote the class of a solution by its two end points, ie. ({\bf B, R}) or ({\bf B, S}), etc.

In addition to end-points, a generic solution may or may not have an {\em A-bounce:} this is a stationary point of one or both of the scale factors which then displays a local minimum or maximum away from the end-points.

By analyzing the behavior of the scale factors near the A-bounces we recognize that the $AdS_d$ or $S^n$ can have at most one A-bounce. This restricts the classes of solutions with the above-mentioned end-points.

Our  analytical and numerical analysis leads to   the following results:

\begin{enumerate}

\item There is only one class of solutions that are everywhere regular: these are solutions that have one regular end-point and one $AdS$-like boundary, i.e.  ({\bf B, R}).

\item If a solution has an A-bounce, then it also has at least one singular end-point. Therefore, we do not find any regular wormhole-like solution.

\item There do not exist solutions in which at both end-points the scale factor of the sphere shrinks to zero sizes, ie. solutions of the type ({\bf R, R}), ({\bf R, S}) and    ({\bf S, S}) do not exist.

\item We find two exact solutions of the Einstein equations: one of them is the global $AdS_{d+n+1}$ space-time; the other is the product solution $AdS_d\times AdS_{n+1}$.

\end{enumerate}

\item {\bf{Space of solutions:}}
The space of solutions is three-dimensional as three initial conditions are needed to solve the equations. From the holographic point of view,  these correspond to the two curvature scales of $AdS_d$ and $S^n$ and one vev parameter of the dual stress-energy tensor. However, one of these parameters can be scaled out and the physics of such solutions depends on two dimensionless parameters. They can be taken as the ratio of curvatures of $AdS_d$ and $S^n$ and the associated ratio for the vev.

We analyze the transition between the above-mentioned solutions in the parameter space. In this space, we can follow how different regular/singular solutions change to each other as we move inside this space. There is a codimension-one subspace (a two dimensional surface) for regular solutions which ends on one side to the product space solution.

\item {\bf{QFT data on the boundary:}}
The Fefferman-Graham expansion near the $AdS$-like boundary (UV boundary) contains three parameters: two of them are the $AdS$ and sphere curvatures $(R^{UV}_{AdS}, R^{UV}_{S})$. Since the dual CFT is conformally invariant, the physics only depends on the ratio of these curvatures. The last parameter ($C$), is related to the vacuum expectation value and corresponds to parts of the vev of the components of the stress tensor. We can construct another dimensionless parameter from $C$ and one of the $AdS$ or sphere curvatures.  Overall, we have two dimensionless ratios that describe the holographic QFT on the conformal boundary of a bulk solution.
The value of $C$ depends on the data of the IR end-point. Here the IR is the location of the regular end-point and the only relevant parameter remaining is the curvature of the $AdS$ slice at this point. The value of $C$ for the product space solution,
$AdS_d\times AdS_{n+1}$,
 diverges and for the global $AdS$ space solution, it is zero as expected. For other regular solutions, it can be a positive or a negative number.

\item {\bf{Free energy:}}
The computation of the free energy for regular solutions shows that among these solutions,  the global $AdS$ solution has the maximum value.
This implies that if one constructs the no-boundary wave-function along the lines of \cite{Hertog} the global $AdS$ solution is the least probable state.

\end{itemize}

All the previous conclusions hold when $n>1$, as in this case, the sphere has non-zero positive curvature. The case $n=1$ needs a separate analysis that is performed in section \ref{s1}. The $S^{1}$ can be interpreted as a Euclidean time, and the structure of the solutions is that of a black hole with a hyperbolic horizon.
Such black holes are known as topological back holes, \cite{top1,top2}. In this case, we only have the following classes of solutions:

1. The regular solutions of ${\bf{(R,B)}}$ type. {This describes the solution outside the horizon of the black hole i.e. stretched from the horizon to the asymptotic boundary.}

2. The singular solutions of ${\bf{(R,A)}}$ type. {This describes the solution behind the horizon of the black hole i.e. stretched from horizon to singularity.}

3. The singular solutions of ${\bf{(A,B)}}$ type. {This describes a solution that is stretched from singularity to boundary (solutions with a naked singularity).}

The regular solutions appear in two classes:

$\bullet$ Black holes with two horizons (one event and one Cauchy horizon). In the limit where the two horizons coincide, we have an extremal black hole solution.

$\bullet$ Solutions with a single horizon. At the boundary of these solutions is the global AdS$_{d+2}$ solution.

Known facts about topological black holes are collected in appendix \ref{topo}.

We finally remark, that the techniques of the conifold ansatz with constant curvature slices can be used to find solutions at higher dimensions while solving only ODEs. It is not clear whether this algorithm captures all negative constant curvature metrics.

 \subsection{Conformal Defects}

There is another context where conifold  geometries  with $AdS\times S$ slices are relevant, namely in the study of conformal defects, \cite{Def1}-\cite{JR2}.

Consider a $D$-dimensional QFT$_D$, with a $d$-dimensional defect in it. If the QFT$_D$ is defined on flat space then its generic symmetry is $ISO(D)$. If it is a CFT$_D$, the symmetry is enhanced
to conformal symmetry, i.e. $O(D+1,1)$. Consider now a $d$-dimensional flat space defect, in QFT$_D$, localized on a $d$-dimensional hyperplane in $R^D$.
 The symmetries that remain unbroken by the defect that is assumed to be a flat  $d$-dimensional hyperplane, are $ISO(d)\times SO(D-d)$.
 If the defect is conformally invariant on the d-dimensional world-volume\footnote{The generic case is that the bulk theory is a QFT$_D$ without conformal invariance, but that the defect theory is tuned to be conformally invariant. Examples of such theories can be found in \cite{JR2}. The most common case, however, studied in the literature is that where the theory in the bulk is a CFT$_D$.}  then $ISO(d)$ is enhanced to $O(d+1,1)$ and the total symmetry becomes $O(d+1,1)\times  SO(D-d)$.

 In a holographic theory such a symmetry will be geometrically realized by a  $AdS_{d+1}\times S^{D-d-1}$ manifold\footnote{ Interestingly, the flat $D$-dimensional metric is conformal to  the metric of $AdS_{d+1}\times S^{D-d-1}$ with the $d$-dimensional defect being identified with the $d$-dimensional boundary of $AdS_{d+1}$.}.

 A special case is a conformal interface that has $d=D-1$. In that case, the symmetry becomes $O(D,1)$ and is geometrically realized by $AdS_D$.
 Moreover, $SO(1)$ is realized by $S^0$ which are two distinct points (and this explains why in this case we have two boundaries).
 The holographic dual of this is given by holographic solutions with the $(D+1)$-dimensional metric to be a conifold with $AdS_{D}$ slices realizing the aforementioned symmetry.

 Similarly, in the case of general $d$, we expect that the holographic ansatz will be  a  $(D+1)$-dimensional conifold with $AdS_{d+1}\times S^{D-d-1}$ slices.
Therefore, the holographic  ansatz we study in this paper is expected to also describe conformal $d$-dimensional defects in a holographic QFT$_D$.
In particular, the structure of the generic solutions is such that their boundary has two components. One is the boundary of the total space, and this is conformal
to $AdS_{d+1}\times S^{D-d-1}$, which is also conformal to flat space\footnote{There is a conical singularity around the defect if the curvatures of  $AdS_{d+1}$ and
$S^{D-d-1}$ are not the same.}. There is another boundary, namely the union of the boundaries of the $AdS_{d+1}$ slices. Insertions on that boundary correspond to defect operators.

The bulk operators are in one-to-one correspondence with the gravitational fields, and their correlators are calculated by putting Dirichlet boundary conditions at the
$AdS_{d+1}\times S^{D-d-1}$ boundary\footnote{When there are non-trivial dynamical degrees of freedom on the defect this ceases to be true.}.  The defect operators are in one-to-one correspondence again with the bulk gravitational fields but their correlators are now determined by putting boundary conditions at  the boundary of $AdS_{d+1}$. Clearly, this picture describes defects that do not carry additional degrees of freedom.

 The special analytic solutions found in this paper are interesting from this point of view. We consider the case $n>1$ that corresponds to defects with codimension $D-d\geq 3$.
 The global $AdS_{D+1}$ solution seems to imply that the defect does not back-react in the induced CFT geometry as the total space is the same as the holographic dual of a CFT without the defect.
 Therefore this seems to correspond to trivial conformal defects associated with the identity operator of the CFT.

The $AdS_{d+1}\times AdS_{D-d}$    solution, on the other hand,  seems to imply a complete decoupling between the defect and its transverse space.
The boundary structure of this solution is different and it has two independent boundaries that in Poincar\'e coordinates are $R^d$ and  $S^{D-d-1}$.
Insertions on these boundaries provide correlators for the defect and its transverse theory. Obviously, these correlators are completely independent.
In particular, all one-point functions vanish.
The study of small graviton fluctuations around this geometry indicates that  there is no flow of energy between defect and bulk.

In the case of $n=1$ or $D-d=2$, again the global $AdS_{D+1}$ solution should correspond to trivial defects.
On the other hand, the product solution is now   $AdS_{d+1}\times \mathcal{M}_2$ where $\mathcal{M}_2$ are the three spaces $EAdS_2^{\pm,0}$ described in section \ref{exact}.
They have one or two $AdS_2$ boundaries. In analogy with extremal black holes whose horizon contains $AdS_2$ factors, we would expect also here similar phenomena: a one-dimensional scale invariance as well as a quantum mode that does not decouple at low temperatures.

Further analysis is needed in order to substantiate such claims.

\subsection{Outlook}

There is one more case of constant negative curvature manifolds that can be written as conifolds that remains to be systematically studied: that where the slices are products of negative curvature manifolds.

The regular solutions found here descend via dimensional reduction on $S^n$ to solutions of Einstein dilaton gravity with a dilaton potential that has confining asymptotics, \cite{GK}.
They imply the correct way of desingularizing the asymptotic singular solutions of the Einstein-dilaton theory. This is an interesting domain as it will teach us about confining theories on $AdS$.

Finally, the implications of our solutions for conformal defects need to be examined. There are several questions in this direction that involve quantitative questions
like correlation functions both in the bulk and the defect as well as the dynamics of symmetries broken by the defect.
In particular, an interesting question involves the construction of non-trivial defect flows in the holographic context.
This is in principle straightforward in the holographic context, as such flows will involve solutions that will depend on two radial coordinates, $u$ and the radial coordinate of the AdS slice.
The relevant boundary conditions are that the solutions are vev only at the $u$-boundary while they have sources on the slice AdS boundary. Special solutions of this type have been considered in \cite{C1}.   We plan to study this further in the near future.

The structure of this paper is as follows:

In section 2 we derive the equations of motion for a metric with a domain wall holographic coordinate and slices which in general are the product of Einstein manifolds. In section 3 we compute the asymptotic expansions near the boundary, singular and regular end-points for $AdS_d\times S^n$ slices. We also explore the possibility of having A-bounces in the scale factors of $AdS_d$ or $S^n$. In section 4, we present two exact solutions of the theory, the global $AdS_{d+n+1}$ and product space solution $AdS_d\times AdS_{n+1}$. In sections 5 and 6, we show all the numerical solutions that we found and how they are related to each other through a three-dimensional space of solutions. In section 7, we extract the boundary CFT data of the regular solutions and identify the dimensionless parameters that characterize the CFT. Using this data, we calculate the on-shell action and the renormalized free energy in section 8. In section 9, we focus on the special case of $AdS_d\times S^1$ and use a suitable coordinate transformation to obtain exact solutions of the equations of motion. Then we discuss their properties. In section 10, we comment on how to generalize our solutions to conifolds of conifolds.

\section{Constant negative curvature solutions with $AdS_d
\times S^n$ slices}

\subsection{The general conifold ansatz}

We consider an  Einstein theory in a $d+1$ dimensional bulk space-time parametrized by coordinates $x^a\equiv (u, x^\mu)$ where $u$ is the holographic coordinate.
The most general two-derivative action is
\be
S= M_P^{d-1} \int d^{d+1}x \sqrt{-g} \big(
R -\Lambda\big) + S_{GHY}\,,
\label{nA2}\ee
where $M_P$ is the $d+1$ dimensional Plank mass. In this action $g_{ab}$ is the bulk metric, $R$ is its associated Ricci scalar and $\Lambda$ is a cosmological constant. The surface term $S_{GHY}$ is the Gibbons-Hawking-York term at the space-time boundary (e.g. the UV boundary if the bulk is asymptotically $AdS$). The  bulk field equations of motion are given by
\begin{gather}
 R_{ab} -{\frac12} g_{ab} (R-\Lambda) =0\,.
\label{nFE1}
\end{gather}

We shall consider a (holographic)  boundary QFT defined
on a space that is a product of Einstein manifolds. The natural bulk
metric ansatz that preserves all the original symmetries of the boundary metric, is given in terms of a domain wall holographic coordinate $u$ and a conifold  ansatz  (for both Euclidean and Lorentzian signatures)
\begin{align}\label{eq:metric}
d s^2 = g_{ab} d{x^a} d{x^b} = d u^2 + \sum_{i=1}^n \mathrm{e}^{2A_i(u)} \zeta^i_{\alpha_i, \beta_i} d{x^{\alpha_i}} d{x^{\beta_i}} \,.
\end{align}
Here the geometry of the constant $u$ slices are products of $n$ Einstein manifolds, each with metric $ \zeta^i_{\alpha_i, \beta_i}$, dimension $d_i$ and coordinates $x^{\alpha_i}$, $\a_i=1,2,...,d_i$.
Each Einstein manifold is associated with a different scale factor $A_i(u)$, which depends on the coordinate $u$ only. Therefore,  every $d$-dimensional slice at constant $u$ is given by the product of $n$ Einstein manifolds of dimension $d_1,...,d_n$. This is the conifold ansatz.

Since $\zeta^i_{\mu \nu}$ are Einstein manifolds, the following relations hold
\begin{align}
\label{eq:Rzeta}
R^{(\zeta^i)}_{\mu \nu} = \kappa_i \zeta^i_{\mu \nu} \sp R^{(\zeta^i)} = d_i \kappa_i \, , \quad
\end{align}
where $\kappa_i$ is the (constant) scalar curvature  scale of the $i$th manifold and no sum on $i$ is implied. We have the identity
\be
\sum_{i=1}^n~d_i=d\,.
\label{z7}
\ee
In the case of maximal symmetry, the scalar curvatures are
\be \label{kappai}
\quad \kappa_i = \left\{
  \begin{array}{c l}\displaystyle{
   \hphantom{-} \frac{(d_i-1)}{\alpha_i^2} } & \quad {dS}_{d_i}~~{\rm or} ~~S^{d_i}\\
   0 & \quad \mathcal{M}_{d_i} \\
{\displaystyle- \frac{(d_i-1)}{\alpha_i^2}} & \quad {AdS}_{d_i}\\
  \end{array} \right. \, \ ,
  \ee
where $\alpha_i$ are associate radii and $\mathcal{M}_{d_i}$ denotes $d_i$-dimensional Minkowski space.

The non-trivial components of Einstein's equation from \eqref{nFE1} are
\begin{gather}
\label{eq:EOM1bis}
\Big(\sum_{k=1}^n d_k \dot{A}_k\Big)^2 -
\sum_{k=1}^n d_k \dot{A}_k^2 - \sum_{k=1}^n \mathrm{e}^{-2A_k}
R^{\zeta^k} + \Lambda  = 0\sp  uu
\\
 \label{eq:EOM4bis}
2(1 - \frac{1}{d}) \sum_{k=1}^n d_k \ddot{A_k} + \frac{1}{d} \sum_{i,
  j=1}^n d_id_j (\dot{A_i} - \dot{A_j})^2 + \frac{2}{d} \sum_{k=1}^n
\mathrm{e}^{-2A_k} R^{\zeta^k}  = 0\sp  ii \\
 \label{eq:EOM5bis}
\ddot{A_i} + \dot{A_i} \sum_{k=1}^n d_k \dot{A_k} - \frac{1}{d_i} \mathrm{e}^{-2A_i} R^{\zeta^i} = \ddot{A_j} + \dot{A_j} \sum_{k=1}^n d_k \dot{A_k} - \frac{1}{d_j} \mathrm{e}^{-2A_j} R^{\zeta^j}\sp  i\not= j
\end{gather}
where the derivatives with respect to $u$  are denoted by a dot.
The details of computations are found in appendix \ref{conii}.
The above equations are the same for both Lorentzian and Euclidean
signatures of the slices, so all our results hold for both cases.

Holographic saddle points are in one-to-one correspondence with the regular
solutions to the equations
\eqref{eq:EOM1bis}--\eqref{eq:EOM5bis}. Hence, in the following, we shall be
interested in the structure and properties of solutions to these equations, specifically for a  negative cosmological constant $\Lambda$.

To check the regularity of the solutions, we analyze scalar invariants of curvatures. For example (see appendix \ref{apk} for more details)
the Ricci scalar is given by:
\be
R = -2 \sum_{i=1}^n d_i \ddot{A_i} - \big(\sum_{i=1}^n d_i \dot{A_i}\big)^2 - \sum_{i=1}^n d_i \dot{A}_i^2 + \sum_{i=1}^n \mathrm{e}^{-2A_i} R^{\zeta^i}\,,
\label{zz6}
\ee
while the Ricci squared scalar reads
\be
R_{ab}R^{ab} = \Big(\sum^{n}_{i=1}d_{i}(\ddot{A}_{i} + \dot{A}^{2}_{i})\Big)^{2} +\sum_{i=1}^{n} d_{i}\Big(\mathrm{e}^{-2A_{i}}\kappa - \big(\ddot{A}_{i} + \dot{A}_{i}\sum^{n}_{j =1}d_{j}\dot{A}_{j}\big)\Big)^{2}\,. \label{zz5}
\ee
Moreover, the Kretschmann scalar,  $\mathcal{K}=R_{abcd}R^{abcd}$ is given by
\begin{align}\label{zz11}
\mathcal{K} &=
\sum^{n}_{i =1}\Big(e^{-4A_{i}}\mathcal{K}^{\zeta^{i}} - 4e^{-2A_{i}} \dot{A}_{i}^{2}R^{\zeta^{i}} - 2d_{i}\dot{A}_{i}^{4} \nn \\
&+ 4d_{i}(\ddot{A}_{i} + \dot{A}^{2}_{i})^{2}\Big) +  \sum^{n}_{i, j =1} 2d_{i}d_{j}\big(\dot{A}_{i}\dot{A}_{j}\big)^{2}\,,
\end{align}
where $\mathcal{K}^{\zeta^{i}}$ is the Kretschmann scalar of the  $\zeta^{i}$ metric.

\subsection{The $AdS_d\times S^n$ slice}
We now specialize the general conifold ansatz to the main subject of investigation of this paper, namely the bulk holographic description of  QFTs living on $AdS_d\times S^n$ space-time. The metric  (\ref{eq:metric}) in this case  is
\be\label{j1}
ds^2=du^2+e^{2A_1(u)}\zeta^{1}_{\a\b} dx^{\a} dx^{\b} + e^{2A_2(u)}\zeta^{2}_{\m\n} dx^{\m} dx^{\n}\,,
\ee
where $\zeta^1$ and $\zeta^2$ are the $AdS_d$ and $S^n$ metrics respectively.

We have set the dimensions of the Einstein manifolds to $d_1=d$ and $d_2=n$. The non-trivial components of Einstein's equation are
\be
 \label{j2}
 \big( d \dot{A_1}+n\dot A_2\big)^2 - d \dot{A}_1^2-n\dot{A}_2^2 - e^{-2A_1}
R_1- e^{-2A_2}R_2  + \Lambda  = 0\,,
\ee
\be
 \label{j3}
 (d+n-1) \big(d \ddot{A_1}+n\ddot{A_2}\big) +dn (\dot{A_1} - \dot{A_2})^2 + e^{-2A_1} R_1 +e^{-2A_2} R_2 = 0\,,
\ee
\be
 \label{j4}
 \ddot{A_1} + \dot{A_1} ( d\dot{A_1}+n\dot{A_2}) - \frac{1}{d} e^{-2A_1} R_1 = \ddot{A_2} + \dot{A_2} ( d\dot{A_1}+n\dot{A_2})- \frac{1}{n} e^{-2A_2} R_2\,,
\ee
where we have defined
$$R_1\equiv R^{\zeta^1}\sp R_2\equiv R^{\zeta^2}\;.$$
To check the regularity of the solutions we need to know the Kretschmann scalar from \eqref{zz11}.
In the geometry (\ref{j1}), it  is given by
\begin{align}\label{kerdn}
\mathcal{K}&=e^{-4A_1}\mathcal{K}_1+e^{-4A_2}\mathcal{K}_2-4 e^{-2A_1}R_1\dot{A}_1^2-4 e^{-2A_2}R_2\dot{A}_2^2+2d(d-1)\dot{A}_1^4\nn\\
&+2n(n-1)\dot{A}_2^4+4n d \dot{A}_1^2\dot{A}_2^2+4d(\ddot{A}_1+\dot{A}_1^2)^2+4n(\ddot{A}_2+\dot{A}_2^2)^2\,,
\end{align}
where $\mathcal{K}_1$ and $\mathcal{K}_2$ are the  Kretschmann scalars for $AdS_d$ and $S^n$ respectively
\begin{align}
\mathcal{K}_1=\frac{2}{d(d-1)}R_1^2\sp
\mathcal{K}_2=\frac{2}{n(n-1)}R_2^2\,.
\end{align}

\section{Regular and singular asymptotic of the solutions}

In the rest of this paper
we parametrize the value of the cosmological constant as
\be\label{pot0}
\Lambda=-\frac{1}{\ell^2}(d+n)(d+n-1)\,.
\ee
The gravitational equations  demand a  constant negative curvature Einstein manifold and read
\be
 \label{eq1}
 \big( d \dot{A_1}+n\dot A_2\big)^2 - d \dot{A_1}^2-n\dot{A_2}^2 - e^{-2A_1}
R_1- e^{-2A_2}R_2    = \frac{1}{\ell^2}(d+n)(d+n-1)\,,
\ee
\be
 \label{eq2}
 (d+n-1) \big(d \ddot{A_1}+n\ddot{A_2}\big) +dn (\dot{A_1} - \dot{A_2})^2 + e^{-2A_1} R_1 +e^{-2A_2} R_2 = 0\,,
 \ee
 \be
 \label{eq3}
 \ddot{A_1} + \dot{A_1} ( d\dot{A_1}+n\dot{A_2}) - \frac{1}{d} e^{-2A_1} R_1 = \ddot{A_2} + \dot{A_2} ( d\dot{A_1}+n\dot{A_2})- \frac{1}{n} e^{-2A_2} R_2\,.
\ee
In this section, we find the expansions of the $AdS_d$ and $S^n$ scale factors ($A_1$ and $A_2$) near the $AdS$ (UV) boundary, the end-points, and the A-bounces\footnote{A-bounces are points where any scale factor $A$ changes direction, i.e. $\dot A=0$.}. Using these expansions we can search and classify various regular and singular bulk solutions and extract the values of sources and vevs of the dual boundary CFTs.

\subsection{Near-boundary expansions}\label{NBE}

The Fefferman-Graham expansion of the \eqref{j1} metric near a UV boundary, which can  be  reached either as  $u\rightarrow +\infty$ or at $u\rightarrow -\infty$, is
\begin{align} \label{FGM}
ds^2 &=du^2+e^{\pm\frac{2u}{\ell}}(ds_{QFT}^2+\cdots)\nn \\
&=du^2+e^{\pm\frac{2u}{\ell}}\Big[e^{2\bar{A}_1}\zeta^{1}_{\a\b} dx^{\a} dx^{\b} + e^{2\bar{A}_2}\zeta^{2}_{\m\n} dx^{\m} dx^{\n}\Big]+\text{sub-leading}\,,
\end{align}
where $\bar{A}_1, \bar{A}_2$ are arbitrary constants.
Therefore,  the holographic CFT will be  living on a boundary with geometry $AdS_d\times S^n$, with metric given by the square  bracket in equation (\ref{FGM}) and with the corresponding curvatures  given by
\be \label{RUV12}
R_1^{UV}=e^{-2\bar{A}_1} R_1\sp R_2^{UV}=e^{-2\bar{A}_2} R_2\,.
\ee
In the expression above,  $R_1$ and $R_2$ are the scalar curvatures of the metrics $\zeta_1$ and $\zeta_2$ of $AdS_d$ and $S^n$, respectively. We  parametrize them by introducing the corresponding curvature radii
\be \label{R1R2}
R_1=-\frac{d(d-1)}{\a_1^2}\sp R_2=\frac{n(n-1)}{\a_2^2}\,,
\ee
where $\a_1$ and $\a_2$ are the associated radii of the $AdS$ and $S$ spaces.

As equations \eqref{eq1}--\eqref{eq3} show, we have two second-order equations plus one
first-order constraint for the two scale factors $A_1(u)$ and $A_2(u)$. This system has three integration constants. Two of them are shifts of $\bar{A}_1$ and $\bar{A}_2$ which can be fixed by demanding that $R_1$ and $R_2$ coincide with the actual curvatures of the manifold on which the UV boundary theory is defined according to the holographic dictionary, i.e. the relations in  \eqref{RUV12}.

The last integration constant enters at sub-leading order in the asymptotic (UV) expansion in \eqref{FGM}, and therefore it corresponds to a vacuum expectation value.
Since in our model, the only non-trivial bulk field is the metric, it must correspond
to parts of the vev of the components of the stress tensor. As we have shown in appendix \ref{SET} for the specific cases of $d=n=2$, this constant,  called $C$, appears in the expectation values of the stress-energy tensor of both $AdS_2$ and $S^2$, as seen in equations \eqref{TEMC1} and \eqref{TEMC2}.

The value of the third constant will be fixed once we impose the regularity in the interior.
Since the dual CFT is conformally invariant, the physics depends only on the ratio of the curvature scales of $AdS_d$ and $S^n$ which is the only dimensionless source parameter of our problem.

Solving the equations of motion (\ref{eq1})--(\ref{eq3}), near the putative boundary either at $u\rightarrow+\infty$ or $u\rightarrow-\infty$ gives expansions for scale factors of $AdS_d$ and $S^n$ spaces. For  $d=n=4$ we find the following expansions,\footnote{The formulae can be generalized to arbitrary $d,n$.}

\bsq
\begin{align}\label{near1}
& A_1(u)\!=\!\bar{A}_1\pm\frac{u}{\ell}-\frac{1}{2^4 3^1 7^1}  (5 \mathcal{R}_1 - 2 \mathcal{R}_2)e^{\mp\frac{ 2u}{\ell}}-\frac{1}{2^9 3^2 7^2} (46 \mathcal{R}_1^2 - 20 \mathcal{R}_1 \mathcal{R}_2 - 17 \mathcal{R}_2^2)e^{\mp\frac{ 4u}{\ell}}\nn \\
&-\big(\frac{1}{2^{12} 3^4 7^3}  (356 \mathcal{R}_1^3 -66 \mathcal{R}_1^2 \mathcal{R}_2 - 171 \mathcal{R}_1 \mathcal{R}_2^2 - 92 \mathcal{R}_2^3)\big)e^{\mp\frac{ 6u}{\ell}}\nn \\
&-\big(\frac{1}{2^{19} 3^4 7^4} (2111 \mathcal{R}_1^4 - 1160 \mathcal{R}_1^3 \mathcal{R}_2 - 1740 \mathcal{R}_1^2 \mathcal{R}_2^2 - 1160 \mathcal{R}_1 \mathcal{R}_2^3 + 2111 \mathcal{R}_2^4) + C \big)e^{\mp \frac{8u}{\ell}} \nn \\
&\pm \frac{1}{2^{16} 3^3 7^3} \big(23\mathcal{R}_1^4 - 52 \mathcal{R}_1^3 \mathcal{R}_2+52\mathcal{R}_1\mathcal{R}_2^3-23\mathcal{R}_2^4\big)\frac{u}{\ell}e^{\mp\frac{ 8u}{\ell}}+\mathcal{O}(e^{\mp\frac{10u}{\ell}})\,,\\
~\nonumber
\\
& A_2(u)\!=\!\bar{A}_2\pm \frac{u}{\ell}+\frac{1}{2^4 3^1 7^1}  (2 \mathcal{R}_1 - 5 \mathcal{R}_2)e^{\mp\frac{2u}{\ell}}+\frac{1}{2^9 3^2 7^2} (17 \mathcal{R}_1^2 + 20 \mathcal{R}_1 \mathcal{R}_2 - 46 \mathcal{R}_2^2)e^{\mp\frac{4u}{\ell}}\nn \\
&+\big(\frac{1}{2^{12} 3^4 7^3}  (92 \mathcal{R}_1^3 +171 \mathcal{R}_1^2 \mathcal{R}_2 +66 \mathcal{R}_1 \mathcal{R}_2^2 - 365 \mathcal{R}_2^3)\big)e^{\mp\frac{6u}{\ell}}\nn \\
&-\big(\frac{1}{2^{19} 3^4 7^4} (2111 \mathcal{R}_1^4 - 1160 \mathcal{R}_1^3 \mathcal{R}_2 - 1740 \mathcal{R}_1^2 \mathcal{R}_2^2 - 1160 \mathcal{R}_1 \mathcal{R}_2^3 + 2111 \mathcal{R}_2^4) - C\big)e^{\mp\frac{8u}{\ell}}  \nn \\
&\mp\frac{1}{2^{16} 3^3 7^3} \big(23\mathcal{R}_1^4 - 52 \mathcal{R}_1^3 \mathcal{R}_2+52\mathcal{R}_1\mathcal{R}_2^3-23\mathcal{R}_2^4\big)\frac{u}{\ell}e^{\mp\frac{8u}{\ell}}+\mathcal{O}(e^{\mp\frac{10u}{\ell}})\,.\label{near2}
\end{align}
\esq
Here $\mathcal{R}_1$ and $\mathcal{R}_2$ are  dimensionless curvature parameters, defined as
\be \label{CR1CR2}
\mathcal{R}_1\equiv \ell^2 R_1 e^{-2\bar{A}_1}=\ell^2 R_1^{UV}\sp
\mathcal{R}_2\equiv \ell^2 R_2 e^{-2\bar{A}_2}=\ell^2 R_2^{UV}\,.
\ee
 The constant $C$ that appeared in the above equations is proportional to the vev of the stress-energy tensor of the boundary CFT that we already discussed above, see also appendix \ref{SET} for more details. A similar argument can be found in \cite{S2xS2} for holographic CFTs on $S^2\times S^2$.

We also  note that the coefficients of $\frac{u}{\ell}e^{\mp\frac{8u}{\ell}} $ in \eqref{near1} and \eqref{near2} reflect the conformal anomaly in $d+n=8$ dimensions. To see more details in $d+n=4$ see appendix \ref{SET} or \cite{Henningson:1998gx, Skenderis:2002wp}.

\subsection{Regular and singular end-points}

We now study the geometry close to an (IR) end-point, i.e. a point   $u=u_0$ where one or both scale factors of the $AdS$ and  $S$ shrink to zero. At this point,  the $u$ direction terminates\footnote{If it is the $AdS_d$ scale factor that shrinks to zero, and the $AdS$ has Minkowski signature, this point is a horizon, \cite{CdL}. However, as we shall see, this kind of end-point is always singular.}. Such an end-point may be regular, or it may be a curvature singularity. In the latter case, from the point of view of holography, the associated solution has to be rejected.

Given such an endpoint, we now work out an expansion of the solution near it and compute the Kretschmann scalar. This will determine if this end-point is regular or singular.

To solve equations of motion near $u=u_0$ ($u\rightarrow u_0^+$), we consider the following expansions for scale factors\footnote{A power-law leading behavior, $A_1=\k_1 (u-u_0)^a+\cdots$ and $A_2=\k_2 (u-u_0)^b+\cdots$ in which $a,b<0$ cannot solve Einstein's equations near $u=u_0$, as one can show along the lines of \cite{S2xS2}. Other non-power-law behaviors do not produce solutions. }
\bsq
\begin{align}\label{gexp1}
 A_1(u) &= \l_1 \log\frac{u-u_0}{\ell}+\frac12 \log a_0  + a_1\frac{u-u_0}{\ell}+ a_2\frac{(u-u_0)^2}{\ell^2}+\mathcal{O}(u-u_0)^3\,,\\ \label{gexp2}
 A_2(u) &= \l_2 \log\frac{u-u_0}{\ell}+\frac12 \log s_0  + s_1\frac{u-u_0}{\ell}+ s_2\frac{(u-u_0)^2}{\ell^2}+\mathcal{O}(u-u_0)^3\,.
\end{align}
\esq
The constants appearing in the above expansions determine the behavior (regularity or singularity) of the end-point at $u=u_0$.

Inserting the first two leading terms in the above expansions into the equations of motion \eqref{eq1}--\eqref{eq3} we obtain
\be
\frac{(d+n-1) (d+n)}{ \ell^2}+\frac{\frac{r_1}{ \ell^2}}{(u-u_0)^{2\l_1}}+\frac{\frac{r_2}{ \ell^2}}{(u-u_0)^{2\l_2}} \label{t2ex3}
-\frac{(d\l_1+n\l_2)^2
-d\l_1^2-n\l_2^2}{(u-u_0)^2}+\cdots=0\,,\\
\ee
\be
d n (\l_1-\l_2)^2-(d+n-1) (d \l_1+\l_2 n)
+\frac{\frac{r_1}{ \ell^2}}{ (u-u_0)^{2\l_1-2}}+\frac{\frac{r_2}{ \ell^2}}{(u-u_0)^{2 \l_2-2}}+\cdots=0\,,\label{t2ex4}\\
\ee
\be
\frac{n \frac{r_1}{ \ell^2}}{(u-u_0)^{2\l_1-2}}-\frac{d \frac{r_2}{ \ell^2}}{(u-u_0)^{2\l_2-2}}
+d n (\l_2-\l_1)(d\l_1+n\l_2-1)+\cdots=0\,,\label{t2ex5}
\ee
where we have defined
\be
r_1\equiv\frac{\ell^2 R_1}{a_0}\sp r_2\equiv\frac{\ell^2 R_2}{ s_0}\,.
\ee

By an exhaustive analysis of the above equations for various regions of $\l_1$ and $\l_2$,  we  find the following possibilities for $\l_1$ and $\l_2$:
\begin{itemize}
\item Singular end-point: ($1>\l_1>0$ and $ 0>\l_2>-1$) or ($0>\l_1>-1$ and $ 1>\l_2>0$).
\item Regular end-point (sphere shrinking): $\l_1=0, \l_2=1$.
For other values of $\l_1$ and $\l_2$, for example, $\l_1>1$ or $\l_2>1$ or both, or for example when $\l_1=1, \l_2=0$ where the $AdS$ is shrinking to zero sizes, we find no solution for equations \eqref{t2ex3}--\eqref{t2ex5}.

\end{itemize}

When solving equations \eqref{t2ex3}--\eqref{t2ex5}, the values of $\l_1$ and $\l_2$ are fixed. Moreover, we find $a_0$ and $s_0$ (and $u_0$) as free parameters and
\bsq
\begin{gather}
a_1=s_1=0\,, \\
a_2=\frac{(d+n) \big(d (2 \l_1-2 \l_2-1)-n+1\big)}{4 d (\l_1-\l_2) (2 d \l_1+2 n\l_2 +1)}\,, \\
s_2=\frac{(d+n) \big(n(2\l_1 -2 \l_2 +1)+d-1\big)}{4 n (\l_1-\l_2) (2 d \l_1+2n \l_2 +1)}\,,
\end{gather}
\esq
and all higher coefficients of the expansion can be similarly determined.

\subsubsection{Singular end-points}\label{singsec}
We may have solutions that one of the scale factors shrinks but the other one blows up when $u\rightarrow u_0^+$. Here  we  find only two possible cases:
\begin{itemize}

\item $1>\lambda_{1}>0\,,\quad 0>\lambda_{2}>-1$:

In this case, the $AdS_d$ scale factor vanishes and the $S^n$ scale factor diverges. We have named this asymptotic $A_0S_{\infty}$.   We obtain
\begin{align}\label{adsshr}
\lambda_1 &=\frac{ \sqrt{d n (d+n-1)}+d}{d (d+n)}>0\sp
\lambda_2= \frac{ n-\sqrt{d n (d+n-1)}}{n (d+n)}<0\,.
\end{align}

\item $1>\lambda_{2}>0\,,\quad 0>\lambda_{1}>-1$:

In this class of solutions, the $AdS_d$ size is growing and the $S^n$ size is shrinking. We have named this asymptotic $A_{\infty}S_{0}$. We have the following solution
\begin{align}\label{sshr}
\lambda_1 &=\frac{d-\sqrt{d n (d+n-1)}}{d (d+n)}\sp
\lambda_2= \frac{ n+\sqrt{d n (d+n-1)}}{n (d+n)}\,.
\end{align}
For both cases above, the Kretschmann scalar is singular as $u\rightarrow u_0$
\begin{align}\label{ksings}
\mathcal{K} &=-\frac{4 \l_1^2 r_1}{\ell^4} \left(\frac{u-u_0}{\ell}\right)^{-2 \l_1-2}-\frac{4 \l_2^2 r_2}{\ell^4} \left(\frac{u-u_0}{\ell}\right)^{-2 \l_2-2}\nn \\
&+\mathcal{O}\Big(\frac{u-u_0}{\ell}\Big)^{-4\l_1}+\mathcal{O}\Big(\frac{u-u_0}{\ell}\Big)^{-4\l_2}\,.
\end{align}
\end{itemize}

\subsubsection{Regular end-points}\label{regsec}

Consider the case when the scale factor of $S^n$ shrinks to zero sizes as $u\rightarrow u_0^+$, but the $AdS_d$ has a finite size at this point, corresponding to ($\l_1=0,\l_2=1$). The position $u_0$ is arbitrary, as it can be changed by a shift in $u$ (which however may change the value of the near-boundary parameters).

Solving the equations of motion using the expansion \eqref{gexp1} and \eqref{gexp2},  we find the following expansions for the scale factors ($\l_1=0, \l_2=1$)
\bsq
\begin{align}\label{rega1}
e^{2A_1(u)} &= a_0 + \frac{a_0 d (d + n) + \ell^2 R_1}{d \ell^2 (1 + n)} (u-u_0)^2 \nn \\
 &-\frac{(a_0 d (d + n) +  \ell^2 R_1) (a_0 d (d - n-4) (d + n) + (d-3) \ell^2 R_1)}{ 3 a_0 d^2 \ell^4 (1 + n)^2 (3 + n)} (u-u_0)^4 \nn \\
 &+ \mathcal{O}(u-u_0)^6\,, \\
e^{2A_2(u)} &= \frac{R_2}{n(n-1)} (u-u_0)^2
+\frac{(a_0 (d - d^2 + n + n^2) - \ell^2 R_1) R_2}{3 a_0 \ell^2 n^2 (n^2-1)} (u-u_0)^4\nn \\
&+ \mathcal{O}(u-u_0)^6\,,\label{rega2}
\end{align}
\esq
which is valid for all values of $d,n>1$. The quantity $a_0$ is a non-zero positive (but otherwise arbitrary) constant.

Computing the Kretschmann scalar \eqref{kerdn} at $u=u_0$ we shows that
\be\label{rega3}
\mathcal{K}=\frac{2(d+n)^2}{\ell^4 n (n+1)}\Big[ (d-2) d+(n+1)^2
+\frac{ (d+n-1) \left(2 \bar a_0 (d-1) d+1\right)}{\bar a_0^2 (d-1) d (d+n)}\Big]+\mathcal{O}(u-u_0)\,,
\ee
where
\be
\bar a_0\equiv \frac{a_0}{ \ell^2 R_1}\,.
\ee
Equation (\ref{rega3}) implies that at this end-point the geometry is regular.
For comparison,
the Krerschmann scalar of an $AdS_{d+n+1}$ space with length scale $\ell$ is constant everywhere and is given by
\be  \label{KADS}
\mathcal{K}_{AdS}=\frac{2(d+n)(d+n+1)}{\ell^4}\,.
\ee
We obtain
\be
\mathcal{K}-\mathcal{K}_{AdS}=\frac{2(d+n)(d+n-1)}{ \ell^4~\bar a_0^2 d(d-1)n(n+1)}\left(d(d-1)\bar a_0+1\right)^2+\mathcal{O}(u-u_0)\,,
\label{KADS2}\ee
which suggests that at $\bar a_0=-\frac{1}{ d(d-1)}$ we obtain $AdS_{n+d+1}$. We shall verify this in section  \ref{global}.

This class of solutions has only two arbitrary parameters, $a_0, u_0$, and is, therefore, a ``tuned" solution as we implemented regularity.

We should note that at this regular end-point, we always have
\be \label{rega4}
\dot A_1\sim (u-u_0)
\sp
\dot A_2\sim \frac{1}{u-u_0}\,,
\ee
and
\be
\begin{cases} \label{crega4}
&\!\!\!\ddot{A}_1 \ge 0 \,,\quad a_0\ge -\frac{\ell^2 R_1}{d(d+n)}\,,
\\[12pt]
&\!\!\!\ddot{A}_1 < 0 \,, \quad \text{otherwise}\,.
\end{cases}
\ee
When $\ddot{A}_1<0$ it might be expected that the $AdS$ space shrinks to zero at some point $u>u_0$.
We shall find such solutions in the next sections.

On the other hand, we may also consider that $AdS_d$ shrinks to zero sizes while the size of $S^n$ is finite. This corresponds to $\l_1=1, \l_2=0$ in the expansions \eqref{gexp1} and \eqref{gexp2}, and from them, the expansions of the scale factors can be written as
\be
e^{2A_1(u)}=a_2\frac{(u-u_0)^2}{\ell^2}+\mathcal{O}(u-u_0)^3\sp e^{2A_2(u)}=s_0+2 s_0 s_1 \frac{u-u_0}{\ell}+\mathcal{O}(u-u_0)^2\,,
\label{rega5}\ee
which by inserting into the equations of motion we find that
\be\label{rega6}
a_2=\frac{R_1}{d(d-1)}<0\,.
\ee
With our initial signature, $e^{A_1}>0$, and therefore this case is not possible.
 {\it We can not have a solution that the $AdS$ scale factor vanishes while the sphere scale is finite}.

\subsection{Solutions with A-bounces and monotonic solutions}\label{SAM}

Except for the shrinking of $AdS_d$ and $S^n$ factors, we can also have places where $\dot A_{1,2}=0$ and then the evolution of scale factors is not monotonic.
We call points where $\dot A_{1,2}=0$ ``A-bounces".
We shall investigate such a regime in this section.

Consider the case in which the arbitrary point $u=u_0$ is an A-bounce.  In general, the expansions of the scale factors around such a point (that is a regular point of the equations) can be written as \footnote{These expansions are the expansion in \eqref{gexp1} and \eqref{gexp2} for $\l_1=\l_2=0$. The constant parameters are denoted by a hat to distinguish them from the end-point parameters.}
\bsq
\begin{align}\label{mon1}
& A_1(u)=\frac12 \log \hat{a}_0 +\hat{a}_1 \frac{u-u_0}{\ell} + \hat{a}_2 \frac{(u-u_0)^2}{\ell^2} +\mathcal{O}(u-u_0)^3\,, \\
& A_2(u)=\frac12 \log \hat{s}_0 +\hat{s}_1 \frac{u-u_0}{\ell} + \hat{s}_2 \frac{(u-u_0)^2}{\ell^2} +\mathcal{O}(u-u_0)^3\,.\label{mon2}
\end{align}
\esq
From equations of motion \eqref{eq1}--\eqref{eq3} we know that all unknown coefficients above can be written as functions of three arbitrary constants. We choose these constants to be $\hat{a}_0, \hat{s}_0$ and $\hat{s}_1$.
From the equations, we obtain
\bsq
\begin{gather}\label{mon3}
\hat{a}_1= \frac{-n d \hat{a}_0 \hat{s}_0 \hat{s}_1\pm \chi}{d (d-1) \hat{a}_0\hat{s}_0 }\,,
\\
\hat{a}_2=
\frac{- \left( d \left(n  (d+n)\hat{s}_0+\ell^2 R_2\right)\hat{a}_0+\ell^2 R_1 \hat{s}_0\right)}{2 (d-1) d \hat{a}_0 \hat{s}_0} \nn \\
+\frac{d n (1-d-2 n)\hat{a}_0  \hat{s}_0 \hat{s}_1^2 \pm(d+1) n  \hat{s}_1\chi}{2 \hat{a}_0 (d-1)^2 d \hat{s}_0}\,, \label{mon4}
\\
\hat{s}_2=
\frac{(d-1)\hat{a}_0  \left(n (d+n) \hat{s}_0 +\ell^2 R_2\right)+n^2 \hat{a}_0  \hat{s}_0 \hat{s}_1^2 \mp n \hat{s}_1 \chi}{2 (d-1) n \hat{a}_0  \hat{s}_0}
\,, \label{mon5}
\end{gather}
\esq
where
\begin{gather}
\chi\equiv\Big(
d \hat{a}_0 \hat{s}_0 \big[(d-1) \big((d+n-1) (d+n)\hat{a}_0 \hat{s}_0+\ell^2(\hat{a}_0  R_2+ \hat{s}_0 R_1)\big)\nn \\
+n (d+n-1)\hat{a}_0 \hat{s}_0 \hat{s}_1^2 \big]\Big)^\frac12\,.\label{mon6}
\end{gather}
The reality of \eqref{mon6} restricts the parameters to
\bsq
\begin{gather}\label{Rmon7}
|\hat{s}_1|\geq \sqrt{
-\frac{(d-1) \left( (d + n)(d + n-1)\hat{a}_0 \hat{s}_0 + \ell^2 R_2\hat{a}_0+\ell^2 R_1 \hat{s}_0\right)}{n (d+n-1)\hat{a}_0 \hat{s}_0 }
}\,, \\
R_1 + \frac{\hat{a}_0 (\ell^2 R_2 + (d + n)^2 \hat{s}_0)}{\ell^2 \hat{s}_0} < \frac{\hat{a}_0 (d + n)}{\ell^2}\,.\label{Rmon8}
\end{gather}
\esq
According to the above expansions, we can divide the solutions of Einstein's equations into two sets of solutions:
\begin{itemize}
\item Solutions with A-bounce: There is at least one point where either $\hat{a}_1$ or $\hat{s}_1$
 or both are zero
\item Monotonic solutions: There is no point where  $\hat{a}_1$ or $\hat{s}_1$ are zero.

\end{itemize}

As we already mentioned, we may have solutions that one or both of the scale factors have an $A$-bounce. At this point, the scale factor reaches a non-zero minimum or a finite maximum. Similar bounces were found in flat RG flows in \cite{exotic}, in which what changed direction (bounced) was the scalar field. Scale factor bounces, or  $A$-bounces in short,  were instead found to be ubiquitous in curved RG-flows with $AdS$ slices,  \cite{Bak,C, ads}.

In the subsequent sections, we shall study the properties of the solutions with A-bounces. A subset of monotonic solutions was studied in section \ref{regsec}. Other monotonic solutions will be studied numerically in section \ref{mono}.

\subsubsection{$AdS_d$ bounce}\label{adsbon}

Consider the case when the scale factor of $AdS_d$ displays a bounce at some radial position $u=u_0$. We call this an {\em $A_1$-bounce} ($\dot A_1=0, \dot A_2 \neq 0 $). This corresponds to consider $\hat{a}_1=0$ in \eqref{mon1}, i.e.
\bsq
\begin{gather}\label{adsb1}
A_1(u)=\frac12\log(\hat{a}_0)+\hat{a}_2\frac{(u-u_0)^2}{\ell^2}+{\cal O}((u-u_0)^3)\,,\\
A_2(u)=\frac12\log(\hat{s}_0)+\hat{s}_1\frac{(u-u_0)}{\ell}+\hat{s}_2\frac{(u-u_0)^2}{\ell^2}+{\cal O}((u-u_0)^3)\,,\label{adsb2}
\end{gather}
\esq
where $\hat{a}_0$ and $\hat{s}_0$ are the sizes of $AdS$ and the sphere at the bounce. Moreover, one finds

\bsq
\begin{gather}\label{vs1}
\hat{s}_1= \pm\frac{\sqrt{(d + n)(d + n-1)+(\hat{r}_1+\hat{r}_2)}}{\sqrt{(n-1) n}}\,,\\ \label{vs2}
\hat{s}_2= -\frac{d^2 n+d n^2+ (n \hat{r}_1+\hat{r}_2)}{2(n-1) n}\,, \\ \label{vs3}
\hat{a}_2= \frac{d^2+d n+ \hat{r}_1}{2 d}\,,
\end{gather}
\esq
and
\be \label{rhat12}
\hat{r}_1\equiv\frac{\ell^2 R_1}{\hat{a}_0}<0\sp \hat{r}_2\equiv\frac{\ell^2 R_2}{\hat{s}_0}>0\,.
\ee
The expansions \eqref{adsb1} and \eqref{adsb2} show the following  properties for solutions of equations of motion with an $A_1$-bounce:
\begin{itemize}
\item
 Since $R_1<0$, then
\be \label{ahat2}
\begin{cases}
\hat{a}_2\ge 0, & \text{for} \quad \hat{a}_0\ge -\frac{\ell^2 R_1}{d (d+n)}\,,
\vspace{0.5cm} \\
\hat{a}_2< 0, & \text{for} \quad -\frac{\ell^2 R_1}{d (d+n)}>\hat{a}_0>{\hat{a}_0^{min}}\,,
\end{cases}
\ee
where the reality of the value of $\hat{s}_1$ in \eqref{vs1} puts a lower bound on $\hat{a}_0$ for any positive value of $\hat{s}_0$
\be \label{a0min}
{\hat{a}_0^{min}}=-\frac{\ell^2 R_1 \hat{s}_0}{\hat{s}_0 (d + n)(d + n-1)+\ell^2 R_2}\,.
\ee
\item $\hat{s}_1 \in \mathbb{R}$ also indicates that at $A_1$-bounce always $\hat{s}_2<0$.
\item
At a specific value
\be \label{acs}
\hat{a}_0=a^c_0\equiv -\frac{\ell^2 R_1}{d (d+n)}\,,
\ee
the bounce disappears and we can find an exact solution
\bsq
\begin{align} \label{exsol3}
e^{2A_1(u)} &=-\frac{\ell^2 R_1}{d (d+n)}\,,\\
e^{2A_2(u)} &=\hat{s}_0\Big(\l_0 \sinh\big[k\frac{u-u_0}{\ell}\big]+\cosh\big[k\frac{u-u_0}{\ell}\big]\Big)^2\,,\nn  \\
&= \hat{s}_0(\l_0^2-1)\sinh^2\Big[k\frac{u-u_0}{\ell}-{\frac12}\log\big(\frac{\l_0-1}{\l_0+1}\big)\Big] \,,\label{exsol5}
\end{align}
\esq
where
\be\label{lam0k}
\l_0= \sqrt{\frac{\ell^2 R_2}{(n-1) (d+n)\hat{s}_0}+1}\,,\qquad k= \sqrt{\frac{d}{n}+1}\,.
\ee
This solution will be discussed in more detail in section \ref{ESS}.

\item
If we have two A-bounces, both for $AdS_d$ and $S^n$ at the same point $u=u_0$ (equivalently, when $\hat{s}_1=0$)
then
\be \label{a0b0}
\hat{a}_0=a_0^b\equiv \frac{-\ell^2 R_1}{(d + n)(d + n-1) }>0\,,
\ee
and
\be\label{ahat2c}
\hat{a}_2=-\frac12 \big(\frac{\ell^2 R_2}{\hat{s}_0}+(d+n)(d+n-2)\big)<0\,,
\ee
which implies that the $AdS$ scale factor has a finite {\em maximum} at this point.
As we shall see, the solutions with this property have two end-points where the  $AdS$ space shrinks to zero sizes.
\end{itemize}

To summarize, the space of solutions with an $A_1$-bounce is parametrized by three free parameters $\hat{a}_0$ and $\hat{s}_0$ and $u_0$ which represent the size of $AdS$ and sphere at the bounce as well as the position of the bounce. The analysis above shows that:

 1)
 From equation \eqref{ahat2} we deduce that the $AdS$ scale factor has at most one A-bounce. To see this, consider two neighboring bounces that one of them is a local maximum ($\hat{a}_2<0$) and the other is a local minimum ($\hat{a}_2>0$). According to \eqref{ahat2} the value of the $AdS$ scale factor ($\hat{a}_0$) for the local minimum should be greater than its local maximum neighbor, which is impossible, therefore we can not have more than one $A_1$-bounce in a solution.

2) There is a lower bound \eqref{a0min}  on $\hat{a}_0$ for an $A_1$-bounce to exist. $\hat{a}_0$ controls  the minimum size of $AdS$ at the $A_1$-bounce.

3) We have a special solution with a constant scale factor of $AdS$. This is an exact solution that will be discussed later in section \ref{ESS}.

4) If both $AdS$ and the sphere have A-bounces at the same point, at this point, $AdS$ has a finite maximum size but the sphere reaches a nonzero minimum.

\subsubsection{$S^n$ bounce}\label{sbon}

An alternative possibility occurs when $S^n$ has a bounce ($A_2$-bounce). This is the case with $\hat{s}_1=0$ in \eqref{mon2}. Then we obtain
\bsq
\begin{gather}\label{bs1}
A_1(u)=\frac12\log(\hat{a}_0)+\hat{a}_1\frac{(u-u_0)}{\ell}+\hat{a}_2\frac{(u-u_0)^2}{\ell^2}+\mathcal{O}(u-u_0)^3\,,\\ \label{bs2}
A_2(u)=\frac12\log(\hat{s}_0)+\hat{s}_2\frac{(u-u_0)^2}{\ell^2}+\mathcal{O}(u-u_0)^3\,,
\end{gather}
\esq
with the following values for the above coefficients
\bsq
\begin{gather}\label{ssa1}
\hat{a}_1= \pm\frac{\sqrt{(d + n)(d + n-1)+(\hat{r}_1+\hat{r}_2)}}{\sqrt{(d-1) d}}\,,\\ \label{ssa2}
\hat{a}_2= -\frac{d^2 n+d n^2+ (d \hat{r}_2+\hat{r}_1)}{2 (d-1) d}\,, \\ \label{sss2}
\hat{s}_2= \frac{n^2+d n+ \hat{r}_2}{2 n }\,.
\end{gather}
\esq
Here unlike the $A_1$-bounce case, we always have
$\hat{s}_2>0$, therefore, we do not expect to have a solution with two shrinking end-points for $S^n$.
However, there is a constraint for the reality of $\hat{a}_1$ in equation \eqref{ssa1}
\begin{align}\label{sich}
\begin{cases}
\hat{a}_0 \leq a_0^b ~~\Rightarrow & 0 < \hat{s}_0 \leq \frac{-\hat{a}_0 \ell^2 R_2}{\ell^2 R_1 +\hat{a}_0 (d + n)(d + n-1) }\,, \vspace{0.5cm}\\
\hat{a}_0 > a_0^b ~~\Rightarrow & 0 < \hat{s}_0\,,
\end{cases}
\end{align}
where $a_0^b$ is given in equation \eqref{a0b0}.
Moreover, the reality of $\hat{a}_1$ in \eqref{ssa1} shows that $\hat{a}_2<0$ at the $A_2$-bounce.

To summarize:

1) Since at an $A_2$-bounce we always have a minimum size for the sphere scale factor ($\hat{s}_2>0$) we can not have a solution with more than one $A_2$-bounce.

2) If there is an $A_2$-bounce, we do not expect to find a solution that has two end-points with a shrinking sphere.

3) According to \eqref{sich} in the space of solutions described by the two parameters  $(\hat{a}_0, \hat{s}_0)$, there is an upper bound on the size of the sphere at the bounce, as far as $\hat{a}_0 \leq a_0^b$.


\section{Exact solutions\label{ESS}}
In this section, we shall find some exact solutions for the equations of motion \eqref{eq1}--\eqref{eq3}.
There are several special cases in which we can solve equations of motion exactly. The expansions in \eqref{rega1} and \eqref{rega2} also can help us to find these solutions.
\begin{itemize}

\item {\bf{$AdS_d \times AdS_{n+1}$ (product space) solution}}

As we can observe from equation \eqref{rega1}, in the special case where
\be \label{a0t}
a_0=a_0^c=-\frac{\ell^2 R_1}{d (d+n)}\,,
\ee
the scale factor of $AdS_d$ is fixed and is independent of the $u$ coordinate. In this situation, the equations of motion are exactly solvable and we find for $n>1$
\be \label{exsol1}
e^{2A_1(u)}=-\frac{\ell^2 R_1}{d (d+n)}\sp
e^{2A_2(u)}=
\Big(c e^{\sqrt{\frac{d+n}{n}}\frac{u }{\ell}}-\frac{\ell^2 R_2 }{4 c (n-1) (d+n)}e^{-\sqrt{\frac{d+n}{n}}\frac{u }{\ell}}\Big)^2\,,
\ee
where $c$ is the constant of integration.
This solution in general has an end-point for the sphere at
\be\label{u0end}
u_{0}= -\frac{1}{2} \ell \sqrt{\frac{n}{d+n}} \log \Big(\frac{4 c^2 (n-1) (d+n)}{\ell^2 R_2}\Big)\,.
\ee
Therefore, we can rewrite the scale factors as
\be \label{exsol2}
e^{2A_1(u)}=-\frac{\ell^2 R_1}{d (d+n)}\sp
e^{2A_2(u)}=\frac{\ell^2 R_2}{(n-1)(d+n)}\sinh^2\Big(\sqrt{\frac{d+n}{n}}\frac{u-u_0}{\ell}\Big)\,,
\ee
which means that the metric describes a product space $AdS_d \times AdS_{n+1}$.

The Kretschmann scalar for this solution is a constant
\be\label{kerprod}
\mathcal{K}= \frac{2 (d+n)^2 (2 d n+d-n-1)}{n(d-1) \ell^4 }\,,
\ee
and differs from the $AdS_{d+n+1}$ value which is
\be  \label{AKADS}
\mathcal{K}=\frac{2(d+n)(d+n+1)}{\ell^4}\,.
\ee
We should remind the reader that we already encountered this solution when we studied the $A_1$-bounces, where at the critical value of $a_0^c$,  the $AdS$ bounce disappeared and we found a similar exact solution in \eqref{exsol5}.
In appendix \ref{EFL} we have considered the fluctuations around this solution to show that the asymptotic behavior changes completely near the boundary of this solution. In particular, the boundary of this solution (in Euclidean signature) has two components: $AdS_d\times S^n \cup S^{d-1}\times AdS_{n+1}$.

\item {\bf{Global $AdS_{d+n+1}$ solution}\label{global}}

Another exact solution for equations of motion is the global $AdS$ solution
\be \label{glob0}
ds^2= du^2 + e^{2\bar{A}_1} \cosh^2\frac{u-u_0}{\ell} ds^2_{AdS_d} + e^{2\bar{A}_2} \sinh^2\frac{u-u_0}{\ell} d\Omega_n^2\,,
\ee
where equations of motion fix the coefficients to
\be \label{glob1}
e^{2\bar{A}_1}=-\frac{\ell^2 R_1}{d(d-1)}\sp
e^{2\bar{A}_2}=\frac{\ell^2 R_2}{n(n-1)}\,.
\ee
We obtain
\be
\bar a_0=-\frac{1}{ d(d-1)}\,,
\ee for this solution, verifying the claim below (\ref{KADS2}).
We also obtain
\be\label{glob2}
R_1^{UV}=4e^{-2\bar{A}_1}R_1=-\frac{4d(d-1)}{\ell^2}\sp
R_2^{UV}=4e^{-2\bar{A}_2}R_2=\frac{4n(n-1)}{\ell^2}\,,
\ee
or the ratio of dimensionless curvatures are fixed by dimensions of $AdS_d$ and $S^n$ spaces
\be\label{glob3}
\frac{\mathcal{R}_1}{\mathcal{R}_2}=-\frac{d(d-1)}{n(n-1)}\,.
\ee
This can be confirmed also in the specific case of $d=n=4$ which we have the UV expansions in \eqref{near1} and \eqref{near2}. Here we realize that the global solution is equivalent to considering the vev $C=0$ and $\mathcal{R}_1=-\mathcal{R}_2=-48$. For a discussion on $AdS$ space in various coordinates including the ones discussed here, see appendix \ref{adscor}.

\end{itemize}

\section{Numerical solutions}
We employ numerical techniques to uncover every potential solution to equations of motion and verify our analytical results obtained from the asymptotics.
The independent equations we solve are ({\ref{j2}) and (\ref{j3}) and they require three constants of integration.

We assume that at a generic point $u=u_0$,  the following expansions of the scale factors satisfy the equations of motion
\bsq
\begin{align}\label{Nmon1}
& A_1(u)=\frac12 \log \hat{a}_0 +\hat{a}_1 \frac{u-u_0}{\ell} + \hat{a}_2 \frac{(u-u_0)^2}{\ell^2} + \mathcal{O}(u-u_0)^3\,, \\
& A_2(u)=\frac12 \log \hat{s}_0 +\hat{s}_1 \frac{u-u_0}{\ell} + \hat{s}_2 \frac{(u-u_0)^2}{\ell^2} +  \mathcal{O}(u-u_0)^3\,.\label{Nmon2}
\end{align}
\esq
Among the constant coefficients in these expansions, we select $\hat{a}_0, \hat{s}_0$ and $\hat{s}_1$ as our free parameters. Using the aforementioned expansions we can read the initial conditions required for solving the equations of motion on both sides of $u=u_0$. This approach will lead us to four classes of end-points for solutions:
\begin{itemize}

\item {\bf{B:}} An $AdS$-like boundary\footnote{In our solutions we have only one boundary which we consider to be located at $u=+\infty$.} where both sphere and $AdS$ sizes diverge. The behavior of the scale factors close to this boundary is given in section \ref{NBE}.

\item {\bf{R:}} A regular end-point where the sphere shrinks to a zero size and the $AdS$ scale factor asymptotes to a constant value. Properties of this end-point are discussed in section \ref{regsec}.

\item {\bf{A:}} This is a singular end-point where the sphere size diverges while the $AdS$ size vanishes, as discussed in section \ref{singsec}.

\item {\bf{S:}}  This is another singular end-point where the sphere size vanishes while the $AdS$ size diverges, as discussed in section \ref{singsec}.

\end{itemize}

According to the above possible end-points, we find the following types of solutions. Each solution is characterized by its end-points:

\begin{itemize}
\item ({\bf B, R})--type: This is a regular class of solutions.


\item ({\bf R, A})--type, ({\bf S, B})--type, ({\bf A, B})--type, ({\bf A, A})--type and ({\bf S, A})--type: These are all singular solutions.
\end{itemize}
In the subsequent sections, we show examples of the above solutions.

{When $S^n$ or $AdS_d$ shrinks to zero size, this signals in Euclidean signature an endpoint of the flow.
From our findings, a sphere can shrink to zero sizes, and the solution is regular there.
But $AdS_d$ cannot shrink to zero sizes and the solution to be regular at that point\footnote{Unless the holographic direction is timelike.}.}

\subsection{Solutions with one regular end-point}\label{numregs}
Among the solutions that we have, two classes of solutions have a regular end-point ({\bf R}). The product space solution $AdS_d\times AdS_{n+1}$ also has a regular end-point.
As we already showed in section \ref{regsec}, the regular end-point at an arbitrary value $u=u_0$ has  the following expansions for scale factors
\bsq
\begin{align}\label{Arega1}
e^{2A_1(u)} &= a_0 + \frac{a_0 d (d + n) + \ell^2 R_1}{d \ell^2 (1 + n)}\Big((u-u_0)^2 \nn \\
& -\frac{a_0 d (d - n-4) (d + n) + (d-3) \ell^2 R_1}{ 3 a_0 d \ell^2 (1 + n) (3 + n)} (u-u_0)^4+ \mathcal{O}(u-u_0)^6\Big)\,, \\
e^{2A_2(u)} &= \frac{R_2}{n(n-1)} (u-u_0)^2
+\frac{(a_0 (d - d^2 + n + n^2) - \ell^2 R_1) R_2}{3 a_0 \ell^2 n^2 (n^2-1)} (u-u_0)^4\nn\\
&+ \mathcal{O}(u-u_0)^6\,.\label{Arega2}
\end{align}
\esq
Here we have a free parameter $a_0=e^{2A_1(u_0)}$. Varying $u_0$ does not give more solutions as such a variation can be undone by a translation in $u$.

For a point at $u=u_0+\e$ with $\e>0$ the initial conditions required to numerically solve the equations of motion are
\bsq
\begin{gather} \label{intr1}
A_1(\e)=\frac12 \log a_0+\mathcal{O}(\e^2)\sp \dot{A}_1(\e)= \frac{a_0 d (d + n) + \ell^2 R_1}{a_0 d \ell^2 (1 + n)}\e+\mathcal{O}(\e^3)\,, \\
A_2(\e)=\frac12 \log\big(\frac{R_2}{n(n-1)}\e^2\big) +\mathcal{O}(\e^2) \sp
\dot{A}_2(\e)=\frac{1}{\e}+\mathcal{O}(\e)\,,\label{intr2}
\end{gather}
\esq
where the higher order terms depend on $a_0$.
 According to the value of the only parameter $a_0$ in this class, we find three different types of answers.\footnote{In the rest of this paper, for the numerical solutions we fix $d=n=4$, $R_1=-1$, $R_2=2$, and $\ell=1$.}

\begin{itemize}

\item {\bf{(R, B)--type:}} This is a solution that starts from a regular end-point at $u=u_0$ and asymptotes to an $AdS$ boundary at $u\rightarrow +\infty$. At the end-point, the scale factor of the sphere is zero but the $AdS$ space has a finite size. This solution exists as far as $a_0>a_0^c$ where
\be \label{acprod}
a^c_0\equiv -\frac{\ell^2 R_1}{d (d+n)}\,.
\ee
An example of this type is sketched in figure \ref{shrink1}.
\begin{figure}[!ht]
\begin{center}
\includegraphics[width=0.50\textwidth]{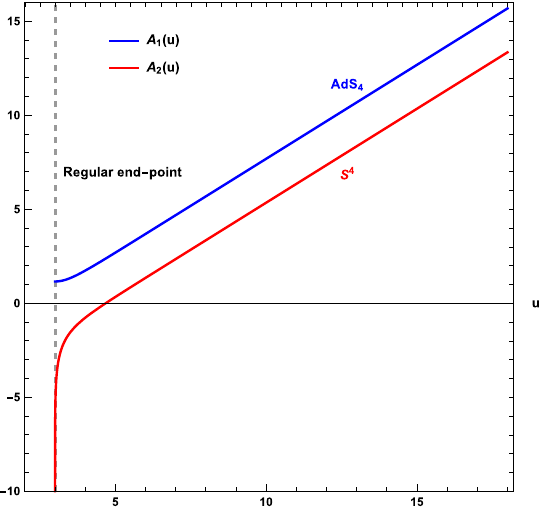}
\caption{\footnotesize{(R, B)--type: The scale factor for $AdS$ (blue curve), and the scale factor of $S$ (red curve), start at a regular end-point (dashed line). At this point, the sphere scale factor shrinks to a zero size but $AdS$ has a finite non-zero size. Both scale factors reach the $AdS$ boundary ($u\rightarrow+\infty$).}}\label{shrink1}
\end{center}
\end{figure}

\item  The product space solution: This is an $AdS_d\times AdS_{n+1}$ solution that we discussed in section \ref{ESS}. While the scale factor of $AdS_d$ is fixed, the scale factor of $S^n \subset AdS_{n+1}$ starts from a zero value at the end-point and reaches the UV boundary at $u\rightarrow +\infty$.
This is a single solution corresponding to choosing $a_0=a_0^c$ from \eqref{acprod}.
Figure \ref{shrink0} shows this solution.
\begin{figure}[!ht]
\begin{center}
\includegraphics[width=0.49\textwidth]{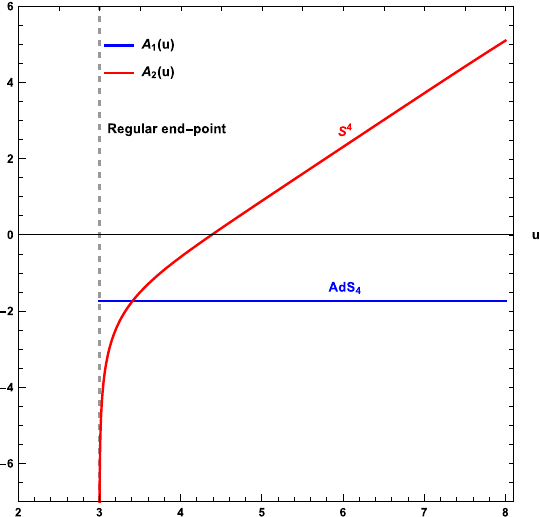}
\caption{\footnotesize{$AdS_d\times AdS_{n+1}$ solution.}}\label{shrink0}
\end{center}
\end{figure}
\vspace{-1.2cm}
\item {\bf{(R, A)--type:}}  If we choose the value of $a_0$ such that according to \eqref{crega4} $\ddot{A}_1(u_0)<0$, or equivalently if $a_0<a_0^c$,  then although we start from a regular end-point at $u=u_0$, the $AdS$ scale factor decreases until it reaches zero at a finite $u>u_0$.
\begin{figure}[!ht]
\begin{center}
\includegraphics[width=0.49\textwidth]{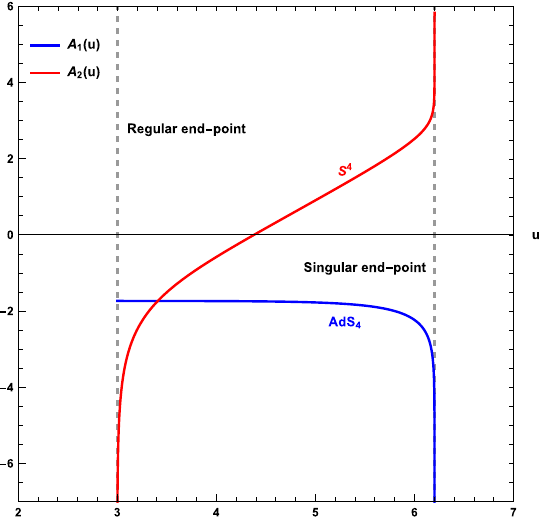}
\caption{\footnotesize{(R, A)--type: A singular solution that starts at a regular end-point (left dashed line) and reaches a singular end-point (right dashed line).}}\label{shrink2}
\end{center}
\end{figure}
At this point, we can check that the scale factors behave as \eqref{gexp1} and \eqref{gexp2} with $\l_1$ and $\l_2$ coefficients in \eqref{adsshr}, so we have a singular end-point here. An example of this singular solution is given in figure \ref{shrink2}.

\end{itemize}

At an arbitrary regular end-point $u_0$, as we decrease the scale factor of $AdS_d$ i.e. $a_0=e^{2A_1(u_0)}$, we observe the transition between the above solutions. This is sketched in figure \ref{UVIR1}.
Figure \ref{KK1} shows how the Kretschmann scalar changes for three different types of solutions. For (R, A)--type at the singular end-point, this scalar is diverging.
\begin{figure}[!ht]
\begin{center}
\begin{subfigure}{0.48\textwidth}
\includegraphics[width=1\textwidth]{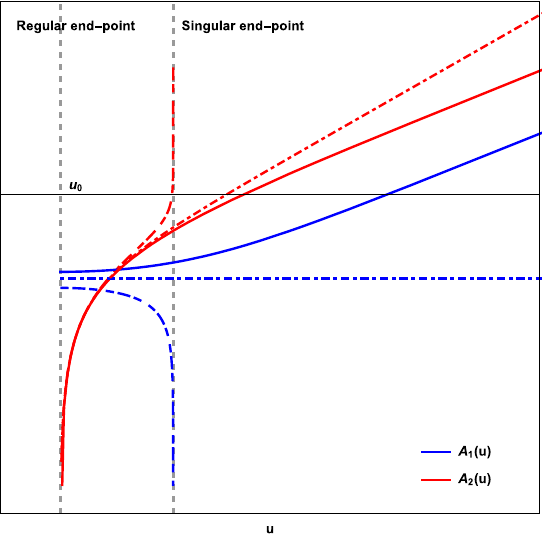}
\caption{}\label{UVIR1}
\end{subfigure}
\begin{subfigure}{0.475\textwidth}
\includegraphics[width=1\textwidth]{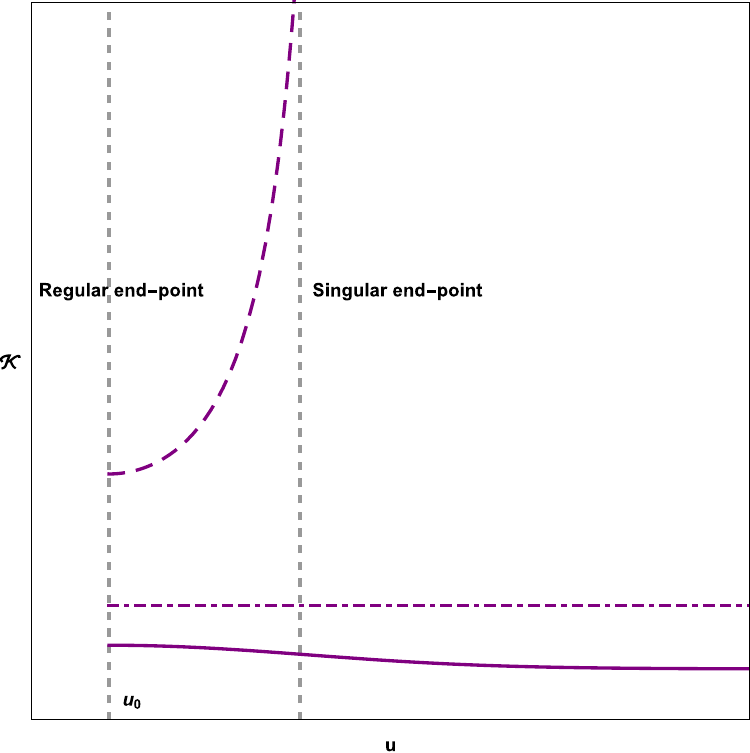}
\caption{}\label{KK1}
\end{subfigure}
\end{center}
\caption{\footnotesize{(a): Transition between solutions as we change the initial value of the $AdS_d$ scalar factor, $a_0$. The solid curves show an example of (R, B)--type. By decreasing $a_0$ and at a specific point $a_0=a_0^c$ we have the $AdS_{d}\times AdS_{n+1}$ solution (dot-dashed curves). Below that point, all solutions are the (R, A)--type. (b): The Kretschmann scalar $\mathcal{K}$ vs $u$. For all solutions in figure (a) we have sketched the Kretschmann scalar.} }
\end{figure}
\subsection{Solutions with A-bounces}
If we assume that there is either an $A_1$-bounce or $A_2$-bounce at an arbitrary point $u=u_0$,  we find three different types of singular solutions. We observe that the existence of an A-bounce is {always} accompanied by one or two singular end-points.

If we consider that at $u=u_0$ there is an $A_1$-bounce then according to expansions of \eqref{adsb1} and \eqref{adsb2} the initial conditions to solve the equations of motion are
\bsq
\begin{gather}\label{ib1}
A_1(u_0)=\frac12 \log(\hat{a}_0)\sp \dot{A}_1(u_0)=0\sp
A_2(u_0)=\frac12 \log(\hat{s}_0)\,,
\\
\dot{A}_2(u_0)=\pm\frac{\sqrt{(d + n)(d + n-1)+(\frac{\ell^2 R_1}{\hat{a}_0}+\frac{\ell^2 R_2}{\hat{s}_0})}}{\ell\sqrt{(n-1) n}}\,.\label{ib2}
\end{gather}
\esq
Here we find two types of solutions. A solution with just one $A_1$-bounce and another solution with one $A_1$-bounce and one $A_2$-bounce:

{\bf{(S, B)--type:}} Figure \ref{AB1} shows a solution with an $A_1$-bounce. On the left-hand side of the bounce, the solution has a singular end-point i.e. the sphere shrinks but the scale factor of $AdS$ diverges. On the right-hand side, there is an $AdS$ boundary at $u\rightarrow +\infty$. This solution corresponds to the plus sign in \eqref{ib2}. With the minus sign, we find the mirror image where the $AdS$ boundary is at $u\rightarrow -\infty$.
\begin{figure}[!ht]
\begin{center}
\includegraphics[width=0.50\textwidth]{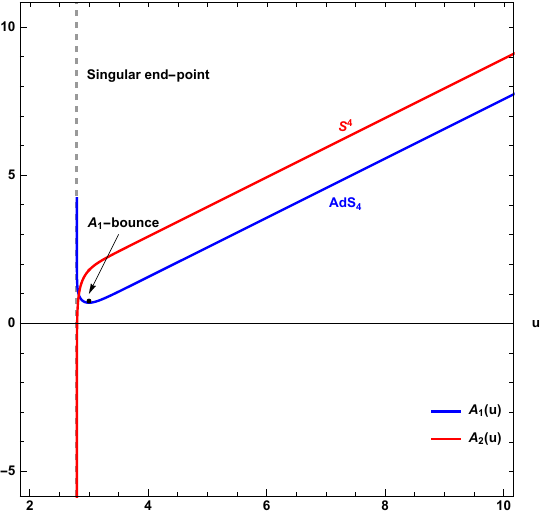}
\caption{\footnotesize{(S, B)--type: Left to the $A_1$-bounce there is a singular IR end-point. Both scale factors reach the UV boundary at $u\rightarrow +\infty$.}}\label{AB1}
\end{center}
\end{figure}
\vspace{-0.5cm}

{\bf{(A, A)--type:}} There is another type of solution with one $A_1$-bounces and one $A_2$-bounce. Here the bounces are not necessarily at the same point in the $u$ coordinate. Figure \ref{AB2} shows a solution of this type. On both sides of these bounces the scale factor of the sphere is diverging but for $AdS$ space it shrinks to zero and so on both sides, we have singular end-points.
\begin{figure}[!ht]
\begin{center}
\includegraphics[width=0.5\textwidth]{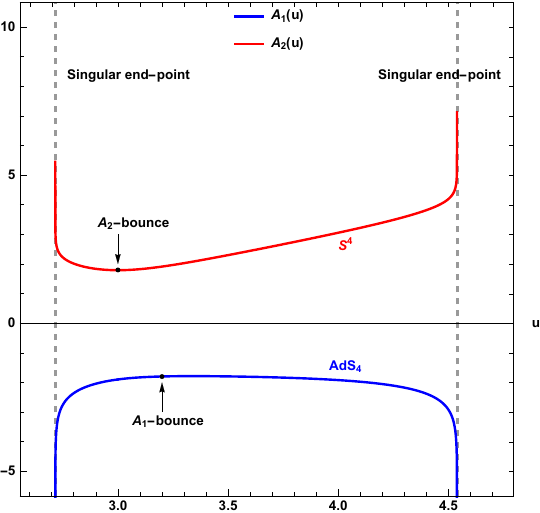}
\caption{\footnotesize{(A, A)--type: An example of solutions with two singular end-points. Both $AdS$ and the sphere has a bounce.}}\label{AB2}
\end{center}
\end{figure}

We can consider that at $u=u_0$ there is an $A_2$-bounce. Looking at the expansions  \eqref{bs1} and \eqref{bs2} we can read the initial conditions required to solve the equations of motion
\bsq
\begin{gather}\label{ib3}
A_1(u_0)=\frac12 \log(\hat{a}_0)\sp
A_2(u_0)=\frac12 \log(\hat{s}_0)\sp \dot{A}_2(u_0)=0\,,
\\
\dot{A}_1(u_0)=\pm\frac{\sqrt{(d + n)(d + n-1)+(\frac{\ell^2 R_1}{\hat{a}_0}+\frac{\ell^2 R_2}{\hat{s}_0})}}{\ell\sqrt{(d-1) d}}\,.\label{ib4}
\end{gather}
\esq
\begin{figure}[!ht]
\begin{center}
\includegraphics[width=0.5\textwidth]{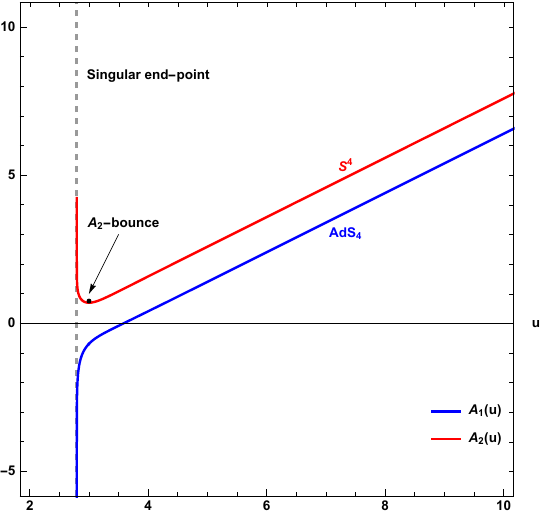}
\caption{\footnotesize{(A, B)--type: Left to the $A_2$-bounce there is a singular end-point where the $AdS$ scale factor is zero but sphere scale factor diverges. Both scale factors reach the $AdS$ boundary at $u\rightarrow +\infty$.}}\label{AB3}
\end{center}
\end{figure}
Once again we may have solutions with just one $A_2$-bounce or solutions with one $A_2$-bounce and one $A_1$-bounce which we already showed in figure \ref{AB2}.

{\bf{(A, B)--type:}}
Figure \ref{AB3} shows an example of solutions with just one $A_2$-bounce. On the left-hand side of the sphere bounce, the solution has a singular end-point
where the scale factor of the sphere diverges but the $AdS$ scale shrinks to zero. On the right-hand side, there is an $AdS$ boundary as $u\rightarrow +\infty$. This solution corresponds to the plus sign in \eqref{ib4} and its mirror image is given by the minus sign.

The behavior of the sphere scale factor that we observe in (A, B)--type and (A, A)--type is consistent with what we already found in section \ref{sbon} where we show that at the $A_2$-bounce always $\ddot{A}_2>0$ because $\hat{s}_2>0$ in \eqref{sss2}.

\subsection{$A_1$-bounce space of solutions}
To see how the different solutions with $A_1$-bounce change under the variation of the initial values at the bounce, we can draw the space of these solutions.

The initial conditions for solutions with an $A_1$-bounce in \eqref{ib1} and \eqref{ib2} depend on two free parameters $\hat{a}_0$ and $\hat{s}_0$. These two parameters describe the coordinates of the space of the solutions with an $A_1$-bounce, see figure \ref{region1}.
\begin{figure}[!ht]
\begin{center}
\includegraphics[width=0.5\textwidth]{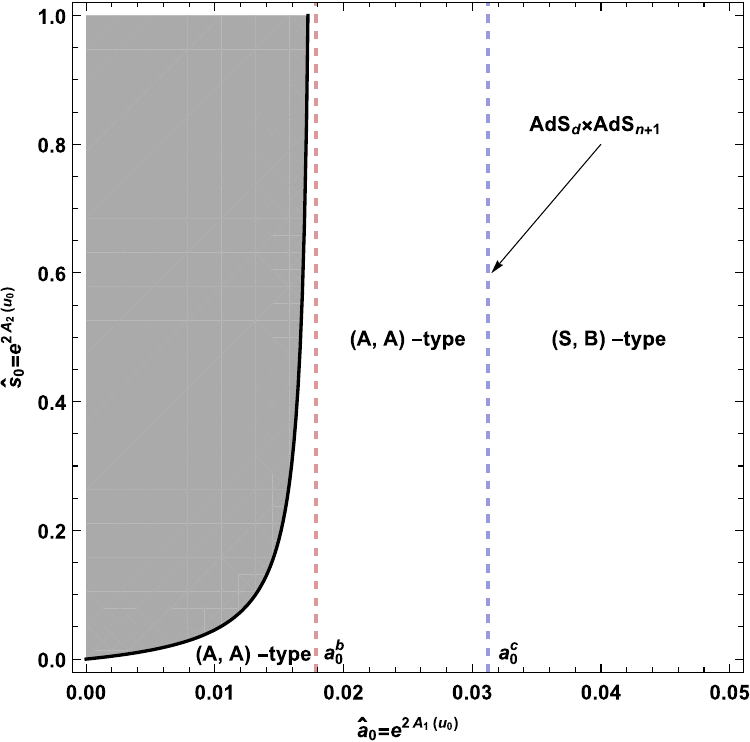}
\caption{\footnotesize{The $A_1$-bounce space of solutions: The (S, B)--type solutions are living on $\hat{a}_0>a_0^c$, the right-hand side of the blue dashed line. The (A, A)--type solutions are limited from right to the $a_0=a_0^c$ and are bounded from left to the gray region. There is no solution in the gray region. Exactly on the blue dashed line we have the product space solutions. On the red dashed line, $AdS_d$ and $S^n$ have a bounce at the same point. Here again, the solutions are the (A, A)--type. }}\label{region1}
\end{center}
\end{figure}
\vspace{-0.8cm}

This space has the following properties:
\begin{enumerate}

\item For values $\hat{a}_0> a_0^c$ where $a_0^c$ is defined in \eqref{acs}, and for all the values of $\hat{s}_0>0$, only solutions of the (S, B)--type can exist.

\item In the place indicated by the blue dashed line at $\hat{a}_0=a_0^c$ in figure \ref{region1}, we have the product space solution.

\item Left to the blue dashed line and right to the gray region only the solutions of (A, A)--type can exist.

\item The boundary of the gray region is given by equation \eqref{a0min}. Inside the gray region, there is no real solution for equations of motion.

\item The red dashed line at $\hat{a}_0=a_0^b$, given in equation \eqref{a0b0}, shows the solutions that have two bounces for $AdS_d$ and $S^n$ at the same point $u=u_0$.
\end{enumerate}
\begin{figure}[!t]
\begin{center}
\begin{subfigure}{0.44\textwidth}
\includegraphics[width=1\textwidth]{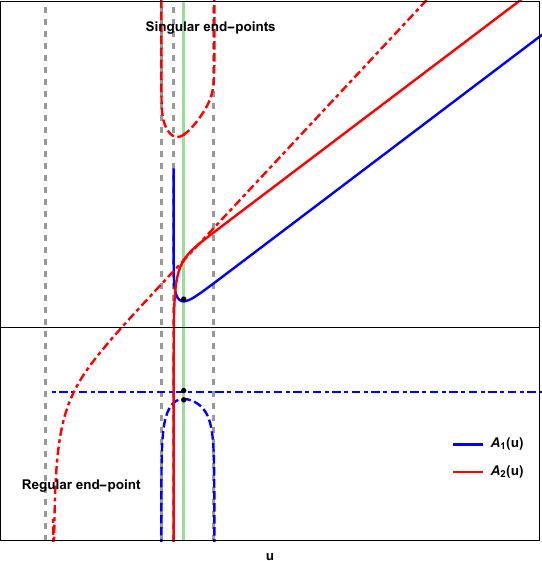}
\caption{}\label{UVB1}
\end{subfigure}
\begin{subfigure}{0.46\textwidth}
\includegraphics[width=1\textwidth]{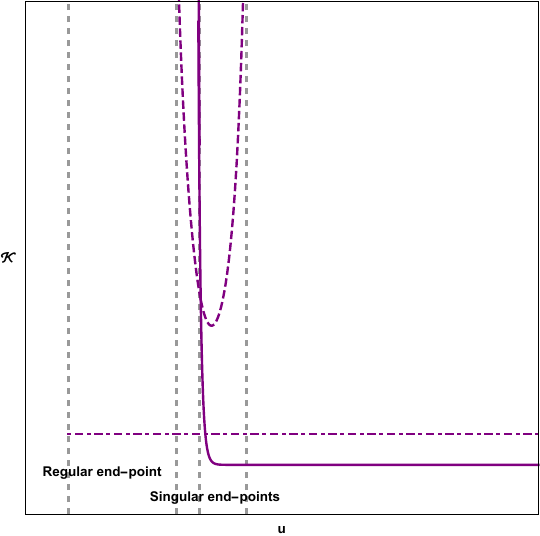}
\caption{}\label{KK2}
\end{subfigure}
\end{center}
\caption{\footnotesize{(a): At a fixed value of $\hat{s}_0=e^{A_2(u_0)}$ ($u_0$ is the location of the green vertical line), and by decreasing the initial value of $\hat{a}_0$ (moving down on the green line) we first observe the solid curves which show an (S, B)--type solution. At a specific value $\hat{a}_0=a_0^c$ a new solution (dot-dashed curves) will appear, This is a product space solution. Below $a_0^c$, the solutions (dashed curves) are the (A, A)--type ones. There is a lower bound for $\hat{a}_0$ given by \eqref{a0min}. (b): The Kretschmann scalar for solutions in figure (a).}}
\end{figure}
\begin{figure}[!ht]
\begin{center}
\begin{subfigure}{0.43\textwidth}
\includegraphics[width=1\textwidth]{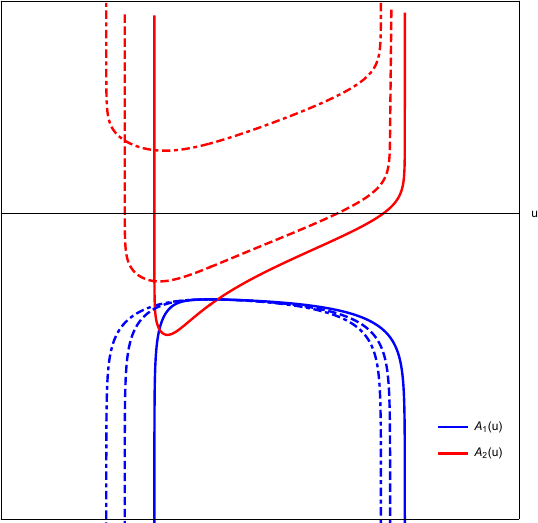}
\caption{}\label{UVB2}
\end{subfigure}
\begin{subfigure}{0.43\textwidth}
\includegraphics[width=1\textwidth]{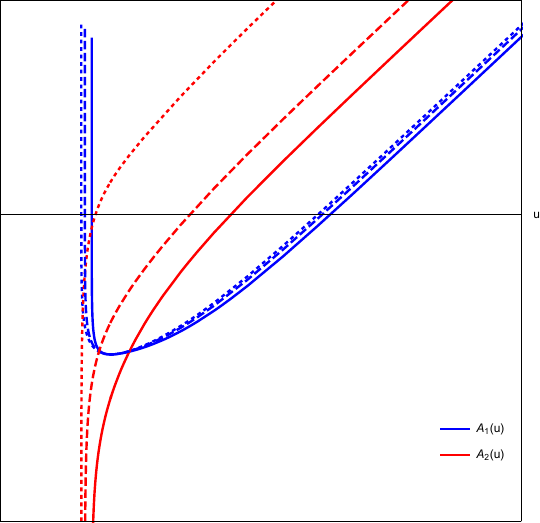}
\caption{}\label{UVB3}
\end{subfigure}
\end{center}
\caption{\footnotesize{Consider solutions with $A_1$-bounce at a fixed $u=u_0$ (the common point through which all the blue curves pass). (a) Shows the transformation of the solutions as we change the value of $\hat{s}_0$  for a fixed $\hat{a}_0<a_0^c$. (b) Shows this transformation for a fixed $\hat{a}_0>a_0^c$.}}
\end{figure}
Consider an arbitrary point in $u=u_0$ where the $A_1$-bounce is happening and then change the value of $\hat{a}_0$ while the value of $\hat{s}_0$ is kept fixed (move horizontally in figure \ref{region1}).
Figure \ref{UVB1} shows the transition between solutions as we change the parameter $\hat{a}_0$ of the $A_1$-bounce.
Figure \ref{KK2} shows how the Kretschmann scalar diverges at the singular end-points of the solutions in figure \ref{UVB1}.


We can also move vertically on the space of solution in figure \ref{region1}, i.e. keep $\hat{a}_0$ fixed and change $\hat{s}_0$.
Depending on which region we are in figure \ref{region1} we either have figure \ref{UVB2} or \ref{UVB3}.

\subsection{$A_2$-bounce space of solutions}
The initial conditions for solutions with an $A_2$-bounce in \eqref{ib3} and \eqref{ib4} depend on two free parameters $\hat{a}_0$ and $\hat{s}_0$. These two parameters describe the coordinates of the space of the solutions with an $A_2$-bounce, see figure \ref{region2}.
\begin{figure}[!ht]
\begin{center}
\includegraphics[width =0.5\textwidth]{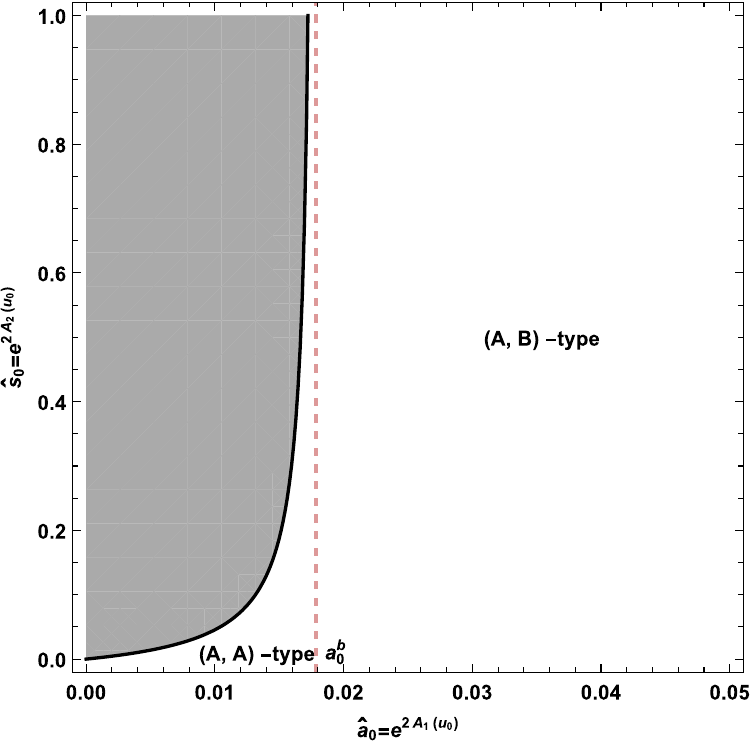}
\caption{\footnotesize{$A_2$-bounce space of solutions: (A, B)--type solutions are living on the right-hand side of the red dashed line which is drawn at $\hat{a}_0=a_0^b$, see equation \eqref{a0b0}. (A, A)--type solutions are limited from the right to the dashed line and are bounded from the left. In the gray region, we do not have any solution. Exactly on the dashed line, both $AdS_d$ and $S^n$ scale factors bounce at the same point $u=u_0$. Here we have the (A, A)--type solutions.}}\label{region2}
\end{center}
\end{figure}
\vspace{-0.5cm}

This space has the following properties:
\begin{enumerate}
\item For every value of $\hat{a}_0> a_0^b$ (right to the red dashed line in figure \ref{region2}) and for all values of $\hat{s}_0>0$, only solutions of  (A, B)--type can exist.

\item Exactly on the dashed line where $\hat{a}_0= a_0^b$ (defined in \eqref{a0b0}) both the $AdS$ and sphere have a bounce at the same point $u=u_0$. At this point as we already discussed we have (A, A)--type solutions.

\item Left to the dashed line and right to the gray region only the (A, A)--type solutions can exist.

\item The reality of solutions forbids the parameters inside the gray region, see equation \eqref{ssa1}.

\end{enumerate}
Figures \ref{SB1}, \ref{SB2} and \ref{SB3} show the transformations and transitions between solutions with $A_2$-bounce as we move inside the space of solutions in figure \ref{region2}.
\begin{figure}[!ht]
\begin{center}
\begin{subfigure}{0.45\textwidth}
\includegraphics[width=1\textwidth]{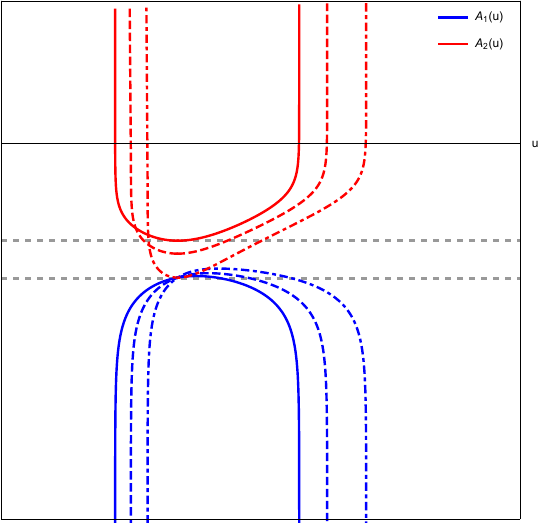}
\caption{}\label{SB1}
\end{subfigure}
\begin{subfigure}{0.45\textwidth}
\includegraphics[width=1\textwidth]{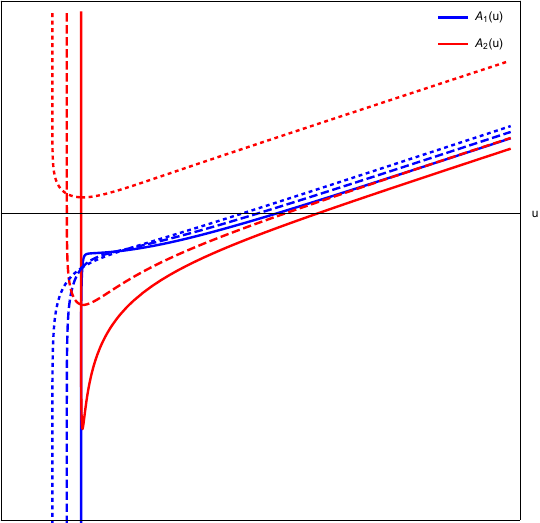}
\caption{}\label{SB2}
\end{subfigure}
\end{center}
\caption{\footnotesize{ (a): For certain values of $\hat{a}_0<a_0^b$ (for example at the lower horizontal dashed line which is the intersection of all the blue curves), the parameter $\hat{s}_0$ (the location of $A_2$-bounces) is bounded by equation \eqref{sich} (the upper dashed horizontal line).  All the solutions in this region are the (A, A)--type. (b): For a fixed  $\hat{a}_0>a_0^b$ the value of $\hat{s}_0$ is unbounded. In this case, all the solutions are the (A, B)--type.}}
\end{figure}
\begin{figure}[!ht]
\begin{center}
\includegraphics[width=0.45\textwidth]{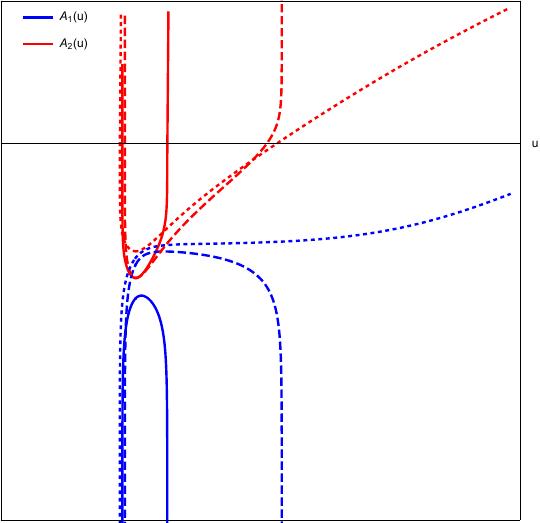}
\caption{\footnotesize{For a fixed value of $\hat{s}_0$ by decreasing the value of $\hat{a}_0$ (moving horizontally to the left in figure \ref{region2}) we see a transition from the (A, B)--type to (A, A)--type solutions.}}\label{SB3}
\end{center}
\end{figure}

\subsection{Monotonic solutions}\label{mono}
As we already discussed in section \ref{SAM}, we may have solutions that do not have any A-bounce. These are solutions with monotonic scale factors. There are two types of solutions with this monotonic behavior:
\begin{itemize}

\item Solutions with one regular end-point which we already found in section \ref{numregs}. The other end-point of these solutions was either at the UV boundary or was a singular end-point.

\item {\bf{(S, A)--type:}} These are solutions with two singular end-points. If we read the initial conditions for a point at $u=u_0$ from the upper signs in \eqref{mon3}--\eqref{mon5}, we find a solution which at the left end-point $u_L<u_0$, the $AdS_d$ scale factor diverges but $S^n$ shrinks.

\begin{figure}[!ht]
\begin{subfigure}{0.49\textwidth}
\centering
\includegraphics[width=1\textwidth]{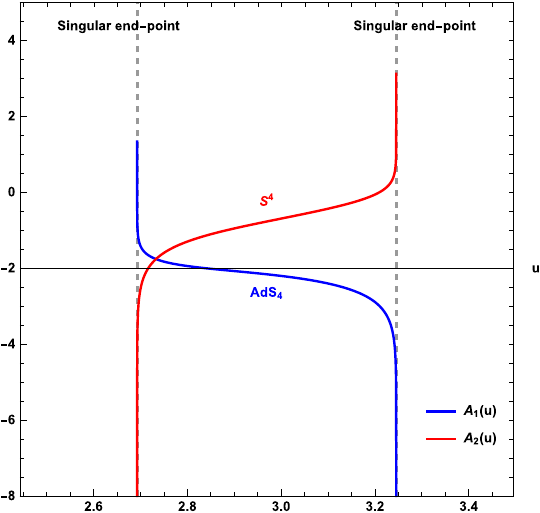}
\caption{\footnotesize{}}\label{TypeS4}
\end{subfigure}
\begin{subfigure}{0.49\textwidth}
\centering
\includegraphics[width=1\textwidth]{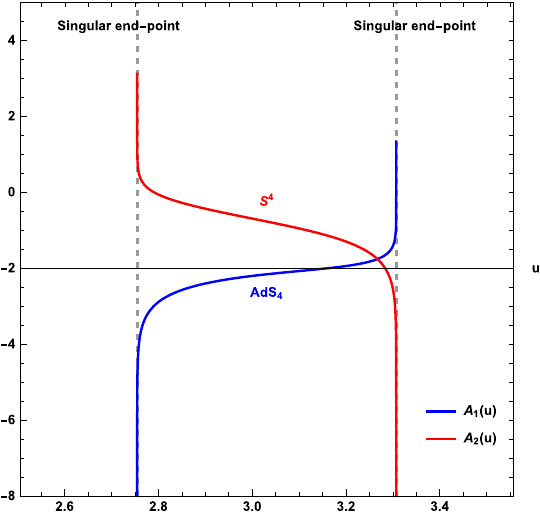}
\caption{\footnotesize{}}\label{TypeS5}
\end{subfigure}
\caption{\footnotesize{(a): An example of (S, A)--type, a monotonic solution with two singular end-points.  Here the free parameters are fixed to $\hat{a}_0=\frac{1}{82}, \hat{s}_0=\frac14$ and $\hat{s}_1=\frac52$. (b): (A, S)--type: A  mirror image of the figure (a) with parameters $\hat{a}_0=\frac{1}{82}, \hat{s}_0=\frac14$ and $ \hat{s}_1=-\frac52$.}}
\end{figure}
 On the right end-point $u_R>u_0$ however, the $AdS_d$ shrinks and $S^n$ diverges, see figure \ref{TypeS4}. There is a mirror image of this solution, (A, S)--type, in which at the left end-point  the$AdS_d$ shrinks but $S^n$ diverges and at the right end-point the $AdS_d$ diverges and $S^n$ shrinks, see figure \ref{TypeS5}. This solution is obtained by choosing the lower signs in \eqref{mon3}--\eqref{mon5} and keeping the values of $\hat{a}_0$ and $\hat{s}_0$ fixed but $\hat{s}_1\rightarrow -\hat{s}_1$.

\end{itemize}

\section{The space of all solutions}
We can observe various types of transitions between different solutions if we carefully describe the space of solutions. To do this, we choose a generic point $u_0$ with the following scale factor expansions
\bsq
\begin{align}\label{Smon1}
& A_1(u)=\frac12 \log \hat{a}_0 +\hat{a}_1 \frac{u-u_0}{\ell} + \hat{a}_2 \frac{(u-u_0)^2}{\ell^2} +\mathcal{O}(u-u_0)^3\,, \\
& A_2(u)=\frac12 \log \hat{s}_0 +\hat{s}_1 \frac{u-u_0}{\ell} + \hat{s}_2 \frac{(u-u_0)^2}{\ell^2} +\mathcal{O}(u-u_0)^3\,.\label{Smon2}
\end{align}
\esq
Assuming these expansions satisfy the equations of motion, we have three free parameters, here for example we select $(\hat{a}_0, \hat{s}_0, \hat{s}_1)$.
These parameters construct the three dimensional space of solutions with $(\hat{a}_0, \hat{s}_0, \hat{s}_1)$ coordinates. Figure \ref{map3d} shows this space for some fixed slices of $\hat{s}_0$.
According to figure \ref{map3d}, we observe the following properties:

\begin{itemize}
\item As mentioned, in some regions of this space we have no solution. For example in the slice with $\hat{a}_0=0.01$ there is a void space inside the blue region. This is coming from the reality of $\chi$ in equation \eqref{mon6} which implies that both conditions \eqref{Rmon7} and \eqref{Rmon8} should be satisfied.

\item For fixed values of $\hat{a}_0$ which $\hat{a}_0<a_0^c=\frac{1}{32}\approx 0.031$, by increasing $\hat{s}_1$ we observe the transition: {\bf (A, B) $\rightarrow$ (A, A) $\rightarrow$ (R, A) $\rightarrow$ (S, A)}. We should emphasize that the {\bf (R, A)}--type solutions are at the boundary of the blue and green regions in figure \ref{map3d}.

\item At the critical value of $\hat{a}_0=a_0^c=\frac{1}{32}$, equation \eqref{a0t}, we should have the product space solution. Fixing $\hat{a}_0$ to this value gives a relation between $\hat{s}_1$ and $\hat{s}_0$ which draws the black curved in figure \ref{map3d}. This curve is the place where the blue region {\bf ((A, A)}--type) and the yellow region ({\bf (S, B)}--type) are terminated. We saw the same behavior in figure \ref{region1} when we studied the $A_1$-bounce space of solutions. Moreover, this curve is the boundary between the green ({\bf (S, A)}--type) and red ({\bf (A, B)}--type) regions.

\item For fixed values of $\hat{a}_0$ which $\hat{a}_0>a_0^c$, by increasing $\hat{s}_1$ we observe the transition: {\bf (A, B) $\rightarrow$ (R, B) $\rightarrow$ (S, B) $\rightarrow$ (S, A)}.

\item There is an orange surface in figure \ref{map3d} which belongs to the regular solutions, the {\bf (R, B)}--type. This surface is the boundary between the yellow, {\bf (S, B)}--type, and the red region {\bf (A, B)}--type. It terminates at the black curve, the product space solution.

\end{itemize}
\begin{figure}[!ht]
\begin{center}
\includegraphics[width=0.9\textwidth]{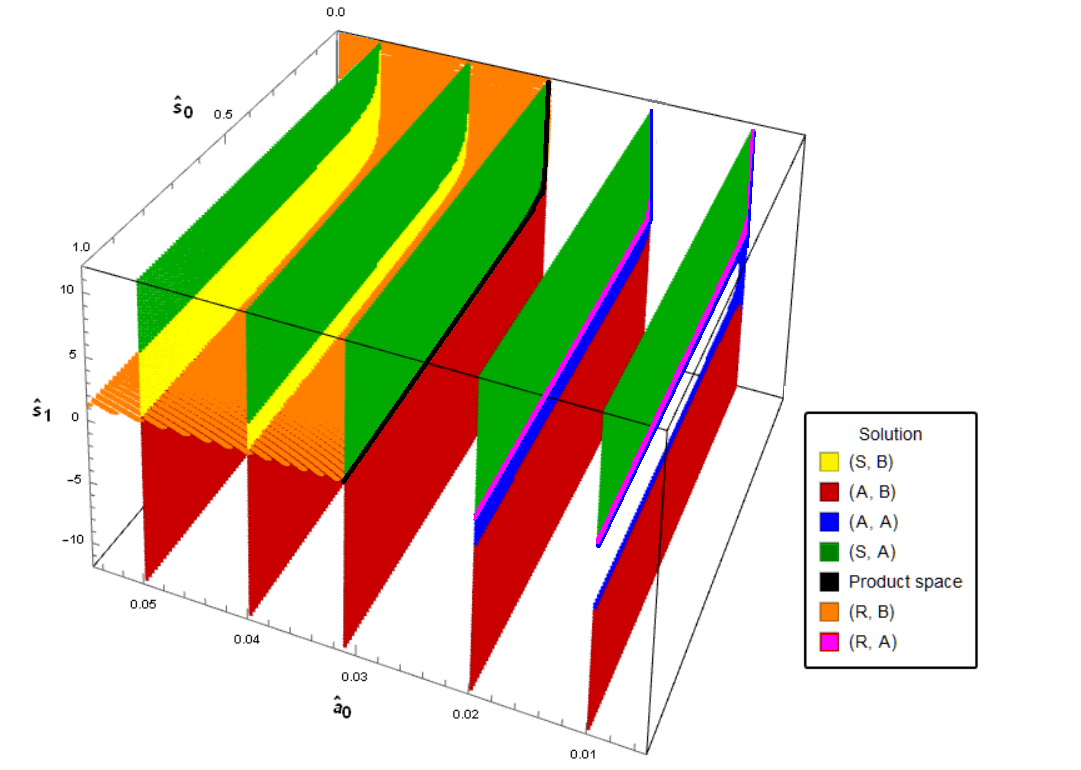}
\caption{\footnotesize{The space of  solutions. To see all possible transitions between different solutions we have sketched five slices of this space. The void space inside the blue region on the slice $\hat{a}_0=0.01$ is a forbidden region where solutions are not real.  Here we have fixed $u_0=3$. The orange surface belongs to the regular solutions, the (R, B)--type, which terminates at the black curve (product space solution). This surface is the boundary between the yellow and red regions. See table \ref{Tab1} for the related links to solutions.}}\label{map3d}
\end{center}
\end{figure}

\begin{table}[!ht]
\begin{center}
\begin{tabular}{|c|c|c|}
\hline
Color & Solution & Figure \\
\hline
Yellow & {\bf (S, B)}--type & \ref{AB1} \\
\hline
Red & {\bf (A, B)}--type & \ref{AB3} \\
\hline
Blue & {\bf (A, A)}--type & \ref{AB2} \\
\hline
Green & {\bf (S, A)}--type & \ref{TypeS4} \\
\hline
Black & $AdS_d \times AdS_{n+1}$ & \ref{shrink0} \\
\hline
Orange & {\bf (R, B)}--type & \ref{shrink1} \\
\hline
Magenta & {\bf (R, A)}--type & \ref{shrink2} \\
\hline
\end{tabular}
\end{center}
\caption{Different solutions in figure \ref{map3d} and their related figures.}\label{Tab1}
\end{table}

The above analysis is performed when we read the initial conditions from \eqref{mon3}--\eqref{mon5} by choosing the upper signs. We can start with the lower signs. The results are similar but we shall find the mirror solutions, i.e. at fixed $(\hat{a}_0, \hat{s}_0)$ we should send $\hat{s}_1\rightarrow -\hat{s}_1$.

\section{The boundary CFT data}
As we observed so far, there are regular and singular solutions that reach the $AdS$ boundary. In this section, we are returning to the holographic correspondence.
In this, we need solutions that are everywhere regular, and therefore the only class to consider is {\bf (R, B)}.

For the regular solutions, we shall compute the near-boundary data that according to the holographic dictionary corresponds to data in the dual CFT.
For example, we shall compute the dimensionless curvatures of $AdS_d$ and $S^n$ spaces at the UV boundary where the CFT is living.
 We are also interested in finding the parameter $C$ which is proportional to the vev of stress-energy tensor of the boundary CFT.

\subsection{Boundary data of (R, B)--type}
There are two free parameters for regular IR end-points, the end-point location $u_0$, and
\be\label{TIRADS}
T^{IR}_{AdS}\equiv  R_1 e^{-2A_1(u_0)}=\frac{ R_1}{a_0}\,,
\ee
with  $a_0$ that has  appeared in the expansions in equations \eqref{rega1} and \eqref{rega2}. On the other hand, we have three free parameters on the UV boundary $R_{AdS}^{UV}=R_1^{UV}, R_S^{UV}=R_2^{UV}$ \eqref{RUV12}, and $C$, which represents the vev of stress-energy tensor of the boundary QFT, see appendix \ref{SET} and equations, \eqref{TEMC1} and \eqref{TEMC2}.
According to the asymptotic expansions of the scale factors, i.e. equations \eqref{near1} and \eqref{near2}, under a shift $u\rightarrow u+u_\infty$ near the boundary we obtain
\be\label{uinf}
R_{AdS,S}^{UV}\sim e^{-\frac{2u_\infty}{\ell}}\sp
C\sim e^{-\frac{8u_\infty}{\ell}}\,,
\ee
so the following dimensionless ratios are independent of $u_\infty$
\be\label{dimless}
\frac{R_{AdS}^{UV}}{R_S^{UV}}\sp \frac{C}{(R_S^{UV})^4 \ell^8}\,.
\ee
Now we find the behavior of these UV parameters in terms of the $T^{IR}_{AdS}$ on the IR side numerically. In the following figures we have fixed
\be\label{fixedv}
R_1=-1\sp R_2=2\sp \ell=1\sp d=n=4\,.
\ee
With the above choices, the critical value of $a_0$ is $a_0^c=\frac{1}{32}$.
We observe the following behaviors for the physical curvatures $R_{AdS}^{UV}$, $R_S^{UV}$ and $C$ as functions of the IR parameter $T^{IR}_{AdS}$:


\begin{figure}[!ht]
\begin{center}
\includegraphics[width=0.6\textwidth]{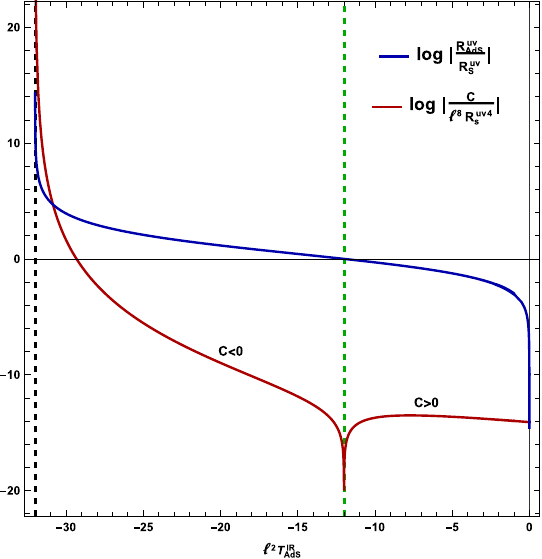}
\caption{\footnotesize{The logarithm of the dimensionless ratios of the UV parameters vs. $\ell^2 T^{IR}_{AdS}$. The black dashed line on the left corresponds to the lower bound $a_0=a_0^c$ of regular IR end-point solutions where $|\ell^2 R^{UV}_{AdS}|\rightarrow +\infty$ and $\ell^2 R^{UV}_{S}\rightarrow 0$ (product space solution). The green dashed line shows the location where $|R^{UV}_{AdS}|=R^{UV}_{S}$ or $C=0$. This point corresponds to the global solution.}}\label{CR1}
\end{center}
\end{figure}

\begin{itemize}

\item As $a_0\rightarrow \infty$ or $T^{IR}_{AdS}\rightarrow 0$ the UV curvature $R_{AdS}^{UV}\rightarrow 0$ but $R_S^{UV}$ has a finite value. At this point, $C$ also has a positive finite value.

\item As far as $R_{S}^{UV}>|R|_{AdS}^{UV}$ we have $C>0$ and visa-verse and at the point where $R_{S}^{UV}=|R|_{AdS}^{UV}$ the value of $C$ vanishes and we have the global solution \eqref{glob0}.

\item At the lowest value for regular solutions i.e. at $a_0=a_0^c$ which is given in \eqref{a0t}, we have the product space solution \eqref{exsol2}. As $a_0$ tends to this point $R_{AdS}^{UV} \rightarrow -\infty$ and
$R^{UV}_S \rightarrow 0$ and $C \rightarrow -\infty$.

\item Below $a_0^c$ we find solutions that have a singular end-point and do not reach the UV boundary at $u\rightarrow +\infty$.

\end{itemize}
The dimensionless ratios of the UV parameters i.e. ${R_{AdS}^{UV}}/{R_S^{UV}}$ and ${C}/{(\ell^2R_S^{UV})^4 }$ in terms of $\ell^2 T^{IR}_{AdS}$
have been shown in figure \ref{CR1}.

\section{The on-shell action and the free energy}
In this section, we find the on-shell action and free energy for regular solutions of the theory. We begin again with the following action
\be \label{on1}
S=M_P^{d+n-1}\int du d^{d+n}x \sqrt{|g|}\Big(R^{(g)}-\frac12 \partial_a\f \partial^a\f -V(\f)\Big)+S_{GHY}\,.
\ee
In this action we have
\be \label{on2}
R^{(g)}=\frac12 (\partial\f)^2-\frac{1+n+d}{1-n-d} V(\f)\sp \partial_a\f \partial^a\f=\dot{\f}^2\sp \sqrt{|g|}=e^{d A_1+n A_2}\sqrt{|\zeta^{1}||\zeta^{2}|}\,.
\ee
Substituting into \eqref{on1} we obtain the on-shell action \footnote{Here we consider solutions in which the boundary (UV) is at $u=+\infty$ while the $S^n$-shrinking end-point (IR) is at $u=u_0$ in \eqref{on5}.}
\be \label{on3}
S_{on-shell}=\frac{2M_P^{d+n-1}}{d+n-1}V_{S^n}V_{AdS_{d}}\int_{u_0}^{+\infty} du \, e^{d A_1+n A_2} V(\f) + S_{GHY}\,,
\ee
where $V_{S^n}$ and $V_{AdS_{d}}$ are the volume of the the sphere and $AdS$ space respectively.
However, we can write  the potential in terms of the scale factors from the equation  of motion \eqref{j2} as follows
\begin{align} \label{on4}
V(\f)&=\frac{(d+n-1)e^{-2(A_1+A_2)}}{d+n}\Big(R_1 e^{2A_2}+R_2 e^{2A_1}\nn \\
&-e^{2(A_1+A_2)}\big(d \ddot{A_1}+n \ddot{A_2}+ (d\dot{A_1}+n\dot{A_2})^2\big)\Big)\,.
\end{align}
Therefore, the on-shell action can be written in terms of the scale factors and their derivatives
\begin{align} \label{on5}
S_{on-shell} &=\frac{2M_P^{d+n-1}}{d+n}V_{S^n}V_{AdS_{d}}\Big(\int_{u_0}^{+\infty} du\,  e^{d A_1+n A_2}\left(R_1 e^{-2 A_1}+R_2 e^{-2 A_2}\right)\nn \\
&-\Big[e^{d A_1+ n A_2}\big(d \dot{A}_1+n \dot{A}_2\big)\Big]_{u_0}^{+\infty}\Big)+ S_{GHY}\,.
\end{align}
The Gibbons Hawking York (GHY) term at the boundary $u=+\infty$, is given as
\be \label{on6}
S_{GHY}=-2M_P^{d+n-1}\Big[\int d^{d+n}x \sqrt{|\gamma|}K\Big]^{u=+\infty}\,,
\ee
where $\gamma_{ij}$ is the induced metric on the $AdS_d\times S^n$ slices and the extrinsic curvature is $K_{ij}=-\frac12 \partial_u \gamma_{ij}$. Therefore we find
\be \label{on7}
K=-d \dot{A}_1 - n \dot{A}_2\sp \sqrt{|\gamma|}=e^{d A_1+n A_2}\sqrt{|\zeta^1||\zeta^2|}\,.
\ee
This gives
\be \label{on8}
S_{GHY}=2M_P^{d+n-1} V_{S^n}V_{AdS_{d}}\Big[e^{d A_1+n A_2}(d \dot{A}_1 + n \dot{A}_2)\Big]^{u=+\infty}\,.
\ee
Moreover, the contribution of the last term in (\ref{on5}) from the $u_0$ endpoint vanishes.
This can be seen  by calculating the derivative of $e^{2A_1(u)}$ and $e^{2A_2(u)}$ in \eqref{rega1} and \eqref{rega2} with respect to the $u$ coordinate.

Using the previous observation and substituting \eqref{on8} in equation \eqref{on5} we obtain
\begin{align} \label{on9}
S_{on-shell} &=\frac{2M_P^{d+n-1}}{d+n}V_{S^n}V_{AdS_{d}}\Big(\int_{u_0}^{+\infty} du\,  e^{d A_1+n A_2}\left(R_1 e^{-2 A_1}+R_2 e^{-2 A_2}\right)\nn \\
&+(d+n-1)\Big[e^{d A_1+ n A_2}\big(d \dot{A}_1+n \dot{A}_2\big)\Big]^{u=+\infty}\Big)\,.
\end{align}
We  introduce two potentials $U_1(u)$ and $U_2(u)$ which satisfy the following differential equations
\bsq
\begin{align}\label{on10a}
& \big((n-2) \dot{A}_2+d \dot{A}_1\big)U_1+\dot{U}_1=-1\,, \\ \label{on10b}
& \big((d-2) \dot{A}_1+n \dot{A}_2\big)U_2+\dot{U}_2=-1\,.
\end{align}
\esq
Then we can use these potentials to write the free energy ($\mathcal{F}=-S_{on-shell}$) as
\begin{align} \label{on11}
\mathcal{F}&=-\frac{2M_P^{d+n-1}}{d+n}V_{S^n}V_{AdS_{d}}\Big(-e^{d A_1+n A_2}\big(\frac{U_2 R_1}{e^{2 A_1}}+\frac{U_1 R_2}{ e^{2 A_2}}\big)\Big|_{u_0}^{+\infty}\nn \\
&+(d+n-1)e^{d A_1+n A_2}\big(d \dot{A}_1+n \dot{A}_2\big)\Big|^{u=+\infty}\Big)\,.
\end{align}
The volume of the $n$-sphere in slices is finite and is given in terms of its curvature by
\be\label{vsn}
V_{S^n}\equiv\frac{V_S}{R_2^{\frac{n}{2}}}=\frac{2\pi^{\frac{n+1}{2}}}{\Gamma(\frac{n+1}{2})}\Big[\frac{(n(n-1))}{R_2}\Big]^{\frac{n}{2}}\,.
\ee
The volume of $AdS_d$ space, on the other hand, is infinite and we should regularize it. Starting from the Poincar\'e coordinates with length scale $\hat{\ell}$
\be\label{poinC}
ds_{AdS_d}^2=\frac{\hat{\ell}^2}{z^2}\big(dz^2+dx_i dx^i\big)\,,
\ee
the volume can be regularized as
\be\label{vadsd1}
V_{AdS_d}=\int_0^L d^{d-1}x \int_{\hat{\epsilon}}^\infty dz \frac{\hat{\ell}^d}{z^d}=\frac{\hat{\ell}^d}{d-1}\frac{L^{d-1}}{\hat{\epsilon}^{d-1}}\,.
\ee
Using the value of $AdS_d$ curvature $R_1=-\frac{d(d-1)}{\hat{\ell}^2}$ we can rewrite the volume as
\be\label{vadsd2}
V_{AdS_d}\equiv\frac{V_A}{|R_1|^{\frac{d}{2}}}=\frac{1}{d-1}\big(\frac{L}{\hat{\epsilon}}\big)^{d-1}\Big[\frac{d(d-1)}{|R_1|}\Big]^{\frac{d}{2}}\,.
\ee
By the above values for the volumes, we can write the free energy as
\begin{align} \label{on12}
\mathcal{F}&=\frac{2M_P^{d+n-1}}{d+n}\frac{V_{S}V_{A}}{|R_1|^{\frac{d}{2}} R_2^{\frac{n}{2}}}\Big(e^{d A_1+n A_2}\big(\frac{U_2 R_1}{e^{2 A_1}}+\frac{U_1 R_2}{ e^{2 A_2}}\big)\Big|_{u_0}^{+\infty}\nn \\
&-(d+n-1)e^{d A_1+n A_2}\big(d \dot{A}_1+n \dot{A}_2\big)\Big|^{u=+\infty}\Big)\,.
\end{align}
To find the free energy, we need to compute $U_1$ and $U_2$  from \eqref{on10a} and \eqref{on10b} at the $AdS$ boundary and the IR end-point $u=u_0$.
For $d=n=4$, we already computed the scale factors near the $AdS$ boundary in equations \eqref{near1} and \eqref{near2}. Moreover, the expansion of these functions near the regular IR end-point is given in \eqref{rega1} and \eqref{rega2}. Therefore, we find the following expansions near the $AdS$ boundary for $U_1$ and $U_2$  as $u\rightarrow +\infty$
\bsq
\begin{align}\label{on13}
U_1 &=-\frac{\ell}{6} +\frac{\ell (8\mathcal{R}_1+\mathcal{R}_2)}{2016} e^{-\frac{2u}{\ell}}  +\frac{\ell (11 \mathcal{R}_1^2-76 \mathcal{R}_1 \mathcal{R}_2+11 \mathcal{R}_2^2)}{338688} e^{-\frac{4u}{\ell}} + \mathcal{B}_1  e^{-\frac{6u}{\ell}} \nn \\
&+ \frac{-26 \mathcal{R}_1^3+216 \mathcal{R}_1^2 \mathcal{R}_2-78 \mathcal{R}_1 \mathcal{R}_2^2+23 \mathcal{R}_2^3}{9483264} u e^{-\frac{6u}{\ell}}+\cdots, \\ \label{on14}
U_2 &=-\frac{\ell}{6} +\frac{\ell (8\mathcal{R}_2+\mathcal{R}_1)}{2016} e^{-\frac{2u}{\ell}}  +\frac{\ell (11 \mathcal{R}_1^2-76 \mathcal{R}_1 \mathcal{R}_2+11 \mathcal{R}_2^2)}{338688} e^{-\frac{4u}{\ell}} + \mathcal{B}_2  e^{-\frac{6u}{\ell}} \nn \\
&+ \frac{-26 \mathcal{R}_2^3+216 \mathcal{R}_2^2 \mathcal{R}_1-78 \mathcal{R}_2 \mathcal{R}_1^2+23 \mathcal{R}_1^3}{9483264} u e^{-\frac{6u}{\ell}}+\cdots,
\end{align}
\esq
and at the end-point, we obtain ($u\rightarrow u_0^+$)
\bsq
\begin{align}\label{on15}
U_1 &= -\frac{12 a_0 \ell^2 \mathfrak{b}_1}{40 a_0+\ell^2 R_1}\frac{1}{(u-u_0)^2}+ \mathfrak{b}_1 -\frac13 (u-u_0)\nn \\
&-\frac{\mathfrak{b}_1 \left(20768 a_0^2+1168 a_0 \ell^2 R_1+17 \ell^4 R_1^2\right)}{250 a_0 \ell^2 \left(40 a_0+\ell^2 R_1\right)}(u-u_0)^2+\cdots, \\ \label{on16}
U_2 &= \frac{126000 a_0^2 \ell^4 \mathfrak{b}_2}{264512 a_0^2+6112 a_0 \ell^2 R_1+53 \ell^4 R_1^2}\frac{1}{(u-u_0)^4} \nn \\
&-\frac{2100 a_0 \ell^2 \mathfrak{b}_2 \left(112 a_0+\ell^2 R_1\right)}{264512 a_0^2+6112 a_0 \ell^2 R_1+53 \ell^4 R_1^2} \frac{1}{(u-u_0)^2}-\frac15 (u-u_0)+\cdots,
\end{align}
\esq
where $\mathcal{B}_1, \mathcal{B}_2, \mathfrak{b}_1$ and $\mathfrak{b}_2$ are constants of integration \footnote{Since the free energy is dimensionless, $[L]^0$, then from \eqref{on12}  $U_1$ and $U_2$ should be $[L]^3$ and also $\mathcal{B}_1$ and $\mathcal{B}_2$. Therefore we expect $\mathcal{B}_{1,2}=\ell^3\mathcal{B}_{1,2}(\mathcal{R}_1^3,\mathcal{R}_2^3,\mathcal{R}_1^2 \mathcal{R}_2, \mathcal{R}_1 \mathcal{R}_2^2)$.}.
\subsection{Regularization}
The on-shell action as defined is infinite due to the infinite volume of the total space.
We now  introduce a regulated boundary  at $u=-\ell \log\epsilon$ and define a dimensionless cut-off
\be \label{on17}
\Lambda\equiv \frac{e^{\frac{A_1+A_2}{2}}}{\ell |R_1 R_2|^{\frac14}}\Big|_{u=-\ell\log\epsilon}=\frac{1}{\epsilon |\mathcal{R}_1 \mathcal{R}_2|^\frac14}\,.
\ee
The free energy can be computed as
\be\label{FUVmIR}
\mathcal{F}=\mathcal{F}^{\Lambda}-\mathcal{F}^{u_0}\,,
\ee
where we have
\begin{align} \label{on18}
\mathcal{F}^{\Lambda} &=- \frac{M_P^{7}\ell^7}{4}V_{S}V_{A}\Big(56 \Lambda^8+\frac43 \Lambda^6 \big(|\frac{\mathcal{R}_1}{\mathcal{R}_2}|^\frac12-|\frac{\mathcal{R}_2}{\mathcal{R}_1}|^\frac12\big)-\frac{\Lambda ^4}{504}  \big(\frac{\mathcal{R}_1}{\mathcal{R}_2}+16 +\frac{\mathcal{R}_2}{\mathcal{R}_1}\big)\nn \\
&
-\frac{\Lambda^2}{84672}\big(-2|\frac{\mathcal{R}_1}{\mathcal{R}_2}|^\frac32-29|\frac{\mathcal{R}_1}{\mathcal{R}_2}|^\frac12+29|\frac{\mathcal{R}_2}{\mathcal{R}_1}|^\frac12+2|\frac{\mathcal{R}_2}{\mathcal{R}_1}|^\frac32\big)\nn \\
&-\frac{\log(|\mathcal{R}_1 \mathcal{R}_2|\Lambda^4) }{37933056 }\big(23(\frac{\mathcal{R}_1}{\mathcal{R}_2})^2-104\frac{\mathcal{R}_1}{\mathcal{R}_2}+432-104\frac{\mathcal{R}_2}{\mathcal{R}_1}+23(\frac{\mathcal{R}_2}{\mathcal{R}_1})^2\big)\nn \\
&-\frac{19}{113799168}\big((\frac{\mathcal{R}_1}{\mathcal{R}_2})^2-10\frac{\mathcal{R}_1}{\mathcal{R}_2}+\frac{4041}{19}-10\frac{\mathcal{R}_2}{\mathcal{R}_1}+(\frac{\mathcal{R}_2}{\mathcal{R}_1})^2\big)-\frac{\mathcal{R}_2\mathcal{B}_1+\mathcal{R}_1\mathcal{B}_2}{\ell \mathcal{R}_1^2 \mathcal{R}_2^2}
\Big)\nn \\
&+\mathcal{O}(\Lambda^{-2})\,,
\end{align}
where
\be\label{onR12}
\mathcal{R}_1=\ell^2 R_1^{UV}\sp
\mathcal{R}_2=\ell^2 R_2^{UV}\,,
\ee
and
\begin{align} \label{on19}
\mathcal{F}^{u_0}= \frac{M_P^{7}}{4}V_{S}V_{A}\Big(
\frac{875 a_0^3 \ell^4 \mathfrak{b}_2}{264512 a_0^2 R_1+6112 a_0 \ell^2 R_1^2+53 \ell^4 R_1^3}-\frac{a_0^3 \ell^2 \mathfrak{b}_1}{R_1^2 \left(40 a_0+\ell^2 R_1\right)}
\Big)\,,
\end{align}
with
\be\label{on20}
a_0=e^{2A_1(u_0)}\,.
\ee
It should be noted that the free energy is independent of the constants of integration for $U_1$ and $U_2$ because it depends on the difference of the UV and IR parts i.e. equation \eqref{FUVmIR}. Therefore, we can choose $\mathfrak{b}_1=\mathfrak{b}_2=0$ and use these conditions to find the values of $\mathcal{B}_1$ and $\mathcal{B}_2$ in the UV.

The free energy that we have found so far depends on the UV cut-off $\Lambda$. To find a renormalized free energy we can add counter-terms on the $AdS$ boundary. The induced metric on this boundary is given by
\be \label{reg1}
ds^2=\gamma_{\m\n}dx^\m dx^\n = e^{2A_1(u)} ds_{AdS_d}^2+e^{2A_2(u)} ds_{S^n}^2\Big|_{u=-\ell\log\epsilon}\,.
\ee
We can read some scalar tensors on the $AdS$ boundary at $u=-\ell \log\epsilon$ as follows
\bsq
\begin{gather}\label{reg2}
\sqrt{\gamma} = e^{d A_1(u)+n A_2(u)}\sqrt{|\zeta^1| |\zeta^2|}\,,\\[5pt] \label{reg3}
R^{(\gamma)} = e^{-2A_1(u)} R_1 + e^{-2A_2(u)} R_2\,,\\
R^{(\gamma)}_{\m\n}R^{(\gamma)\m\n} =e^{-4A_1(u)} \frac{R_1^2}{d} + e^{-4A_2(u)} \frac{R_2^2}{n}\,.\label{reg4}
\end{gather}
\esq
The counter-terms required to cancel the $\Lambda$-dependent terms in \eqref{on18} are given by
\begin{align}\label{counter8}
S^{ct}&=-\frac{M_P^7}{\ell} \int d^{8}x \sqrt{\gamma}\Big(14+\frac{\ell^2}{6} R^{(\gamma)}+\frac{\ell^4}{144} (R^{(\gamma)}_{\m\n}R^{(\gamma)\m\n}-\frac27 R^{(\gamma)2} )\nn \\
&+\frac{\ell^6}{677376}(31 R^{(\gamma)3}-140 R^{(\gamma)} R^{(\gamma)}_{\m\n}R^{(\gamma)\m\n})-\frac{\ell^8}{303464448}(193 R^{(\gamma)4}\nn \\
& - 1960 R^{(\gamma)2} R^{(\gamma)}_{\m\n}R^{(\gamma)\m\n} + 5488 (R^{(\gamma)}_{\m\n}R^{(\gamma)\m\n})^2)\log(\omega |\mathcal{R}_1 \mathcal{R}_2|\Lambda^4)
\Big)\,,
\end{align}
where $\omega$ is a constant and defines our scheme of free energy. Defining $\mathcal{F}^{ct}=-S^{ct}$ we find the regularized free energy as follow
\begin{align}\label{regfree}
\mathcal{F}^{ren}&=\mathcal{F}+\mathcal{F}^{ct}\nn \\
&=-M_P^7\ell^7 V_S V_A\Big(\frac{267 \mathcal{R}_1^4 - 1004 \mathcal{R}_1^3 \mathcal{R}_2 - 2738 \mathcal{R}_1^2 \mathcal{R}_2^2 - 1004 \mathcal{R}_1 \mathcal{R}_2^3 + 267 \mathcal{R}_2^4}{910393344 \mathcal{R}_1^2 \mathcal{R}_2^2}\nn \\
&
+\frac{ \left(23 \mathcal{R}_1^4-104 \mathcal{R}_1^3 \mathcal{R}_2+432 \mathcal{R}_1^2 \mathcal{R}_2^2-104 \mathcal{R}_1 \mathcal{R}_2^3+23 \mathcal{R}_2^4\right) }{151732224 \mathcal{R}_1^2 \mathcal{R}_2^2}\log\omega \nn \\
& -\frac{\mathcal{R}_2\mathcal{B}_1+\mathcal{R}_1\mathcal{B}_2}{4\ell \mathcal{R}_1^2 \mathcal{R}_2^2}
\Big)\,.
\end{align}
\subsection{Fixing the scheme}
As it was already shown, an exact solution of equations of motion is the globally $AdS_{d+n+1}$ solution \eqref{glob0}. This solution among the regular solutions is a special case, for which $\mathcal{R}_1=-\mathcal{R}_2$. For this solution, we can either compute the free energy directly from \eqref{on9} or compute the potentials $U_1$ and $U_2$. For example, we find
\bsq
\begin{align}\label{U1G}
U_1 &=
\frac{1}{192\text{sinh}^2(u-u_0)\text{cosh}^4(u-u_0)} \Big(3 \big(c_1+\sinh (2 (u-u_0))\nn \\
&-\sinh (4 (u-u_0))+4 u\big)-\sinh (6 (u-u_0))\Big)\,, \\
U_2 &= \frac{1}{192\text{sinh}^2(u-u_0)\text{cosh}^4(u-u_0)} \Big(3 \big(c_2+\sinh (2 (u-u_0))\nn \\
&+\sinh (4 (u-u_0))-4 u\big)-\sinh (6 (u-u_0))\Big)\,, \label{U2G}
\end{align}
\esq
where $c_1$ and $c_2$ are constants of integration.
The free energy before renormalization can be read as
\be \label{freeglob}
\mathcal{F}=-M_P^7 \ell^7 V_S V_A \big(14\Lambda^8-\frac{\Lambda^4}{144}-\frac{1}{55296}\log (4\sqrt{3}\Lambda)\big)\,,
\ee
which is obviously independent of $c_1$ and $c_2$.
This result can be confirmed by using the results in equations \eqref{on18} and \eqref{on19} with IR boundary conditions $\mathfrak{b1}=\mathfrak{b}_2=0$ when we choose $\mathcal{R}_1=-\mathcal{R}_2=-48$ for the global solution.

To renormalize \eqref{freeglob} we can use the counter-terms in \eqref{counter8} with an appropriate scheme
\be \label{schm1}
\omega= e^{-\frac{89}{42}}\,,
\ee
which finally gives the free energy of the global $AdS$
\be\label{FG0}
\mathcal{F}^{Global}=0\,.
\ee
\begin{figure}[!t]\hspace{1.7cm}
\includegraphics[width =10cm]{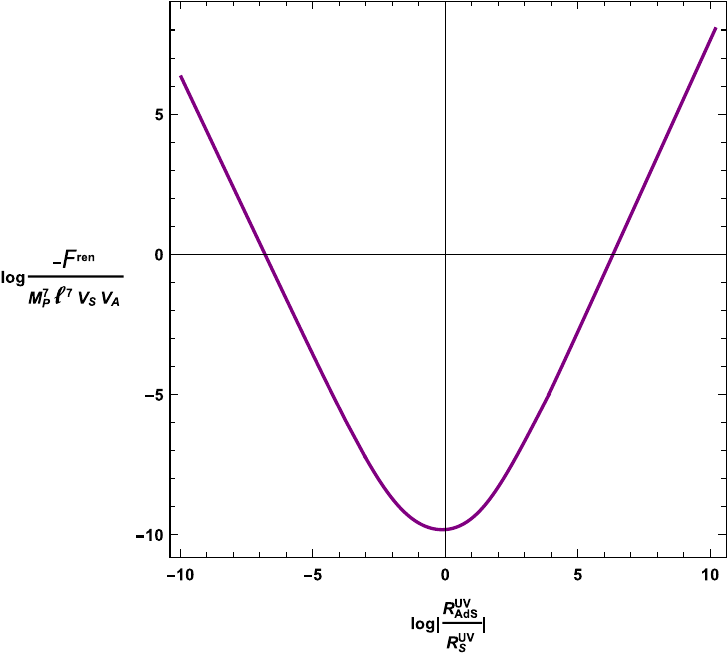}
\caption{\footnotesize{The logarithm of the renormalized free energy vs. the logarithm of the dimensionless ratio of UV curvatures. The free energy is maximum ($\mathcal{F}^{ren}\leq 0$) for global $AdS_{d+n+1}$ solution ($|R_{AdS}^{UV}/R_S^{UV}|=1$).}}\label{free1}
\end{figure}
Now in this scheme, we can compute the free energy of all the regular solutions in the theory.
This is given by
\begin{align} \label{FregF}
\mathcal{F}^{ren}
&=M_P^7\ell^7 V_S V_A\Big(\frac{89 \mathcal{R}_1^4 - 1114 \mathcal{R}_1^3 \mathcal{R}_2 + 28807 \mathcal{R}_1^2 \mathcal{R}_2^2 - 1114 \mathcal{R}_1 \mathcal{R}_2^3 + 89 \mathcal{R}_2^4}{3186376704 \mathcal{R}_1^2 \mathcal{R}_2^2}\nn \\
& +\frac{\mathcal{R}_2\mathcal{B}_1+\mathcal{R}_1\mathcal{B}_2}{4\ell \mathcal{R}_1^2 \mathcal{R}_2^2}
\Big)\,.
\end{align}
The logarithm of the renormalized free energy in terms of the dimensionless ratio of UV curvatures is sketched in figure \ref{free1}.

We should state that all different regular bulk solutions correspond to different UV sources and therefore to distinct CFTs. The distinct dual theories involve the same CFT but on $AdS\times S$
with a different ratio of radii of curvatures.
There is therefore no competition between these saddle points.
We observe, however, that the free energy (on shell action) is maximal  for the globally $AdS$ solution\footnote{The logarithm of the action in figure \ref{free1}
should be $-\infty$ at the
 $AdS$ solution. It is not because of a lack of perfect numerical accuracy.}.

\section{Solutions with $AdS_d\times S^1$ slices\label{s1}}

There is a special case of $S^n$, namely $n=1$,  that is not covered by our previous analysis, as $S^1$ has no curvature ($R_2=0$).
In this section, we study this case.
As in the previous cases studied, whatever we say is valid if replace $AdS_d$ with any $d$-dimensional constant negative curvature manifold.

 The equations of motion \eqref{eq1}--\eqref{eq3} simplify to
\be
 \label{ciq1}
 \big( d \dot{A_1}+\dot A_2\big)^2 - d \dot{A}_1^2-\dot{A}_2^2 - e^{-2A_1}
R_1 = \frac{1}{\ell^2}d(d+1)\,,
\ee
\be
 \label{ciq2}
 d \big(d \ddot{A_1}+\ddot{A_2}\big) +d (\dot{A_1} - \dot{A_2})^2 + e^{-2A_1} R_1 = 0\,,
\ee
\be
 \label{ciq3}
 \ddot{A_1} + \dot{A_1} ( d\dot{A_1}+\dot{A_2}) - \frac{1}{d} e^{-2A_1} R_1 = \ddot{A_2} + \dot{A_2} ( d\dot{A_1}+\dot{A_2})\,.
\ee
By solving $\dot{A}_2$ and $\ddot{A}_2$ from equation \eqref{ciq1} and \eqref{ciq2} and then inserting in \eqref{ciq3} we find the following equation for $A_1(u)$
\be \label{cieq1}
d e^{2 A_1} \left(2 \ell^2 \ddot{A}_1+(d+1) (\ell^2 {\dot{A}_1}^2-1)\right)-\ell^2 R_1=0\,.
\ee
This equation can be integrated to obtain
\be \label{cieq2}
 e^{(d+1)  A_1}  \dot{A}_1^2-\frac{ e^{(d-1)  A_1} }{d (d-1)  \ell^2}\big(d(d-1)  e^{2  A_1}+\ell^2 R_1\big)+\a_1=0\,,
\ee
where $\a_1$ is a constant of integration.
Given $A_1$, $A_2$ can be obtained from
\be
d(d-1)\dot A_1^2+2d\dot A_1\dot A_2=\frac{1}{\ell^2}d(d-1)+R_1 e^{-2A_1}\,.
\label{cieq3a}\ee

\subsection{Asymptotics}

Performing the same analysis as in previous sections we find the following properties for solutions with $AdS_d\times S^1$ slices:
\begin{itemize}
\item{\bf{Near boundary expansions:}}
Solving the equations of motion (\ref{ciq1})--(\ref{ciq3}), near the putative boundary either at $u\rightarrow+\infty$ or $u\rightarrow-\infty$ gives expansions for scale factors of $AdS_d$ and $S^1$ spaces. For example, for $d=3$ we find the following expansions:
\bsq
\begin{align}\label{ci0a}
A_1(u)&= \bar{A}_1\pm \frac{u}{\ell}-\frac{\mathcal{R}_1}{24}e^{\mp\frac{2u}{\ell}}-\big(\frac{\mathcal{R}_1^2}{1152}+\frac{C}{2}\big)e^{\mp\frac{4u}{\ell}}+\mathcal{O}(e^{\mp\frac{6u}{\ell}})\,, \\
A_2(u)&= \bar{A}_2\pm \frac{u}{\ell}+\frac{\mathcal{R}_1}{24}e^{\mp\frac{2u}{\ell}}-\big(\frac{\mathcal{R}_1^2}{1152}-\frac{3C}{2}\big)e^{\mp\frac{4u}{\ell}}+\mathcal{O}(e^{\mp\frac{6u}{\ell}})\,,\label{ci0b}
\end{align}
\esq
where $\mathcal{R}_1=\ell^2 R_1 e^{-2\bar{A}_1}$ is the dimensionless curvature parameter.

\item {\bf{Singular end-points:}} Considering  the expansions in \eqref{gexp1} and \eqref{gexp2}, the only singular end-point possibility is when the $AdS_d$ scale factor vanishes while the scale factor of the circle diverges i.e.
\bsq
\begin{align}\label{ci1a}
 A_1(u) &= \frac{2}{d+1} \log\frac{u-u_0}{\ell}+\frac12 \log a_0  + \mathcal{O}(u-u_0)\,,\\ \label{ci1b}
 A_2(u) &= \frac{1-d}{1+d} \log\frac{u-u_0}{\ell}+\frac12 \log s_0  +\mathcal{O}(u-u_0)\,.
\end{align}
\esq
To see this, it is easy to put $n=1$ in equation \eqref{adsshr} which gives $\l_1=\frac{2}{d+1}$ and $ \l_2=\frac{1-d}{d+1}$ while equation \eqref{sshr} gives $\l_1=0$ and $\l_2=1$ which describes a regular end-point.

\item {\bf{Regular end-points:}} Solving equations of motion \eqref{ciq1}--\eqref{ciq3} by inserting the expansions \eqref{gexp1} and \eqref{gexp2} for $\l_1=0$ and $\l_2=1$ we find the following scale factors near the regular end-point ($S^1$ shrinks but $AdS_d$ has a finite size)
\bsq
\begin{align}\label{ci2a}
e^{2A_1(u)} &= a_0 + \frac{a_0 d (d + 1) + \ell^2 R_1}{2d \ell^2 } (u-u_0)^2 \nn \\
 &-\frac{(a_0 d (d + 1) +  \ell^2 R_1) (a_0 d (d - 5) (d + 1) + (d-3) \ell^2 R_1)}{ 48 a_0 d^2 \ell^4} (u-u_0)^4 \nn \\
 &+ \mathcal{O}(u-u_0)^6\,, \\
e^{2A_2(u)} &= \frac{c_0}{\ell^2} (u-u_0)^2
+\frac{(a_0 (2+d - d^2) - \ell^2 R_1) c_0}{6 a_0 \ell^4 } (u-u_0)^4+ \mathcal{O}(u-u_0)^6\,,\label{ci2b}
\end{align}
\esq
where $c_0$ is an arbitrary positive constant.
These scale factors can be read also from \eqref{rega1} and \eqref{rega2} by replacing $n=1$ and $\frac{R_2}{n-1}=\frac{c_0}{\ell^2}$. Similar to the discussion (at the end of section \ref{regsec}) for the general $S^n$ case we cannot have a regular end-point where $AdS_d$ shrinks while $S^1$ is finite.

\item {\bf{Bounces:}} The analytic computations show that only the circle can have an A-bounce and the scale factor of $AdS_d$ is always monotonic. To see this, starting from the expansions \eqref{adsb1} and \eqref{adsb2} for an $AdS_d$ bounce, the only possible solution for coefficients is when $\hat{a}_2=\hat{a}_3=\cdots=0$ i.e. the scale factor of $AdS_d$ is constant. On the other hand  at the $S^1$ bounce, we have
\bsq
\begin{gather}\label{ci3a}
A_1(u)=\frac12\log(\hat{a}_0)+\hat{a}_1\frac{(u-u_0)}{\ell}+\hat{a}_2\frac{(u-u_0)^2}{\ell^2}+\mathcal{O}(u-u_0)^3\,,\\ \label{ci3b}
A_2(u)=\frac12\log(\hat{s}_0)+\hat{s}_2\frac{(u-u_0)^2}{\ell^2}+\hat{s}_3\frac{(u-u_0)^3}{\ell^3}+\mathcal{O}(u-u_0)^4\,,
\end{gather}
\esq
with the following coefficients
\bsq
\begin{gather}\label{ci4a}
\hat{a}_1= \pm\frac{\sqrt{d(d+1)+\frac{\ell^2 R_1}{\hat{a}_0}}}{\sqrt{d(d-1) }} \sp
\hat{a}_2= -\frac{d^2 +d +\frac{\ell^2 R_1}{\hat{a}_0}}{2  d(d-1)}\,, \\ \label{ci4b}
\hat{s}_2= \frac{d+1}{2}\sp
\hat{s}_3=-\frac{d (d+1) \sqrt{d^2+d+\frac{\ell^2 R_1}{\hat{a}_0}}}{6 \sqrt{ d(d-1)} }\,.
\end{gather}
\esq
Knowing all the properties above we have the following classes of solutions:

1. The regular solutions of ${\bf{(R,B)}}$ type. {This describes the solution outside the horizon of the black hole i.e. stretched from the horizon to the asymptotic boundary.}

2. The singular solutions of ${\bf{(R,A)}}$ type. {This describes the solution behind the horizon of the black hole i.e. stretched from horizon to singularity.}

3. The singular solutions of ${\bf{(A,B)}}$ type. {This describes a solution that is stretched from singularity to boundary (solutions with a naked singularity).}

In this case, there are also several analogs of the product space solution. One of them contains an $AdS_2$ wormhole.
We have plotted this solution in figure \ref{worm}. It will be described analytically in the next subsection.

\begin{figure}[!ht]
\begin{center}
\includegraphics[width =8cm]{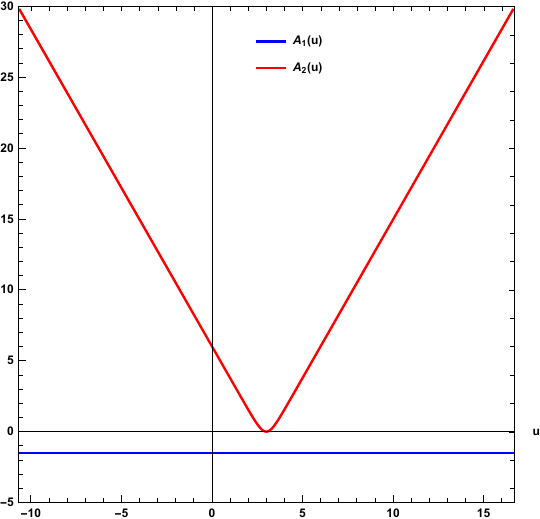}
\caption{\footnotesize{A wormhole solution with geometry given in equation \eqref{ci6}.}}\label{worm}
\end{center}
\end{figure}

\end{itemize}

\subsection{Exact solutions\label{exact}}

To proceed, from equations (\ref{ciq1})--(\ref{ciq3})  we must distinguish two cases:
\begin{itemize}
\item{$\dot A_1=0$}

From equations of motion \eqref{ciq1}--\eqref{ciq3} we find
\be \label{cip1}
e^{2A_1}=-\frac{\ell^2 R_1}{d(d+1)}\sp \ddot{A}_2+\dot{A}_2^2-\frac{d+1}{\ell^2}=0\,.
\ee
We perform the  following change of variable
\be \label{cip2}
A_2(u)=\log r(u)\,,
\ee
so that $0\leq r<+\infty$.
The equation of motion in \eqref{cip1} becomes
\be \label{cip3}
\ddot{r}-\frac{d+1}{\ell^2} r =0\longrightarrow \dot{r}^2=\frac{d+1}{\ell^2}(r^2+k)\,,
\ee
where $k$ is the constant of integration. Therefore the relation between metrics in two coordinates $u$ and $r$  is given by
\begin{align} \label{cip4}
ds^2 &= du^2+ e^{2A_2(u)} d\theta^2  +e^{2A_1(u)} ds_{AdS_d}^2\nn \\
&=\frac{\ell^2}{d+1}\frac{dr^2}{r^2+k}  +r^2 d\theta^2  -\frac{\ell^2 R_1}{d(d+1)} ds_{AdS_d}^2  \,.
\end{align}
where we have normalized the angle $\theta$ to have period $2\pi$. Any rescaling of that period via a rescaling of $r$ corresponds to a rescaling of the constant $k$.
This metric describes a product space $\mathcal{M}_2\times AdS_{d}$. Depending on the value of $k$ we have different geometries for $\mathcal{M}_2$:

1. For $k>0$, the radius of $S^1$ shrinks to zero sizes as $r\rightarrow 0$ but the geometry is regular at this point if
\be
k=\frac{\ell^2}{ d+1}\,.
\label{kk}\ee
Otherwise, there is a conical singularity at $r=0$.

2. For $k=0$, the geometry is the Euclidean $AdS_2$ ($EAdS_2$) space in Poincar\'e coordinates but with one of them compact.

3. For $k<0$, the geometry needs a better coordinate system than we now describe.

We return to the $u$ coordinate and we have the following geometries respectively:

\begin{itemize}

\item For $k>0$
we obtain
\be
 \label{ci5}
ds^2= du^2+
k \sinh^2\Big[\frac{\sqrt{d+1}}{\ell}(u-u_0)\Big] d\theta^2-\frac{\ell^2 R_1}{d (d+1)} ds^2_{AdS_d}\,,
\ee
with $u\geq 0$. The parameter $u_0$ translates into an arbitrary radius for $\theta$.
This has a generic conical singularity.

 The regular metric has
 $k$ given in (\ref{kk}).
and in such a  case  $\mathcal{M}_2$ is the Euclidean hyperboloid given by
\be
-(x^0)^2+(x^1)^2+(x^2)^2=-\frac{\ell^2}{d+1}\,.
\ee
We denote this space with $EAdS_2^+$.

\item $k=0$. The associated metric in the $u$ coordinate is
\be
\label{ci6}
ds^2= du^2+
\exp{\Big[\frac{ 2\sqrt{d+1}}{\ell}(u+c)\Big]} d\theta^2-\frac{\ell^2 R_1}{d (d+1)} ds^2_{AdS_d}\sp u\in \mathbb{R}\,.
\ee
Again the parameter $c$ translates into an arbitrary radius for the coordinate $\theta$.
This metric, if $\theta$  is non-compact is $EAdS_2$ in Poincar\'e coordinates, and it is diffeomorphic to the hyperboloid. When $\theta$ is compact, this is not true
anymore.
We denote this space as $EAdS_2^0$.

\item $k<0$.  In this case the metric (\ref{cip4}) extends from $r\in [-\sqrt{|k|},+\infty)$.
This gives the metric in the $u$ coordinate
\be
ds^2= du^2+
|k| \cosh^2\Big[\frac{\sqrt{d+1}}{\ell}(u-u_0)\Big] d\theta^2-\frac{\ell^2 R_1}{d (d+1)} ds^2_{AdS_d}\,,
\label{ci7}
\ee
with $u>u_0$.
However, this metric is not geodesically complete and one has to extend $u$ to all real values.
The manifold now is a wormhole with $S^1$ boundaries that is the usual $EAdS_2$. We shall denote it as $EAdS^-_2$
The constant $u_0$ allows an arbitrary radius for the $S^1$.

\end{itemize}
All the $\mathcal{M}_2$ metrics above also exist when $\theta$ is taking values in the real line.

In all three cases above the Kretschmann scalar of the total $(d+2)$-dimensional manifold is the same and is equal to
\be \label{cik}
\mathcal{K}=\frac{2 (d+1)^2 (3 d-2)}{(d-1) \ell^4}\,.
\ee
We can obtain Minkowski signature solutions by analytically continuing $\theta\to it$.
In this case $EAdS_2^{+,0}$ becomes the $AdS_2$ black hole while $EAdS^-_2$ becomes $AdS_2$.

\item{$\dot A_1\not =0$.}

In this case
$A_2(u)$ can be obtained from
\be \label{cieq3}
A_2(u)=
\int \frac{\alpha_1  (d-1) \ell^2 e^{-(d+1) A_1}+2}{2 \ell^2 \dot{A}_1} du + \a_2\,.
\ee
Here $\a_2$ is another constant of integration.
By defining
\be \label{cieq4}
A_1(u)=\log r(u)\,,
\ee
equation \eqref{cieq2} becomes
\be \label{cieq5}
(\frac{dr}{du})^2\equiv f(r)= -\alpha_1  r^{1-d}+\frac{R_1}{d(d-1) }+\frac{r^2}{\ell^2}\,,
\ee
and equation \eqref{cieq3} gives
\be \label{cieq6}
A_2(r(u))= \frac12 \log f(r)+\frac{1}{2}\log(\ell^2 d (d-1))+\a_2\,.
\ee
Therefore the metric in these two coordinates are related as follows
\begin{align} \label{cieq7}
ds^2 &= du^2 + e^{2A_1(u)} ds_{AdS_d}^2+ e^{2A_2(u)} d\theta^2 \nn \\
&=\frac{dr^2}{f(r)}+r^2 ds_{AdS_d}^2 + \ell^2 d(d-1) e^{2\a_2} f(r) d\theta^2\,,
\end{align}
where $f(r)$ is defined in equation \eqref{cieq5}. The last metric describes the (well-known) topological black holes with a negative cosmological constant, \cite{top1,top2}, (they are reviewed in appendix \ref{topo}).

The function $f(r)$ in (\ref{cieq5}) has the following properties ($d\geq 2$).

\begin{enumerate}

\item $f\to +\infty$ as $r\to+\infty$.

\item $r\to 0$ is always a curvature singularity of the metrics in (\ref{cieq7}) and the Kretchmann scalar is given by
\be
\mathcal{K}=
 \alpha_1 ^2 (d^2-1) d^2 r^{-2 (d+1)}+\frac{2 (d+1)(d+2)}{\ell^4}\,.
\ee

\item $f\to +\infty$ as $r\to 0^+$ when $\alpha_1<0$, and $f\to -\infty$ as $r\to 0^+$ when $\alpha_1>0$. As we show below when $\alpha_1=0$ the space is $AdS_{d+2}$ provided $\alpha_2$ is chosen appropriately.

\item At a fixed value of $R_1$, there is a value $\a_1^{crit}<0$ so that if $\a_1<\a_1^{crit}$, then  always $f>0$ as $r\in [0,+\infty)$.
All solutions with $\a_1<\a_1^{crit}$ have a bad naked singularity.

\item When  $\a_1=\a_1^{crit}$ then there is a single positive double zero of $f$. In this case, the Minkowski signature solution has an extremal horizon. The geometry near this extremal horizon is AdS$_2\times$AdS$_d$.

\item When $\a_1^{crit}<\a_1<0$ then $f$ has two positive zeroes with $f$ being negative in between the zeroes. The structure of such black holes is similar to Reissner-N\"orstrom ones. In particular, the inner horizon is a Cauchy horizon. If the hyperbolic slice is a finite volume manifold, then such black holes have finite entropy.

\item When $\a_1>0$, then $f$ has a single positive zero. Beyond this zero (at small values of $r$), $f<0$.

\item $r\to\infty$ is a regular conformal boundary of the metrics in (\ref{cieq7}).

\item The relevant (Euclidean) solutions are all segments between a zero or a divergence of $f$, while $f\geq 0$.

\item Most of these solutions are singular. The only potentially regular Euclidean solutions are those between $r=+\infty$ and the first non-trivial zero $r_*$ for $f$.
 The regular solutions are obtained by adjusting the constant $\a_2$ as
 \be
 e^{2\a_2}=\frac{4}{ d(d-1)\ell^2 (f'(r_*))^2}\,.
 \ee
\end{enumerate}

\end{itemize}

We conclude that in this case we have two families of regular solutions in the Euclidean case:

$\bullet$ The solutions with   $\a_1^{crit}<\a_1<0$ that become RN-like black holes upon analytic continuation of $\theta\to i\theta$.

$\bullet$ The solutions with $\a_1>0$, that become Schwarzschild-like black holes upon analytic continuation of $\theta\to i\theta$.

Minkowski signature solutions can also be obtained by giving to the $AdS_d$ a Minkowski signature.

The ``ground" state solution, is the extremal solution with $\a_1=\a_1^{crit}$. It has zero temperature, but the asymptotic circle can have any radius.
All other solutions have a fixed asymptotic circle radius that is correlated with their temperature.
At a fixed asymptotic circle radius the non-extremal black holes have a lower free energy compared to the extremal one, \cite{top2}.

\subsection{ The global $AdS_{d+2}$ solution}

The solution obtained from \eqref{cieq2} and \eqref{cieq3} when we choose $\a_1=0$  is
\be \label{ci8}
ds^2= du^2 -\frac{\ell^2 R_1}{d(d-1)} \cosh^2\frac{u-u_0}{\ell} ds^2_{AdS_d} -e^{2\a_2}\ell^2 R_1 \sinh^2\frac{u-u_0}{\ell} d\theta^2\,.
\ee
Choosing
\be
e^{2\a_2} R_1=-1\,,
\ee
this metric is the metric of $AdS_{d+2}$ in global coordinates.
By the following change of variables
\be  \label{ci9}
-\ell^2 k \cosh^2\frac{u-u_0}{\ell}=r^2\sp \theta=\sqrt{\frac{k}{e^{2\a_2}\ell^2 R_1}} i\,t \quad;\quad k\equiv\frac{R_1}{d(d-1)}\,,
\ee
the metric becomes
\be  \label{ci10}
ds^2=-f(r) dt^2+\frac{1}{f(r)}dr^2+r^2 ds_{AdS_d}^2 \quad;\quad f(r)=\frac{r^2}{\ell^2}+k\,.
\ee
This is the solution that has been discussed in appendix \ref{topo} when $M=0$.

\subsection{Relations between parameters in two coordinates}
Let us for simplicity consider $d=3$ and $\dot{A}\neq 0$. Solving equation \eqref{cieq5} gives us
\be \label{ru1}
r^2(u)=
\frac{1}{24} \left(\ell^2 e^{-2 \sqrt{6} c_1-\frac{2 u}{\ell}} \left(144 \a_1+\ell^2 R_1^2\right)+e^{2 \sqrt{6} c_1+\frac{2 u}{\ell}}-2 \ell^2 R_1\right)\,,
\ee
where $c_1$ is the constant of integration. Moreover, there is another solution that can be found from \eqref{ru1} by replacing $u\rightarrow -u$. At large values of $r$ or when $u\rightarrow +\infty$ we can find the expansions of scale factors in \eqref{ci0a} and \eqref{ci0b} by using equations \eqref{cieq4} and \eqref{cieq6} together with \eqref{cieq5} if we choose
\begin{align} \label{ru2}
\bar{A}_1= \sqrt{6} c_1 -\frac12\log 24\sp \bar{A}_2=\a_2+\sqrt{6} c_1-\log 2\sp \a_1=-\frac{4C}{\ell^2} e^{4\bar{A}_1}\,,
\end{align}
where $\bar{A}_1, \bar{A}_2$ and $C$ are parameters in $u$ coordinate while $\a_1, \a_2$ and $c_1$ are in $r$ coordinate.

Now let us consider a solution which is regular at $u=u_0$ and has a boundary at $u\rightarrow +\infty$. The regularity at $u=u_0$ implies that the scale factor of $A_1$ is constant but $e^{A_2}\rightarrow 0$. In $r$ coordinate this translates to a solution where at $r=r_h$
\be \label{ru3}
f(r_h)=-\a_1 r_h^{-2}+\frac{R_1}{6}+\frac{r_h^2}{\ell^2}=0\,.
\ee
This describes a topological black hole with a horizon at $r=r_h$, see appendix \ref{topo}.
Now we can compare the expansions near the regular end-points. The expansions in $u$ coordinate are given in \eqref{ci2a} and \eqref{ci2b}. The two free parameters $a_0$ and $c_0$ then are given by
\be
a_0= r_h^2\sp c_0=e^{2\a_2}\frac{(12 r_h^2+\ell^2 R_1)^2}{6 r_h^2}\,.
\ee
We can read the values of the sources and vevs in terms of the black hole solutions
\be
\mathcal{R}_1=\ell^2 R_1 e^{-2\bar{A}_1}=24\ell^2 R_1 e^{-2\sqrt{6}c_1}\sp C=-\frac14 \a_1\ell^2 e^{-4\bar{A}_1}= -144\omega M \ell^2 e^{-4\sqrt{6}c_1}\,,
\ee
where $\omega$ and $M$, the mass of the black hole, are defined in \eqref{16a} and \eqref{19}. We can choose $\bar{A}_1=\bar{A}_2=0$ for simplicity then we have $c_1=\frac{\log 24}{2\sqrt{6}}$ and $\a_2=-\frac12 \log 6$ and
\be
\mathcal{R}_1=\ell^2 R_1 \sp C= -\frac{\omega M \ell^2}{4}\,.
\ee
Knowing the above parameters, we now compute the free energy of the regular solutions in both coordinates. The Euclidean action is given by
\begin{align} \label{ru4}
I_{E}&=\frac{M_P^{3}}{2}V_{S}V_{AdS_3}\Big(e^{A_1+ A_2} U R_1\Big|_{u_0}^{+\infty}-3 e^{3A_1+ A_2}\big(3 \dot{A}_1+ \dot{A}_2\big)\Big|^{u=+\infty}\nn \\
&+e^{3A_1+A_2}(3\dot{A}_1+\dot{A}_2)\Big|_{u_0}\Big)\,,
\end{align}
where the last term is non-zero, unlike the $n>1$ cases. In this equation $V_{AdS_3}\sim 1/|R_1|^{\frac32}$ is the volume of three dimensional slice and
\be
V_S=\int_0^\b d\theta\,,
\ee
where $\beta$ is the length of $S^1$.
In \eqref{ru4}, $U$ is a scalar field given by
\be \label{ru5}
(\dot{A}_1+\dot{A}_2) U + \dot{U}+1=0\,,
\ee
and by using the expansions \eqref{ci0a} and \eqref{ci0b} we find that as $u\rightarrow +\infty$
\be \label{ru6}
U(u)=-\frac{\ell}{2}+ \mathcal{B} e^{-2\frac{u}{\ell}}+(C\ell-\frac{\ell^5 R_1}{576 e^{4\bar{A}_1}})e^{-4\frac{u}{\ell}}+\mathcal{O}(e^{-6\frac{u}{\ell}})\,.
\ee
Moreover, near the regular end-point $u=u_0$ equations \eqref{ci2a} and \eqref{ci2b} give
\be
U(u)=\frac{\mathfrak{b}}{u-u_0}-(\frac12+\frac{2\mathfrak{b}}{3\ell^2})(u-u_0)+\mathcal{O}(u-u_0)^3\,.
\ee
We can consider $\mathfrak{b}=0$, so the contribution to the free energy of the first term of \eqref{ru4} at $u=u_0$ is zero.
On the other hand, we can solve $U$ in $r$ coordinate exactly, which we find
\be \label{ru7}
\big(\frac{R_1}{3}+4\frac{r^2}{\ell^2}\big)U+2 r f U'+2 r f^{\frac12}=0\,,\rightarrow
U(r)= \frac{2 c_2-\sqrt{6} \ell r^2}{2 \sqrt{6 r^4+\ell^2 r^2 R_1-6 \a_1 \ell^2 }}\,,
\ee
where $c_2$ is another constant of integration. $U(r)$ is diverging at $r=r_h$ because of \eqref{ru3}. To have a regular function at this point we should have
\be
c_2=\sqrt{\frac32}\ell r_h^2\,.
\ee
If we expand the solution \eqref{ru7} near the boundary at $r\rightarrow +\infty$ and change $r\rightarrow u$ by using \eqref{ru1} we obtain
\be
\mathcal{B}=
\left(4 \sqrt{6} c_2+\ell^3 R_1\right) e^{-2 \sqrt{6} c_1}= \ell\left(12  r_h^2+\ell^2 R_1\right) e^{-2 \sqrt{6} c_1}\,.
\ee
Returning to \eqref{ru4}, if we do the proper counter-terms we can compute the renormalized action as following
\begin{align}
I_E^{ren}&= \frac12 M_P^3 V_S V_{AdS_3}\big( e^{\bar{A}_1+\bar{A}_2}R_1 \mathcal{B}-\frac{1}{\ell}a_0^\frac32 c_0^\frac12\big)\nn \\
&=
M_P^3 V_S V_{AdS_3}\sqrt{6} e^{\a_2}\ell \big(-\frac{r_h^4}{\ell^2}+\frac16 R_1 r_h^2+\frac{1}{48} R_1^2 \ell^2\big)\,.
\end{align}
The last relation is the known result of free energy for topological black holes with negative cosmological constant and $r_h=r_+$, see appendix \ref{topo} for more details. By using the definition of temperature, we find that
\be
\beta = \frac{2\pi }{(e^{A_2})'}\Big|_{u=u_0}\rightarrow T=\frac{c_0^\frac12}{2\pi \ell}\,.
\ee
The free energy of the black hole is given by
\be
\mathcal{F}=\frac{I_E^{ren}}{\b}=(M-M_{crit})-T S\,,
\ee
where $M, M_{crit}$ and $S$ are mass, critical mass, and entropy of the black hole respectively, and are given in equations \eqref{19}, \eqref{21} and \eqref{22}.

\section{On general Einstein manifold solutions with constant negative curvature.}

The general solutions  with constant negative curvature we have found in this paper, and many previous ones provide a hierarchical construction
of such solutions as conifolds of conifolds of conifolds etc.

A few examples are as follows:

In two dimensions, the solutions to Einstein's equations with a negative cosmological constant, up to diffeomorphisms consist of the family of manifolds $\mathcal{M}_2$ we described in the previous section.

In three dimensions, the solutions to Einstein's equations with a negative cosmological constant, up to diffeomorphisms consist of the two-parameter family of rotating $AdS_3$-Schwarszschild black holes (that includes also $AdS_3$.

Consider now solutions in four dimensions with a negative cosmological constant.
The maximally symmetric solution is $AdS_4$ and in global coordinates, it has $S^1\times S^2$ slices. The Euclidean symmetry is $O(4,1)$.
In this same slicing belongs also the $AdS_4$-Schwarszschild black hole with generic symmetry $O(2)\times O(3)$.
The difference between these two solutions is that, in the first, it is $S^2$ that shrinks to zero size ending the geometry while in the second,  it is the $S^1$ that shrinks to zero size ending the geometry.

There are however further solutions where the slices are $S^3$, \cite{C}, with generic symmetry $O(4)$,  as well as conifold solutions with $S^1\times S^1 \times S^1$ solutions (tori) that correspond to
$AdS_4$-Schwarszschild black holes with the toroidal horizon and generic symmetry $O(2)^3$.
There are also the $AdS_2\times S^1$ solutions studied here with generic symmetry $O(2)\times O(2,1)$. In the place of $AdS_2$ above we can have any of our $\mathcal{M}_2$ solutions.
We can also replace $S^1$ with $R$.

All of these four-dimensional solutions are generically distinct and provide a large class of solutions with four-dimensional constant negative curvature.
The structure of their boundaries differs. Some solutions are diffeomorphic to each other, but most are distinct manifolds.
We do not know if they exhaust all solutions with the negative cosmological constant.

We now move to the next dimension which is five and describe the various conifold solutions to the constant negative curvature equations.
The slices can be $(S^1)^4$, $S^1\times S^3$, $S^2\times S^2$, $(S^1)^2\times S^2$, $AdS_2\times S^2$, $AdS_2\times (S^1)^2$, $AdS_3\times S^1$ and $AdS_2\times AdS_2$ and $AdS_4$.
For example, $(S^1)^4$ are the five-dimensional black holes with the toroidal horizon,   $S^1\times S^3$ are the five-dimensional black holes with $S^3$ horizon, and $(S^1)^2\times S^2$ are the five-dimensional black holes with $S^1\times S^2$ horizon. The $S^2\times S^2$ solution was analyzed in \cite{S2xS2} and exhibited Effimov phenomena.
 In the $AdS_2\times S^2$ and  $Ads_2\times (S^1)^2$ solutions $AdS_2$ stands for the one-parameter family of $AdS_2$ black holes.
 The  $AdS_3\times S^1$ solutions contained in the slice the full two-parameter family of BTZ black holes. Finally,  $AdS_2\times AdS_2$ slices have not been systematically studied so far but we expect these solutions to have two boundaries as neither $AdS_2$ can shrink regularly to zero size.

  This algorithm clearly generalizes to higher dimensions.
The structure of the boundaries of such solutions is variable.

\section*{Acknowledgements}
\addcontentsline{toc}{section}{Acknowledgements}

We would like to thank  C. Behan, T. Brennan, M. Chernodub, J. Gauntlett, C. Herzog, A. Konechny, A. Lerda, V. Niarchos, M. Roberts, C. Rosen, J. Russo, A. Stergiou,  E. Tonni and A. Tseytlin for helpful conversations.

This work was supported in part by CNRS grant IEA 199430.
The work of A. G. is supported by Ferdowsi University of Mashhad
under grant 2/60036 (1402/03/06).

\newpage
\appendix

\begin{appendix}
\renewcommand{\theequation}{\thesection.\arabic{equation}}
\addcontentsline{toc}{section}{Appendices}
\section*{APPENDIX}

\section{Product space ansatz for the slice\label{conii}}
Consider the following ansatz, a block diagonal $(d+1)$-dimensional metric
\be
ds^2 = g_{ab} d{x^a} d{x^b} = du^2 + \sum_{i=1}^n \mathrm{e}^{2A_i(u)} \zeta^i_{\alpha_i, \beta_i} d{x^{\alpha_i}} d{x^{\beta_i}}\,,
\label{i1}
\ee
where $\zeta^i_{\alpha_i \beta_i}$ is the $d_i$-dimensional metric of the $i$th Einstein manifold, $\alpha_i$ and $\beta_i$ take values in the $d_i$ coordinates of this manifold. Each Einstein manifold is associated with a different scale factor, all depending on the coordinate $u$ only. Note that every $d$-dimensional slice at constant $u$ is given by the product of $n$ Einstein manifolds of dimension $d_1, \ldots, d_n$.

For this ansatz, the Ricci tensor reads
\bsq
\begin{gather}
R_{uu}  = -\sum_{k=1}^n d_k \big(\ddot{A_k} + {\dot{A}_k}^2\big)\,, \label{11}\\
R_{u\alpha}  = 0 \quad \qfor \quad \alpha \neq u\,, \label{9}\\
R_{\alpha_i \beta_i}  = -\big(\ddot{A_i} + \dot{A_i} \sum_{k=1}^n d_k \dot{A_k}\big) g_{\alpha_i \beta_i} + R^{\zeta^i}_{\alpha_i \beta_j}\,, \label{10}\\
R_{\alpha_i \beta_i}  = 0 \quad \qfor \quad i \neq j\,,
\label{i2}
\end{gather}
\esq
where $R^{\zeta^i}_{\alpha_i \beta_i}$ is the Ricci tensor of the $d_i$-dimensional Einstein metric $\zeta_{\alpha_i \beta_i}^i$. Thus, the Ricci scalar is
\begin{equation}
R = -2 \sum_{k=1}^n d_k \ddot{A_k} - \big(\sum_{k=1}^n d_k \dot{A_k}\big)^2 - \sum_{k=1}^n d_k \dot{A}_k^2 + \sum_{k=1}^n \mathrm{e}^{-2A_k} R^{\zeta^k}\,,
\label{i3}\end{equation}
where $R^{\zeta^k}$ is the Ricci scalar of the metric $\zeta^k$.

We consider in general an  Einstein-dilaton theory in a $d+1$ dimensional bulk space-time.
The most general two-derivative action is
\be
S= M_P^{d-1} \int d^{d+1}x \sqrt{-g} \Big(
R - {\frac12}g^{ab}\pa_a\f \pa_b\f- V(\f)
\Big)\,.
\label{inA2}\ee
The energy-momentum tensor $T_{\mu \nu} = \partial_\mu \varphi \partial_\nu \varphi - g_{\mu \nu}(\frac{1}{2} \partial_a \varphi  \partial^a \varphi + V)$ would be as follow
\bsq
\begin{gather}
T_{uu}  = \frac{1}{2} \dot{\varphi}^2 - V\,, \\
T_{\alpha_i \beta_j}  = -g_{\alpha_i \beta_j} \big(\frac{1}{2} \dot{\varphi}^2 + V \big)\,.
\label{i4}
\end{gather}
\esq
Finally, the Einstein tensor reads
\bsq
\begin{align}
G_{uu} & = \frac{1}{2} \big(\sum_{k=1}^n d_k \dot{A_k}\big)^2 - \frac{1}{2} \sum_{k=1}^n d_k \dot{A}_k^2 - \frac{1}{2} \sum_{k=1}^n \mathrm{e}^{-2A_k} R^{\zeta^k}\,, \label{8}\\
G_{\alpha_i \beta_i} & = \Big(-\big(\ddot{A_i} + \dot{A_i} \sum_{k=1}^n d_k \dot{A_k}\big) + \sum_{k=1}^n d_k \ddot{A_k} + \frac{1}{2} \big(\sum_{k=1}^n d_k \dot{A_k}\big)^2 \nonumber \\
& \mathrel{\phantom{=}} {} + \frac{1}{2} \sum_{k=1}^n d_k \dot{A}_k^2 - \frac{1}{2} \sum_{k \neq i} \mathrm{e}^{-2A_k} R^{\zeta^k}\Big) g_{\alpha_i \beta_i} + G_{\alpha_i \beta_i}^{\zeta^i}\,,
\label{i5}
\end{align}
\esq
where $G^{\zeta^i}_{\alpha_i \beta_i}$ is the Einstein tensor of the metric $\zeta^i$. Since $\zeta^i$ is an Einstein metric, we have
\be
G^{\zeta^i}_{\alpha_i \beta_i} = \big(\frac{1}{d_i} - \frac{1}{2}\big) R^{\zeta^i} \zeta_{\alpha_i, \beta_i} = \big(\frac{1}{d_i} - \frac{1}{2}\big) \mathrm{e}^{-2A_i} R^{\zeta^i} g_{\alpha_i, \beta_i}\,.
\label{i6}
\ee
Hence, we can rewrite the $\alpha_i \beta_i$ component of the Einstein tensor as
\begin{equation}
G_{\alpha_i \beta_i} = \Big(-\big(\ddot{A_i} + \dot{A_i} \sum_{k=1}^n d_k \dot{A_k} - \frac{1}{d_i} \mathrm{e}^{-2A_i} R^{\zeta^i} \big) + \sum_{k=1}^n d_k \ddot{A_k} + \sum_{k=1}^n d_k \dot{A}_k^2 + G_{uu}\Big) g_{\alpha_i \beta_i}\,.
\label{i7}\end{equation}
Therefore, the equations of motions, given by $2G_{\mu \nu} = T_{\mu \nu}$, read
\bsq
\begin{gather}
\label{ieq:EOM1bis}
 \big(\sum_{k=1}^n d_k \dot{A_k}\big)^2 - \sum_{k=1}^n d_k \dot{A}_k^2 - \sum_{k=1}^n \mathrm{e}^{-2A_k} R^{\zeta^k} - \frac{1}{2} \dot{\varphi}^2 + V  = 0\,, \\
\label{ieq:EOM2bis}
-\big(\ddot{A_i} + \dot{A_i} \big(\sum_{k=1}^n d_k \dot{A_k}\big) - \frac{1}{d_i} \mathrm{e}^{-2A_i} R^{\zeta^i} \big) + \sum_{k=1}^n d_k \ddot{A_k} + \sum_{k=1}^n d_k \dot{A}_k^2 + \frac12\dot{\varphi}^2  = 0\,, \\
\label{ieq:EOM3bis}
\ddot{\varphi} + \sum_{k=1}^n d_k \dot{A_k} \dot{\varphi} - \partial_\f V  = 0\,.
\end{gather}
\esq
Multiply \eqref{ieq:EOM2bis} by $d_i$, sum over $i$, divided by $d$ and reorganize we obtain
\be
\label{ieq:EOM4bis}
2(1 - \frac{1}{d}) \sum_{k=1}^n d_k \ddot{A_k} + \frac{2}{d} \sum_{i<j}d_id_j (\dot{A_i} - \dot{A_j})^2 + \frac{2}{d} \sum_{k=1}^n \mathrm{e}^{-2A_k} R^{\zeta^k} + \dot{\varphi}^2 = 0\,,
\ee
the symmetric version of \eqref{ieq:EOM2bis}.

These equations are valid for any choice of $n$ Einstein metrics $\zeta^i_{\alpha_i \beta_i}$. However, the existence of solutions is not guaranteed for an arbitrary choice. The difference between \eqref{ieq:EOM2bis} for different indices yields constraints on the scale factors and the curvatures
\be \label{ieq:EOM5bis}
\ddot{A_i} + \dot{A_i} \sum_{k=1}^n d_k \dot{A_k} - \frac{1}{d_i} \mathrm{e}^{-2A_i} R^{\zeta^i} = \ddot{A_j} + \dot{A_j} \sum_{k=1}^n d_k \dot{A_k} - \frac{1}{d_j} \mathrm{e}^{-2A_j} R^{\zeta^j}\,,
\ee
for all $i$ and $j$. One can see that these constraints are satisfied by $A_i = A(u)$ and $R^{\zeta^i} = d_i \kappa$ for all $i$, where $\kappa$ is a constant and $A(u)$ is a function of $u$. In this case, the equations of motion \eqref{ieq:EOM1bis}--\eqref{ieq:EOM3bis} reduce to
\bsq
\begin{gather}
 d(d-1) \dot{A}^2-e^{-2A} R^{\zeta}-\frac12 \dot{\f}^2+V=0\,, \label{ieq:EOM1} \\
 2(d-1)\ddot{A}+\dot{\f}^2+\frac{2}{d}e^{-2A}R^{\zeta}=0\,, \label{ieq:EOM2} \\
 \ddot{\f}+d \dot{A} \dot{\f}-\partial_\f V=0\,. \label{ieq:EOM3}
\end{gather}
\esq
This could be foreseen since under these conditions there is only one scale factor and the product space is an Einstein manifold, since \eqref{ieq:EOM1}--\eqref{ieq:EOM3} are valid for any Einstein metric $\zeta_{\mu \nu}$.

\subsection{The curvature invariants}\label{apk}
To check the regularity of the solutions we should calculate $R^2$, $R_{ab}R^{ab}$, and $R_{abcd}R^{abcd}$ for the metrics above.
The first two are straightforward to compute using \eqref{11}--\eqref{i2}.
The Ricci squared is
\bea
R_{ab}R^{ab} & =& R_{uu}R^{uu} + R_{\alpha_i \beta_i}R^{\alpha_i \beta_i} \nonumber\\
&= &\Big(\sum^{n}_{i=1}d_{i}(\ddot{A}_{i} + \dot{A}^{2}_{i})\Big)^{2} +\sum_{i=1}^{n} d_{i}\Big(\mathrm{e}^{-2A_{i}}\kappa - \big(\ddot{A}_{i} + \dot{A}_{i}\sum^{n}_{j =1}d_{j}\dot{A}_{j}\big)\Big)^{2}\,.
\eea
The Ricci scalar reads
\be
R = -2 \sum_{i=1}^n d_i \ddot{A_i} - \big(\sum_{i=1}^n d_i \dot{A_i}\big)^2 - \sum_{i=1}^n d_i \dot{A}_i^2 + \sum_{i=1}^n \mathrm{e}^{-2A_i} R^{\zeta^i}\,.
\label{iriccisc}
\ee
The non-zero Riemann tensor components are given by
\bsq
\begin{align}
 R_{\alpha_{i} uu \beta_{i}} &= e^{2A_{i}}\zeta^{i}_{\alpha_{i}\beta_{i}}\big(\ddot{A}_{i} + \dot{A}^{2}_{i}\big) = - R_{u\alpha_{i}u\beta_{i}}= - R_{\alpha_{i}u\beta_{i}u}\,.\\ \label{iRabcd}
R_{\alpha_{i}\beta_{j}\gamma_{k}\delta_{l}} &=  e^{2(A_{i} + A_{j})}\dot{A}_{i}\dot{A}_{j}\big(\delta_{il}\delta_{jk}\zeta^{i}_{\alpha_{i}\delta_{i}}\zeta^{j}_{\beta_{j}\gamma_{j}}- \delta_{ik}\delta_{jl}\zeta^{i}_{\alpha_{i}\gamma_{i}}\zeta^{j}_{\beta_{j}\delta_{j}}\big)\nn \\
&+e^{2A_{i}}\delta_{ij}\delta_{kl}\delta_{ik}R_{\alpha_{i}\beta_{i}\gamma_{i}\delta_{i}}^{\zeta^{i}}\,,
\end{align}
\esq
and one can see that the Riemann tensor is pairwise diagonal.
So one can calculate  the Kretschmann scalar  as a sum of all non-zero components of  the Riemann tensor
\bea
\mathcal{K} = 4K^{2}_{1} + K^{2}_{2} +2K^{2}_{3}\,,
\eea
where
\bea
K_{1} = R_{u\alpha_{i}}^{\,\,\,\,\,\,\,u\beta_{i}} = \delta_{\alpha_{i}}^{\,\,\,\,\,\beta_{i}}(\ddot{A}_{i} +\dot{A}^{2}_{i})\,.
\eea
Equation \eqref{iRabcd} with $i=k$ and $j=l$ gives
 \bea
 K_{2} = R_{\alpha_{i}\beta_{j}}^{\,\,\,\,\,\,\,\,\gamma_{i}\delta_{j}} = \dot{A}_{i}\dot{A}_{j}(-\delta_{\alpha_{i}}^{\,\,\,\gamma_{i}}\delta^{\,\,\,\delta_{j}}_{\beta_{j}})\,,
 \eea
and with $i=j=k=l$ gives
 \bea
 K^{2}_{3} =(R_{\alpha_{i}\beta_{j}}^{\,\,\,\,\,\,\,\gamma_{i}\delta_{i}})^{2}= e^{-4A_{i}}\mathcal{K}^{\zeta^{i}} - 4e^{-2A_{i}} (\dot{A}_{i})^{2}R^{\zeta^{i}} - 2d_{i}(d_i-1)(\dot{A}_{i})^{4}\,.
 \eea
Finally the Kretschmann scalar is
\begin{align}
\mathcal{K} &=
\sum^{n}_{i =1}\Big(e^{-4A_{i}}\mathcal{K}^{\zeta^{i}} - 4e^{-2A_{i}} (\dot{A}_{i})^{2}R^{\zeta^{i}} - 2d_{i}(\dot{A}_{i})^{4} \nn \\
&+ 4d_{i}(\ddot{A}_{i} + \dot{A}^{2}_{i})^{2}\Big) +  \sum^{n}_{i, j =1} 2d_{i}d_{j}\big(\dot{A}_{i}\dot{A}_{j}\big)^{2}\,,
\end{align}
where $\mathcal{K}^{\zeta^{i}}$ is the Kretschmann scalar related to $\zeta^{i}$.

\section{Various global coordinates on $AdS_{d+n+1}$ and its Euclidean version}\label{adscor}

In this appendix, we consider various coordinate systems of $AdS_{d+n+1}$, both standard global coordinates as well as coordinates adapted to $AdS_d\times S^n$ slices and their Euclidean versions. They will be important as benchmarks for the space of solutions we shall find.

\subsection{Standard global coordinates on $AdS_{d+n+1}$}
We consider the embedding equation  that defines $AdS_{d+n+1}$
\be
-(x^0)^2-(x^{(-1)})^2+\sum_{i=1}^{d+n}(x^i)^2=-\ell^2\,.
\label{jj14}
\ee
We start with the standard global coordinates.
First, we parameterize
\bsq
\begin{gather}
x^i=r_2 n^i\sp  i=1,2,\cdots d+n \sp n^in^i=1 \sp r_2\geq 0\,,
\label{j20} \\
x^0=r_1\cos\theta\sp x^{(-1)}=r_1\sin\theta\sp r_1\geq 0\,.
\label{j21}
\end{gather}
\esq
The Minkowski signature metric in $(2,d+n)$ dimensions becomes
\be
ds^2=-(dx^0)^2-(dx^{(-1)})^2+dx^idx^i=-dr_1^2-r_1^2d\theta^2+dr_2^2+r_2^2d\Omega_{d+n-1}^2\,,
\label{jj22}
\ee
and the constraint in (\ref{jj14}) can be written as
\be
-r_1^2+r_2^2=-\ell^2\ar r_1^2-r_2^2=\ell^2\,.
\label{j23}
\ee
We now introduce new coordinates
\be
r_1=\ell\cosh(\rho)\sp r_2=\ell\sinh(\r)\sp \ell\geq 0\sp \rho\geq 0\,,
\label{j24}
\ee
and rewrite the metric in (\ref{jj22}) as
\be
ds^2=-d\ell^2-\ell^2\cosh^2(\r)d\theta^2+\ell^2 d\r^2+\ell^2\sinh^2(\r)d\Omega_{d+n-1}^2\,.
\label{j25}\ee
$AdS_{d+n+1}$ is obtained from the metric above by setting $\ell$ to be constant
\be
ds^2_{n+d+1}=\ell^2\Big(-\cosh^2(\r)d\theta^2+d\r^2+\sinh^2(\r)d\Omega_{d+n-1}^2\Big)\,.
\label{j26}\ee
The usual $AdS$ is obtained by extending the ``time" $\theta$ from $[0,2\pi]$ to the whole real line.
We summarize the embedding map of $AdS$ in global coordinates to the $(2,d+n)$ Minkowski space
\be
x^0=\ell\cosh(\r)\cos\theta ~,~ x^{(-1)}=\ell\cosh(\r)\sin\theta ~,~ x^i=\ell\sinh(\r) ~n^i\sp \rho\geq 0\,.
\label{j27}
\ee
The global boundary of $AdS$ is $\rho\to \infty$ that corresponds to
\be
r_1=\Big({(x^{0})^2+(x^{(-1)})^2}\Big)^\frac12 \to \infty\sp r_2=\Big(\sum_{i=1}^{n+d}x^ix^i\Big)^\frac12 \to \infty\,,
\ee
with their ratio $\frac{r_1}{ r_2}$ fixed.
Indeed the topology of the boundary is $S^1\times S^{d+n-1}$.

\subsection{Coordinates fibered over $AdS_{d}\times S^n$}

We now introduce new coordinates for the same space.
We can separate the variables in $x^{\m}$ with $\m=-1,0,1,2,...,d$ and $y^i$, $i=1,2,...,n$
and  we parametrize the first set by $AdS_d$ and the second by $S^n$
\be
x^{\m}=r m^{\m}\sp m\cdot m=-1\sp y^i=\rho n^i\sp n\cdot n=1\,,
\label{j15}
\ee
where $r, \r\geq 0$, and the flat $(2,d+n)$ metric is
\be
ds^2=-dr^2+r^2 ds^2_{AdS_{d}}+d\rho^2+\rho^2d\Omega_n^2\,.
\label{j16}\ee
The $AdS_{n+d+1}$ constraint is
\bsq
\begin{gather}\label{j17}
-r^2+\r^2=-\ell^2\ar \\
r=\ell\cosh(u)\sp  \r=\ell\sinh(u)\sp  u>0\,.
\label{j14}
\end{gather}
\esq
The induced metric becomes
\be
ds^2_{n+d+1}=\ell^2\Big(du^2+\cosh^2(u)ds^2_{AdS_d}+\sinh^2(u)d\Omega_n^2\Big)\sp u\geq 0\,,
\label{j19}\ee
and has one patch $u>0$.\footnote{
In general, if we have a CFT on $AdS_d\times S^n$ the physics should depend on the ratio of the two radius scales. Here this ratio is set to one.}

There is a related coordinate system where we map
\be
\sinh(u)=\tan(\phi)\sp du=\frac{d\phi}{ \cos(\phi)}\sp \phi\in \big[0, \frac{\pi}{2}\big]\,,
\label{j28}
\ee
and the new metric becomes
\be
ds^2=\frac{\ell}{\cos^2(\phi)}\Big(d\phi^2+\sin^2(\phi)d\Omega^2_{n}+ds^2_{AdS_{d}}\Big)\,,
\label{j29}
\ee
which is conformal to $AdS_{d}\times S^{n+1}$. However as  $ \phi\in \big[0, \frac{\pi}{ 2}\big]$, we have only one hemisphere of $S^{n+1}$.

We now write explicitly (\ref{j15})
\bsq
\begin{gather}
x^0=r_3\cosh(\r)\cos\theta\sp x^{(-1)}=r_3\cosh(\r)\sin\theta\sp x^i=r_3\sinh(\r) ~n^i\,,
\label{j30a} \\
i=1,2,\cdots, d-1\sp n\cdot n=1\sp \rho\geq 0\,,
\label{j30}\\
y^i=r_4 m^i\sp i=1,2,\cdots,n+1\sp m\cdot m=1\,,
\label{j31a}\\
r_3=\ell\cosh(u)\sp r_4=\ell\sinh(u)\sp u\geq 0\,.
\label{j31}
\end{gather}
\esq
Consider first the limit $u\to\infty$. In this case
\be
A~~:~~\Big((x^{0})^2+(x^{(-1)})^2\Big)^\frac12\to \infty ~,~ \Big(\sum_{i=1}^{d}(x^i)^2\Big)^\frac12 \to \infty ~,~  \Big(\sum_{i=1}^{n}(y^i)^2\Big)^\frac12\to \infty\,,
\label{j33}
\ee
in the same way but this is part of the original boundary of $AdS_{d+n+1}$.
It does not contain the limit where $\big(\sum_{i=1}^{d}(x^i)^2\big)^\frac12\to \infty$ and $ \big(\sum_{i=1}^{n}(y^i)^2\big)^\frac12$ remains finite or when
$ \big(\sum_{i=1}^{n}(y^i)^2\big)^\frac12\to \infty $ and $\big(\sum_{i=1}^{d}(x^i)^2\big)^\frac12$  remains finite.
Therefore the part of the boundary obtained by $u\to\infty$ is $S^1\times S^{d-1}\times S^{n-1}\subset S^1\times S^{d+n-1}$.
This piece also includes the special limit
\be
~~B~~:~~\Big((x^{0})^2+(x^{-1})^2\Big)^\frac12 \to \infty ~,~ \Big(\sum_{i=1}^{d}(x^i)^2\Big)^\frac12\rightarrow{\rm finite} ~,~  \Big(\sum_{i=1}^{n}(y^i)^2\Big)^\frac12\to \infty\,,
\label{j34}
\ee
or equivalently
\be
u\to \infty\sp \rho\to 0\sp \rho e^{u} \rightarrow {\rm finite}\,.
\label{j35}
\ee
Now the boundary of $AdS_d$ is when $\rho\to\infty$. In terms of the embedding coordinates
\be
~~C~~:~~\Big((x^{0})^2+(x^{-1})^2\Big)^\frac12\to \infty ~,~ \Big(\sum_{i=1}^{d}(x^i)^2\Big)^\frac12 \to \infty ~,~ \Big(\sum_{i=1}^{n}(y^i)^2\Big)^\frac12 \rightarrow {\rm finite}\,.
\label{j36}\ee
This completes the missing piece of the boundary of $u\to\infty$.
The topology of the three boundary pieces is
\be
A+B=S^1\times S^{d-1}\times S^{n-1}\sp C=S^1\times S^{d-1}\sp \partial (AdS_{d+n+1})=A\cup C\,.
\label{j37}
\ee

\subsection{The special case $n=0$}

Consider now the special case $n=0$.
In that case, the parametrization is
\bsq
\begin{gather}
x^0=r_3\cosh(\r)\cos\theta\sp x^{(-1)}=r_3\cosh(\r)\sin\theta \sp x^i=r_3\sinh(\r) ~n^i\,,
\label{j32}
\\
i=1, 2, \cdots, d-1 \sp n\cdot n=1\sp \rho\geq 0\,,
\label{j38}\\
y=r_4\sp r_3=\ell\cosh(u)\sp r_4=\ell\sinh(u)\sp u\in R\sp y\in R\,.
\label{j33n}
\end{gather}
\esq
The metric is now
\be
ds^2_{d+1}=\ell^2\left(du^2+\cosh^2(u)ds^2_{AdS_d}\right)\spp ~ u\in R\,.
\label{j39}\ee
Consider first the limit $u\to\pm \infty$. This is embedding coordinates correspond to
\be
A~~:~~\Big((x^{0})^2+(x^{(-1)})^2\Big)^\frac12\to \infty \sp \Big(\sum_{i=1}^{d}(x^i)^2\Big)^\frac12\to \infty\sp  y\to \pm\infty\,,
\label{j40}
\ee
and $A$ is topologically $S^1\times S^{d-1}\times S^0$. This also includes
\be
B~~:~~\Big((x^{0})^2+(x^{-1})^2\Big)^\frac12\to \infty \sp \Big(\sum_{i=1}^{d}(x^i)^2\Big)^\frac12 \rightarrow {\rm finite}\sp  y\to \pm\infty\,.
\label{j41}
\ee
This is obtained as the two limits
\be
u\to \pm \infty\sp \rho\to 0\sp e^{\pm u}\rho \rightarrow {\rm finite}\,.
\label{j42}
\ee
Here $C$ is
\be
C~~:~~\Big((x^{0})^2+(x^{-1})^2\Big)^\frac12\to \infty \sp \Big(\sum_{i=1}^{d}(x^i)^2\Big)^\frac12\to \infty\sp  y\rightarrow {\rm finite}\,,
\label{j44}
\ee
and is located at the boundary $\rho\to \infty$ of the slice.

It should be stressed that all coordinate systems above, are by construction, global.

\section{Analytic solutions for other signatures}

In this appendix, we present two more analytic solutions that exist if the signature of the metric is changed.

\subsection{The uniform solution}

This solution  is obtained by setting
\be
e^{2A_1(u)}=\e_1 ~e^{2A(u)}\sp e^{2A_2(u)}=\e_2~e^{2A(u)}\,,
\label{r3}\ee
where $\e_{1,2}$ are constants.
Then, the equations (\ref{j2}), (\ref{j3})  and (\ref{j4}) without the scalar field become
\begin{gather}
(d+n-1)(d+n)\big(\dot A^2-\frac{1}{ \ell^2}\big)=(\bar R_1+\bar R_2)e^{-2A}\sp \bar R_{1,2}\equiv \frac{R_{1,2}}{ \e_{1,2}}\,,
\label{r4}\\
(d+n-1)(d+n)\ddot A+(\bar R_1+\bar R_2)e^{-2A}=0\,,
\label{r5}\\
\bar R_1=\frac{d}{ n}\bar R_2\,,
\label{r6}
\end{gather}
and we have set $V=-\frac{(d+n-1)(d+n)}{ \ell^2}$.
Adding the two first equations we obtain
\be
\ddot{(e^A)}-\frac{e^A}{ \ell^2}=0\,,
\label{r7}\ee
with general solution
\be
e^A=C_1 e^{-\frac{u}{\ell}}+C_2 e^{\frac{u}{\ell}}\,,
\label{r8}\ee
Then  equation (\ref{r4})  becomes
\be
\dot A^2 e^{2A}-\frac{e^{2A}}{ \ell^2}=\frac{(\bar R_1+\bar R_2)}{(d+n-1)(d+n)}=\frac{\bar R_2}{ n(d+n-1)}\,,
\label{r9}\ee
which implies
\be
C_1C_2=-\frac{\ell^2 \bar R_2}{ 4n(d+n-1)}\,.
\label{r10}\ee
We can therefore write the general solution as
\be
e^{A}=e^{A_0}\left[e^{-\frac{u}{ \ell}}-\frac{\ell^2 \bar R_2 }{ 4e^{2A_0}n(d+n-1)}e^{\frac{u}{ \ell}}\right]\,.
\label{r11}\ee
The behavior of this solution as $u\to -\infty$ does not depend on the various arbitrary constants that appear in this solution.
However, such constants affect other properties of the solution.

Since $R_1<0$ and $R_2>0$, in order for (\ref{r6}) to have a non-trivial solution we must take $\e_1<0,\e_2>0$ or vice versa.
In the first case $\bar R_{1,2}>0$, while in the second case $\bar R_{1,2}<0$

$\bullet$ If $\bar R_2>0$ then the scale factor vanishes at a finite value $u=u_0$. This is a curvature singularity of the metric.
Moreover, in this case, the whole $AdS_d$ part of the metric has a minus sign.

$\bullet$ If $\bar R_2<0$ then the scale factor is regular and there is a second $AdS$ boundary at $u\to +\infty$. The solution describes a regular wormhole.
In such a case the $S^n$ part of the metric has a negative sign.

\subsection{The constant $A_2$ solution}

If we set $A_2=\bar A_2$ constant the equations for $A_1$ become
\be
d(d-1) (\dot{A_1})^2 - e^{-2A_1}
R_1-\bar R_2  -\frac{(d+n-1)(d+n)}{ \ell^2}  = 0\sp  \bar R_2\equiv e^{-2\bar A_2}R_2\,,
\ee
\be
 (d+n-1)d (\ddot{A_1}+(\dot{A_1})^2 ) -d(d-1) (\dot{A_1})^2  + e^{-2A_1} R_1 +\bar R_2 = 0\,.
\ee
Again we can deduce that
\be
\ddot{e^{A_1}}-\frac{d+n }{ d\ell^2}e^{A_1}=0\,,
\ee
with general solution
\be
e^{A_1}=C_1 e^{-\frac{u }{ \hat\ell}}+C_2e^{\frac{u }{ \hat\ell}}\sp \hat \ell\equiv \sqrt{\frac{d }{ d+n}}~\ell\,.
\ee
The first equation becomes
\be
d(d-1)\left(\frac{d }{ du}e^{A_1}\right)^2-\left(\bar R_2+\frac{(d+n-1)(d+n) }{ \ell^2}\right)e^{2A_1}=R_1\,,
\ee
which is satisfied if we choose $A_2$ and $C_2$  so that
\be
\bar R_2=-\frac{n(d+n)}{ \ell^2}\sp C_2=-\frac{\ell^2 R_1 }{ 4(d+n)(d-1)C_1}\,,
\label{r2}\ee
and the solution is
\be
e^{A_1}=e^{\bar A_1}\left[e^{-\frac{u }{ \bar\ell}}-\frac{\ell^2 \bar R_1 }{ 4(d+n)(d-1)}e^{\frac{u }{ \bar\ell}}\right]\sp \bar R_1\equiv e^{-2\bar A_1}R_1\,,
\ee
where we set $C_1=e^{\bar A_1}$.
For this solution to exist we must take the contribution of the sphere to the metric to be with a negative  signature so that
(\ref{r2}) be satisfied.

\section{The stress-energy tensor}\label{SET}
The vev of the stress-energy tensor is related to the constant $C$ that appears in the Fefferman-Graham expansion of the metric near the boundary, for example, see the expansions \eqref{near1} and \eqref{near2}. To show this, here for simplicity we restrict ourselves to the $d=n=2$ case.  For an asymptotically $AdS$ space-time the metric near the boundary can be brought into the form
\be\label{se0}
ds^2=du^2+\ell^2 e^{-\frac{2u}{\ell}} g_{ij}(u,x) dx^i dx^j\,,
\ee
where $g_{ij}$ has the following expansion near the boundary when $u\rightarrow +\infty$
\be \label{gij}
g_{ij}(u,x) = g^{(0)}_{ij}(x)+ e^{\frac{2u}{\ell}}g^{(2)}_{ij}(x)+ e^{\frac{4u}{\ell}}\big(g^{(4)}_{ij}(x)+\frac{2u}{\ell} h^{(4)}_{ij}(x)\big)+\cdots\,,
\ee
where $g^{(0)}_{ij}(x)$ corresponds to the boundary condition for the metric. Since we have the second-order equations of
motion, the two independent functions are $g^{(0)}_{ij}(x)$ and $g^{(4)}_{ij}(x)$ which the latter is related to the expectation value of the stress-energy tensor of the dual theory.
 The other functions, $g^{(2)}_{ij}(x)$ and $h^{(4)}_{ij}(x)$ are determined in terms of $g^{(0)}_{ij}(x)$
\bsq
\begin{gather}\label{se1}
g^{(2)}_{ij} = \frac12 R_{ij}-\frac{1}{12} R g^{(0)}_{ij}\,,
\\ \label{se2}
g^{(4)}_{ij} = \frac18 g^{(0)}_{ij}\big[(Tr g^{(2)})^2-Tr[(g^{(2)})^2]\big]+\frac12 (g^{(2)})^2_{ij}-\frac14 Tr[g^{(2)}] g^{(2)}_{ij}+T_{ij}\,, \\ \label{se3}
h^{(4)}_{ij} =\frac{1}{16\sqrt{g^{(0)}}}\frac{\delta}{\delta g^{(0){ij}}}\int d^4x\sqrt{g^{(0)}}(R_{ij}R^{ij}-\frac13 R^2)\,,
\end{gather}
\esq
where the integrand in the last term, is the conformal anomaly in $d+n=4$ dimensions \cite{Henningson:1998gx, Skenderis:2002wp}.

To read the $T_{ij}$ we first compute $g^{(2)}_{ij}$ and the first part of $g^{(4)}_{ij}$ by using
\be \label{se4}
g^{(0)}_{ij} dx^i dx^j = e^{2\bar{A}_1}\zeta^{(1)}_{\a\b} dx^{\a} dx^{\b} + e^{2\bar{A}_2}\zeta^{(2)}_{\m\n} dx^{\m} dx^{\n}\,,
\ee
and we find (we set $\bar{A}_1=\bar{A_2}=0$)
\bsq
\begin{gather} \label{se5}
g_{\a\b}^{(2)} = \frac{1}{144}(2R_1-R_2)^2 \zeta^{(1)}_{\a\b} \sp  g_{\m\n}^{(2)} = \frac{1}{144}(R_1-2R_2)^2 \zeta^{(2)}_{\m\n}\,, \\
g_{\a\b}^{(4)} = \frac{1}{576}(R_1+R_2)^2 \zeta^{(1)}_{\a\b}+T_{\a\b}\sp g_{\m\n}^{(4)} = \frac{1}{576}(R_1+R_2)^2 \zeta^{(2)}_{\m\n}+T_{\m\n}\,, \label{g41}
\end{gather}
\esq
On the other hand, we can compute the scale factors from equations of motion. The results for $d=n=2$ are given by
\bsq
\begin{align} \label{se6}
A_1 &= \log a_0 - \frac{u}{\ell} + a_2 e^{\frac{2u}{\ell}}+  a_4 e^{\frac{4u}{\ell}} +  a_5 \frac{u}{\ell} e^{\frac{4u}{\ell}}+\cdots\,, \\
A_2 &= \log s_0 - \frac{u}{\ell} + s_2 e^{\frac{2u}{\ell}}+  s_4 e^{\frac{4u}{\ell}} +  s_5 \frac{u}{\ell} e^{\frac{4u}{\ell}}+\cdots\,,\label{se7}
\end{align}
\esq
with coefficients $(a_0=e^{\bar{A}_1}, s_0=e^{\bar{A}_2})$
\bsq
\begin{align} \label{se9}
a_2 &= -\frac{\ell^2 }{24} \big(\frac{2 R_1}{a_0^2} - \frac{R_2}{s_0^2}\big)\sp
s_2 = \frac{\ell^2}{24}\big(\frac{R_1}{a_0^2} - \frac{2R_2}{s_0^2}\big)\,, \\
a_4 &= -\frac{\ell^4\big(5 a_0^4 R_2^2 - 8 a_0^2 R_1 R_2 s_0^2 + 5 R_1^2 s_0^4\big)}{2304 a_0^4 s_0^4}-C\,, \\
s_4 &= -\frac{\ell^4\big(5 a_0^4 R_2^2 - 8 a_0^2 R_1 R_2 s_0^2 + 5 R_1^2 s_0^4\big)}{2304 a_0^4 s_0^4}+C\,, \\
s_5 &= -a_5 = -\frac{\ell^4}{192} \big(\frac{R_1^2}{a_0^4} - \frac{R_2^2}{s_0^4}\big)\,.\label{se10}
\end{align}
\esq
Similar to $d+n=8$ in \eqref{near1} and \eqref{near2} the $\frac{u}{\ell}e^{\frac{4u}{\ell}}$ terms in \eqref{se6} and \eqref{se7} are the conformal anomalous terms in $d+n=4$.

From the  above expansions we can read $g_{ij}^{(4)}$ from the near boundary expansion \eqref{gij}
\bsq
\begin{align}\label{g42}
g_{\a\b}^{(4)} &= \Big[\frac{\ell^4}{1152} (11 R_1^2 - 8 R_1 R_2 - R_2^2)-2C \Big] \zeta^{(1)}_{\a\b}\,, \\
g_{\m\n}^{(4)} &= \Big[\frac{\ell^4}{1152} (11 R_2^2 - 8 R_1 R_2 - R_1^2)+2C \Big] \zeta^{(2)}_{\m\n}\,.\label{g43}
\end{align}
\esq
By comparing the results of \eqref{g41} with \eqref{g42} and \eqref{g43} we can read the stress-energy tensor components as (again we assume $\bar{A}_1=\bar{A}_2=0$)
\bsq
\begin{align}\label{TEMC1}
T_{\a\b}= \frac{1}{384} \left(3 R_1^2-4 R_1 R_2-R_2^2-768 C\right)\zeta^{(1)}_{\a\b}\,, \\
T_{\m\n}= \frac{1}{384} \left(3 R_2^2-4 R_1 R_2-R_1^2+768 C\right)\zeta^{(2)}_{\m\n}\,.\label{TEMC2}
\end{align}
\esq
It turns out that $T_{ij}$ and therefore $C$ is proportional to the vev of the stress-energy tensor of the boundary CFT
\be\label{EXPT}
T_{ij}=\frac{1}{4(M_P \ell)^3} \langle T_{ij} \rangle\,.
\ee
Therefore we can write
\be \label{EXPT1}
\langle T_{ij} \rangle=4(M_P \ell)^3\Bigg[\frac{T_{CFT}}{4}\Bigg(\begin{matrix} \zeta^{(1)}_{\a\b}&0\\0&\zeta^{(2)}_{\m\n} \end{matrix}\Bigg)+\hat{T}_{CFT}\Bigg(\begin{matrix} \zeta^{(1)}_{\a\b}&0\\0&-\zeta^{(2)}_{\m\n} \end{matrix}\Bigg)\Bigg]\,,
\ee
where the trace part ${T}_{CFT}$ and traceless part  $\hat{T}_{CFT}$ are defined
\bsq
\begin{gather}\label{EXPT2}
T_{CFT}=\frac{1}{96} \left(R_1^2-4 R_1 R_2+R_2^2\right)\,, \\
\hat{T}_{CFT}=\frac{1}{96}\big(\frac12 R_1^2-\frac12 R_2^2-48 C\big)\,.\label{EXPT3}
\end{gather}
\esq
\section{Perturbations around the product space solution}\label{EFL}

We consider a solution that is a perturbation around the product space solution. For simplicity, in notation, we choose a new variable
\be
z\equiv\sqrt{\frac{d+n}{n}} \frac{(u-u_0)}{\ell}\,.
\ee
The scale factors are defined as follows
\be \label{PE0}
A_1(z)=A^{(0)}_1(z)+\delta A_1(z)\sp
A_2(z)=A^{(0)}_2(z)+\delta A_2(z)\,,
\ee
where
\bsq
\begin{gather} \label{PE1}
A^{(0)}_1(z)=\frac12 \log\Big[-\frac{\ell^2 R_1}{d (d+n)}\Big]\,,\\
A^{(0)}_2(z)=\frac12\log\Big[\frac{\ell^2 R_2}{(n-1)(d+n)}\sinh^2(z)\Big]\,,\label{PE2}
\end{gather}
 \esq
are the product space scale factors.
We can insert \eqref{PE0} into the equation of motion \eqref{j2} and read $\delta A'_2(z)$ and $\delta A''_2(z)$. Then by substituting these derivatives into either \eqref{j3} or \eqref{j4} we find the following differential equation for $\delta A_1(z)$
\be \label{PE3}
-2n\delta A_1+n\coth(z)\delta A'_1 +\delta A''_1=0\,.
\ee
The solution for this equation is
\begin{align}\label{PE4}
\delta A_1 &=\frac{C_1}{ \cosh^{\frac{m+n}{2}}(z)} \, _2F_1\left(\frac{m+n}{4},\frac{1}{4} (m+n+2);\frac{n+1}{2};\tanh ^2(z)\right)\nn \\
&+ \frac{C_2 \tanh ^{1-n}(z)}{ \cosh^{\frac{m+n}{2}}(z)} \, _2F_1\left(\frac{1}{4} (m-n+2),\frac{1}{4} (m-n+4);\frac{3-n}{2};\tanh ^2(z)\right)\,,
\end{align}
where $C_1$ and $C_2$ are two constants of integration and we have defined
\be
m\equiv\sqrt{n(n+8)}\,.
\ee
Expanding around the end-point $u=u_0$ or equivalently $z=0$ we read
\be
\delta A_1= C_1 \big(1+\frac{n z^2}{n+1}+\mathcal{O}(z^4)\big) + C_2 z^{-n} \big(z-\frac{n (n+5) z^3}{6 (n-3)}+\mathcal{O}(z^4)\big)\,.
\ee
This expansion shows that in order the scale factor of $AdS$ i.e. $e^{2A^{(0)}_1(u)+2\delta A_1(u)}$ be finite as $z\rightarrow 0$ we should choose
\be
C_2=0\,.
\ee
We can also expand $\delta A_1$ near the UV boundary as $z\rightarrow +\infty$
\be
\delta A_1= C_1 \frac{(m-2) 2^{n-2} \Gamma \left(\frac{m}{2}-1\right) \Gamma \left(\frac{n+1}{2}\right)}{\sqrt{\pi } \Gamma \left(\frac{m+n}{2}\right)}\left(e^{z}\right)^{\frac{m-n}{2}}+\cdots\,,
\ee
which means that although the fluctuations are small near the IR end-point but grow exponentially as $z$ moves toward the UV boundary.

The equation of motion for $\delta A_2$ in terms of the new variable $z$ is given by
\be \label{PE5}
\coth (z) \left(d \delta A'_1+(n-1) \delta A'_2\right)-d \delta A_1+(n-1)\text{csch}^2(z) \delta A_2 =0\,.
\ee
The solution is obtained by
\be \label{PE6}
\delta A_2= C_3 \coth(z)+\coth(z)\int_{1}^{z}
\frac{d}{n-1}
\tanh (w) \left(\tanh(w) \delta A_1 - \delta A'_1\right) dw\,,
\ee
where $C_3$ is another constant of integration.
Equation \eqref{PE6} is hard to solve but to see the series expansion of $\delta A_2$ we can solve \eqref{PE5} near $z=0$. The series is
\be
\delta A_2=-C_1 (\frac{d}{3(n+1)}z^2+
\frac{ d (8 n-3)}{45 (n+1) (n+3)} z^4+\cdots)\,.
\ee
Here the constant of integration $C_3$ in \eqref{PE6} is related to $C_1$ to have a regular solution for the scale factor of the sphere $A_2$.
Moreover, near the UV as $z\rightarrow +\infty$ we have
\be
\delta A_2=C_1 \frac{ 2^{n-2} d\,(m-n-2) \Gamma \left(\frac{m}{2}\right) \Gamma \left(\frac{n-1}{2}\right)}{\sqrt{\pi } (n-m) \Gamma \left(\frac{m+n}{2}\right)}
\left(e^{z}\right)^{\frac{m-n}{2}}+\cdots\,,
\ee
however, here unlike the $\delta A_1$, the fluctuations remain small compared to the leading term which is growing like $e^{2z}$ because
\be
2> \frac{m-n}{2}\sp \text{for} \quad n>0\,.
\ee

\section{Topological Black holes with a negative cosmological constant\label{topo}}

We consider solutions to the Einstein equation  with a negative cosmological constant
\be
G_{\m\n}=\frac{d(d+1)}{ \ell^2}\,,
\label{F10}\ee
in $d+2$ dimensions, with an ansatz
\be
ds^2=-f(r)dt^2+\frac{dr^2}{ f(r)}+r^2h_{ij}(x)dx^i dx^j\,.
\label{F11}\ee
\be
f = k -\frac{\omega_d M}{ r^{d-1}} +
\frac{r^2}{ \ell ^2}\sp
\omega_{d} =
\frac{16 \pi G
}{d \,\textit{Vol}(h)}\sp \textit{Vol}(h)=\int d^{d}x \sqrt{h}\,,
\label{12}\ee
and $h_{ij}$ is a constant curvature metric
\be
R_{ij}(h)=(d-1)k~ h_{ij}\,.
\label{13}\ee
If the constant curvature manifold $h$ is maximally symmetric, then
\be
R_{ijkl}(h) = k(h_{ik} h_{jl} - h_{il} h_{jk})\,.
\label{14}\ee
The $M=0$ solution above is isomorphic to a maximally symmetric constant curvature space satisfying
\be
R_{\m\n\r\s}=-\frac{1}{ \ell^2}
(g_{m\r}g_{\n\s}-g_{\m\s}g_{\n\r})\,.
\label{15}\ee
Therefore, the solution with $M=0$ is locally isometric to AdS$_{d+2}$ space, but the topology depends on the sign of $k$.
The boundary is conformally equivalent to $AdS_{d}\times S^1$ if we take the $t$ coordinate to be an angle $t\in[0,2\pi]$.

The metric above is invariant under the following rescaling
\be
t\rightarrow \frac{t}{ \l}\sp r\to \l ~r\sp k\to \l^2 k\sp M\to \l^{d+1}M\sp h_{ij}\to \frac{h_{ij}}{ \l^2}\,.
\label{16}\ee
By choosing $\l=\frac{1}{ \sqrt{|k|}}$ when $k\not=0$ the metric can be written as
\be
f = \e -
\frac{\omega_d M}{ r^{d-1}} +\frac{r^2}{ \ell ^2}\sp\e=0,\pm 1\sp\omega_{d} =
\frac{16 \pi G}{ d\textit{Vol}(h)}\,\, \,,\,\, \textit{Vol}(h)=\!\!\int d^{d}x \sqrt{h}\,,
\label{16a}
\ee
and $h_{ij}$ is a constant curvature metric with
\be
R_{ij}(h)=(d-1) h_{ij}\,.
\label{17}\ee
In the maximally symmetric case, the horizon surface can be an $S^{d}$ or any quotient, $T^3$ or any quotient, or a compact quotient of $AdS_3$.

We now take $d=3$ and set $k=-|k|$.
The equation for the horizon position is
\be
-|k|\ell^2 r^2 -
{\omega_3 M\ell^2 } +
{r^4}=0~~~\to~~~ r_{\pm}^2=\ell^2 \frac{|k|\pm \sqrt{k^2+4\frac{\omega_3 M}{ \ell^2} }}{ 2}\,.
\label{18}\ee
As $r=0$ is curvature singularity, we are interested in solutions with $r>0$.

We can distinguish the following cases
\begin{enumerate}

\item $\omega_3 M<- \frac{k^2\ell^2}{ 4}\equiv \omega_3 M_{crit}$.

In this case, the solutions are complex and there is no horizon.

\item  $\omega_3 M=- \frac{k^2\ell^2}{ 4}\equiv \omega_3 M_{crit}$.

We have a double root $r_+=r_-=\ell\sqrt \frac{|k|}{ 2}$.
This is an extremal horizon.

\item  $0>\omega_3 M>-\frac{k^2\ell^2}{ 4}$. There are two distinct real roots with $r_+$ the largest. This is a case similar to the RN black holes and $r_-$ is a Cauchy horizon.

\item $M=0$.  In this case $r_+=\ell\sqrt{|k|}$ while $r_-=0$. Now $r=0$ is not anymore a curvature singularity and the solution is now locally $AdS_5$

\item $M>0$.
In this case, there is a single root $r_+>0$ and the black hole structure is as in Schwarzschild.

Parametrizing
\be
M =
\frac{r_+^{d-1}}{ \omega_d}
\left(\frac{r_+^2}{ \ell^2}-|k|\right)\,,
\label{19}\ee
the temperature  is given in general $d$ as
\be
T =\frac{(d+1)r_+^2-(d-1)|k|\ell^2}{ 4\pi \ell^2 r_+}\,.
\label{20}\ee
When $T=0$,
\be
r_+^2=\frac{(d-1)}{d+1}|k|\ell^2\equiv r_{crit}^2\,,
\ee
and
\be
M=M_{crit}=-\frac{2}{ (d+1) \omega_d}\left(\frac{d-1}{d+1}\right)^{\frac{d-1}{ 2}}|k|^{\frac{d+1}{2}}\ell^{d-1}\,.
\label{21}\ee
Using the extremal solution $M=M_{crit}$ as the reference solution, we can write the energy and entropy as
\be
E=M-M_{crit}\sp S=\frac{\textit{Vol}(h)~r_+^{d}}{ 4G}\,.
\label{22}\ee
The specific heat is
\be
\frac{\pa E}{ \pa T}=\frac{4\pi r_+^{d-1}}{ \omega_d}\frac{(d+1) r_+^2-(d-1)|k|\ell^2}{ (d+1) r_+^2-(d-1)|k|\ell^2}=\frac{4\pi r_+^{d-1}}{ \omega_d}\frac{r_+^2-r_{crit}^2}{ r_+^2+r_{crit}^2}\,.
\label{23}\ee
It is clear that for $M>M_{crit}$ all solutions are thermodynamically stable.

More details, as well as the analysis of the thermodynamics and possible phase transitions, can be found in \cite{top1,top2}.

\end{enumerate}

\end{appendix}


\end{document}